\documentclass[12pt,a4paper]{book}

\usepackage{fancyhdr}
\usepackage{amsmath}
\usepackage{amssymb}
\usepackage{pstricks}
\usepackage{pst-3d}
\usepackage{epsfig}
\usepackage[english]{minitoc}
\usepackage{makeidx}

\def\Unitmatrix{\mbox{1\hspace{-.25em}I}}
\def\Z{\mbox{Z\hspace{-.3em}Z}}

\newcommand{\myappendix}[0]{
\appendix
\renewcommand{\chaptername}{Appendix}
\fancyhead{}
\fancyhead[CE,CO]{\textbf{\leftmark}}
\fancyhead[LE,RO]{\textbf{\thepage}} }

\makeindex

\begin{document}



\thispagestyle{empty}
\setlength{\voffset}{-1cm}
\setlength{\textheight}{25cm}
\setlength{\footskip}{0cm}
\setlength{\paperwidth}{0cm}

\noindent
\epsfig{file=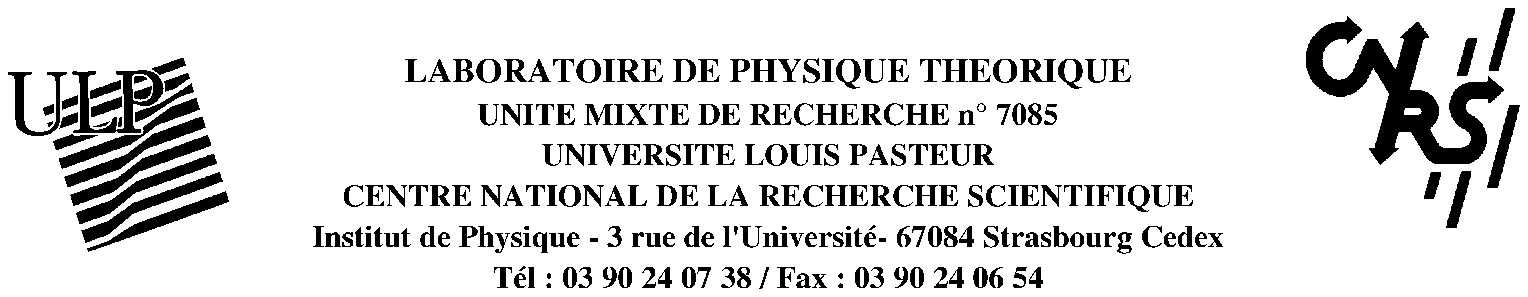}

\vspace*{1.5cm}
\noindent
\rule{\textwidth}{.2cm}
\begin{center}
\Huge
\textbf{
Effects of a \emph{strict} site-occupation constraint in the
description of quantum spin systems at finite temperature
}
\end{center}
\rule{\textwidth}{.2cm}

\vspace*{1cm}
\begin{center}
{\LARGE
\bfseries
Th\`ese}

\vspace*{1cm}
pour l'obtention du grade de 
DOCTEUR DE L'UNIVERSIT\'E LOUIS PASTEUR STRASBOURG I \\
Discipline : Physique Th\'eorique \\

\vspace*{0.5cm}
Pr\'esent\'ee par \\

\vspace*{0.5cm}
{\LARGE
Raoul Dillenschneider}\\

\vspace*{0.5cm}
Le 8 Septembre 2006 \`a Strasbourg \\

\vspace*{1cm}
\textbf{Directeur de th\`ese} \\
Jean Richert, Directeur de Recherche \\

\vspace*{0.5cm}
\textbf{Jury} \\
\begin{tabular}{lcl}
Rapporteur interne & : & Daniel Cabra, Professeur \\
Rapporteur externe & : & Fr\'ed\'eric Mila, Professeur \\
Rapporteur externe & : & Pierre Pujol, Ma\^{i}tre de conf\'erences \\
Examinateur & : & Malte Henkel, Professeur \\
Examinateur & : & Michel Rausch de Traubenberg, Ma\^{i}tre de conf\'erences \\
\end{tabular}

\end{center}

\newpage
\setlength{\voffset}{0cm}
\setlength{\textheight}{23cm}
\setlength{\footskip}{30pt}
\setlength{\paperwidth}{597pt}
\thispagestyle{empty}
\newpage

\chapter*{Abstract}
\thispagestyle{empty}


We study quantum spin systems described by Heisenberg-like models at finite 
temperature with a strict site-occupation constraint imposed by a procedure 
originally proposed by V. N. Popov and  S. A. Fedotov
\cite{Popov-88}. We show that the strict 
site-occupation constraint modifies quantitatively the behaviour of physical
quantities when compared to the case for which this constraint is fixed in the
average by means of a Lagrange multiplier method. The relevance of the N\'eel 
state with the strict site-occupation contraint of the spin lattice is studied.
With an exact site-occupation the transition temperature of the 
antiferromagnetic N\'eel and spin liquid order parameters are twice as large 
as the critical temperature one gets with an average Lagrange multiplier 
method. We consider also a mapping of the low-energy spin Hamiltonian into a 
$QED_3$ Lagrangian of spinons. In this framework we compare the dynamically 
generated mass to the one obtained by means of an average site-occupation 
constraint.

\vspace*{2cm}

\noindent
{\bfseries \Huge R\'esum\'e}
\thispagestyle{empty}

\vspace{1.25cm}

\noindent
Nous \'etudions des syst\`emes de spin quantiques \`a temp\'erature finie
avec une contrainte d'occupation stricte des sites au moyen d'une proc\'edure
introduite par V. N. Popov et S. A. Fedotov
\cite{Popov-88}. Nous montrons que cette contrainte modifie 
le comportement d'observables physiques par rapport au cas o\`u cette
contrainte est fix\'ee de fa\c{c}on moyenne par la m\'ethode des 
multiplicateurs de Lagrange. La pertinence de l'\'etat de N\'eel est 
\'etudi\'ee en pr\'esence de la contrainte stricte d'occupation des sites du 
r\'eseau de spin. 
La temp\'erature de transition des param\`etres d'ordre antiferromagn\'etique
de N\'eel et d'\'etat de liquide de spins sont doubl\'es par rapport \`a ceux 
obtenu par la m\'ethode moyenne des multiplicateurs de Lagrange. Nous 
consid\'erons l'Hamiltonien de basse \'energie d\'ecrit par un Lagrangien de 
$QED_3$ pour les spinons. Dans ce contexte la masse g\'en\'er\'ee dynamiquement
est compar\'ee \`a celle obtenue par la m\'ethode d'occupation moyenne de site.

\newpage
\thispagestyle{empty}
\newpage

\chapter*{}
\thispagestyle{empty}

\vspace*{15cm}

\begin{flushright}
\textit{\bfseries
\`{a} mes parents Huguette et Bertrand\\
\`{a} mon fr\`{e}re Vivien
}
\end{flushright}

\newpage
\thispagestyle{empty}
\newpage

\chapter*{Acknowledgements}
\thispagestyle{empty}

First and foremost I would like to thank and express my gratitude to my 
supervisor Jean Richert for having been so considerate and motivating in all 
aspects of my research during these three years. I am greatly honored to have 
had the opportunity to work with him.

\vspace{0.5cm}
I am greatly thankful to Prof. Bertrand Berche for having permit me to work 
with Jean Richert and giving me his help.

\vspace{0.5cm}
I am grateful to the members of my Ph.D. jury : Prof. Daniel Cabra, Prof.
Fr\'ed\'eric Mila, Pierre Pujol, Prof. Malte Henkel and Michel Rausch de 
Traubenberg for finding time to be on my jury and also for helping me
to render this manuscript more meticulous.

\vspace{0.5cm}
A special thank to the teachers from the Universit\'e Henri Poincar\'e at
Nancy who taught me the beautiful physics and developed my wish to become a
researcher like them.

\vspace{0.5cm}
I would like to thank all people with whom I had discussions about physics 
subjects giving me the possibility to do this work.

\vspace{0.5cm}
I would like to thank my friends : Elodie Offner, Ang\'elique Dieudonn\'e, 
Luc Strohm, Jean-Christophe Brua, Gabriel Delhaye, Lo\^{\i}c Joly, 
Tarek Khalil, Frank Stauffer and Adrien Tanas\u{a}. 
It was and will always be sea, s\dots and fun !

\vspace{0.5cm}
Last but not the least I owe a lot to my parents Bertrand and Huguette
for their education, guidance and support.

\newpage
\thispagestyle{empty}
\newpage


\frontmatter

\fancyhead{}
\fancyhead[LE,RO]{\textbf{\thepage}}
\fancyhead[RE,LO]{\textbf{Contents}}
\dominitoc
\tableofcontents


\mainmatter

\fancyhead[RE]{\textbf{\leftmark}}
\fancyhead[LO]{\textbf{\rightmark}}
\fancyhead[LE,RO]{\textbf{\thepage}}
\fancyfoot{}


\chapter{R\'esum\'e}

\section{La proc\'edure de Popov et Fedotov (PFP)}

Apr\`{e}s la d\'ecouverte de la supraconductivit\'e par Bednorz et M\"{u}ller
\cite{BednorzMuller-86} un effort th\'eorique important fut consacr\'e \`{a}
la recherche d'une explication \`{a} ce ph\'enom\`{e}ne qui se manifeste \`{a}
une temp\'erature critique \'elev\'ee. De nombreux supraconducteurs furent 
d\'ecouverts parmi lesquels on retrouve les cuprates avec 
$La_{2-x} Sr_{x} Cu O_4$ et $Bi_2 Sr_2 Ca Cu_2 O_8$ qui sont des exemples de 
supraconducteurs dop\'es en trous, les ruth\'enates $Sr_2 Ru O_4$, les 
m\'etaux \`{a} fermions-lourds tels que $U Pt_3$ et $Zr Zn_2$, ainsi que les 
mat\'eriaux organiques comme $\kappa-(BEDT-TTF)_2-Cu[N(CN)_2]Br$.

Tous les cuprates partagent la m\^{e}me structure atomique
compos\'ee de couches de $Cu O_2$, tenues pour responsables de la formation 
de paires de Cooper et de la supraconductivit\'e, et intercal\'ees de couches
de substance dopante et/ou non-dopante.

Dans ce manuscrit nous nous concentrerons sur les couches $Cu O_2$ et plus
pr\'ecisement sur la phase isolante antiferromagn\'etique des supraconducteurs
haute temp\'erature. En effet, la phase isolante des mat\'eriaux comme
les cuprates peut \^{e}tre mod\'elis\'ee par le mod\`{e}le de Heisenberg.

L'\'etude de la phase isolante est motiv\'ee par le fait que les corr\'elations
sous-jacentes \`{a} la phase antiferromagn\'etique isolante pourrait se
prolonger dans la phase supraconductrice sous l'effet du dopage. En
d'autre termes nous attendons que la phase supraconductrice garde quelques
traits de caract\`{e}re de la phase antiferromagn\'{e}tique isolante. 

Notre travail est centr\'e sur l'\'{e}tude des effets de la contrainte
stricte d'occupation de site de spin caract\'erisant la phase isolante dans
la description de syst\`{e}me de spin quantique \`{a} temp\'{e}rature finie.
Cette contrainte consiste \`{a} imposer exactement un spin $S=1/2$ pour chaque
site de r\'eseaux.

L'impl\'ementation d'une telle contrainte fut introduite par Popov et Fedotov
\cite{Popov-88}. Contrairement \`{a} d'autre m\'ethodes bas\'ees sur 
l'utilisation de multiplicateurs de Lagrange 
\cite{ArovasAuerbach-88,Auerbach-94,ReadSachdev-91}, conduisant \`{a} une
occupation moyenne, la \textbf{p}roc\'edure de \textbf{P}opov et 
\textbf{F}edotov (\textbf{PFP}) \'evite ce traitement
approximatif au moyen de l'introduction d'un potentiel chimique imaginaire.
Nous portons notre int\'er\^{e}t \`{a} l'analyse et l'application de la
PFP sur des syst\`{e}mes antiferromagn\'etiques de spins \`{a} temp\'erature
finie, et nous la comparons aux r\'esultats obtenus par la m\'ethode des
multiplicateurs de Lagrange.

\section{Propri\'et\'es magn\'etiques du mod\`{e}le de Heisenberg par la PFP}

Le chapitre \ref{Chapter3} pr\'esente et discute l'application du champ
moyen et du d\'eveloppement en nombre de boucles pour la d\'etermination de
propri\'et\'ees physique de syst\`{e}mes antiferromagn\'etique de type
Heisenberg pour des dimensions D du r\'eseau \cite{DRepjb-05}.

Des travaux r\'ecents sur des syst\`{e}mes de spins quantiques discutent
l'existence possible d'\'{e}tats de liquide de spin, et pour deux 
dimensions d'espace de la comp\'etition ou transition de phase entre \'etats
liquide de spin et antiferromgn\'etique de N\'eel 
\cite{GhaemiSenthil-05,Morinari-05,SenthilFisher-05,SenthilFisher-04}.
Il est \'egalement connu que les supraconducteurs (cuprates) pr\'esentent
une phase antiferromagn\'etique \cite{LeeNagaosaWen-04}.

Dans le chapitre \ref{Chapter3} nous focalisons notre attention sur la
phase dont le champ moyen est de type N\'eel dans la description des 
syst\`{e}mes quantiques de spin repr\'esent\'es par le mod\`{e}le de 
Heisenberg.
Plus pr\'ecis\'ement, nous pr\'esentons une \'etude d\'etaill\'ee de
l'aimantation et de la susceptibilit\'e magn\'etique pour ces syst\`{e}mes
de r\'eseaux de spin en dimension $D$ et \`{a} temp\'erature finie.
Le but de ce travail est d'\'etudier la pertinence de l'\emph{ansatz} de N\'eel
comme approximation de champ moyen en utilisant la PFP et pour les intervalles
de temp\'eratures $0<T<T_c$ o\`{u} $T_c$ est la temp\'erature critique.
Dans l'objectif d'obtenir une r\'eponse sur ce point nous calculons les
contributions des fluctuations quantiques et thermiques au-del\`{a} de
l'approximation de champ moyen et sous la contrainte d'occupation stricte
d'un seul spin par site de r\'eseau 
\cite{Azakov-01,DRepjb-05,Dillen-05,KFO-01,Popov-88}.
Une comparaison est effectu\'ee entre nos r\'esultats et ceux obtenus par la 
th\'eorie des ondes de spins sur les mod\`{e}les de Heisenberg et $XXZ$.

Il est montr\'{e} que la PFP introduit un grand 
d\'ecalage de la temp\'erature critique par rapport \`{a} celle 
obtenue par la m\'ethode ordinaire des multiplicateurs de Lagrange. 
En effet, il apparait un doublement de la temp\'erature de transition entre 
un \'etat antiferromagn\'etique de N\'eel et la phase paramagn\'etique 
\cite{Azakov-01,Dillen-05}.
Nous montrons dans le chapitre \ref{Chapter3}, et c'est l\`{a} une des 
contribution originales de ce travail de th\`{e}se, qu'\`{a} basse
temp\'erature l'aimantation ainsi que la susceptibilit\'e magn\'etique sont
\'egalles en valeur \`{a} celles obtenues par la th\'eorie des ondes de spin,
comme le montre la figure \ref{fig:MagP3D100} du chapitre \ref{Chapter3}.
La figure \ref{fig:MagP3D100} repr\'esente l'aimantation en champ moyen
(en traits tiret\'es), l'aimantation avec fluctuation en utilisant la PFP
(en traits plein) et enfin l'aimantation calcul\'ee par la th\'eorie des
ondes de spins (en traits pointill\'es), pour des syst\`{e}mes 
tri-dimensionnels. La superposition des courbes \`{a} basse temp\'erature
apparait clairement sur cette figure. Le d\'ecalage entre l'aimantation 
moyenne et celle obtenue en consid\'erant les fluctuations est due \`{a}
l'effet intrins\`{e}que des fluctuations quantiques et thermiques sur la
moyenne statistique des orientations de spin. La PFP n'est pas responsable
de ce d\'ecalage.

\`{A} plus haute temp\'erature les contribution des fluctuations de nature
quantiques et thermiques croissent en une singularit\'e au voisinage de la
temp\'erature critique. L'hypoth\`{e}se que le champ moyen de N\'eel contribue
pour une majeure partie \`{a} l'aimantation et \`{a} la susceptibilit\'e n'est
plus valide. En approchant de la temp\'erature critique $T_c$ l'aimantation en
champ moyen tend vers z\'ero et les fluctuations croissent de plus en plus
\`{a} l'ordre d'une boucle. Ceci r\'esulte en une divergence de la valeur 
th\'eorique pr\'evue pour l'aimantation globale du syst\`{e}me de spins.
Ce comportement est commun aux mod\`{e}le de Heisenberg et $XXZ$. 
Nous en d\'eduisons que la brisure de sym\'etrie induite par le choix de 
l'\'etat de N\'eel n'est pas impliqu\'ee dans l'apparition des divergences 
calcul\'ees.

L'influence des fluctuations d\'ecroit avec l'augmentation de la dimension
$D$ du syst\`{e}me due \`{a} la r\'eduction des fluctuations par rapport au 
champ moyen.

En dimension $D=2$ l'aimantation v\'erifie effectivement le th\'eor\`{e}me
de Mermin et Wagner \cite{MerminWagner-66}. Pour $T \neq 0$, les fluctuations
sont plus importantes que la contribution du champ moyen.
Ainsi dans une description plus r\'ealiste de la physique il est n\'ecessaire 
de prendre en consid\'eration d'autre champ moyen. 
En effet, Ghaemi et Senthil \cite{GhaemiSenthil-05} ont montr\'e, \`{a} l'aide 
d'un mod\`{e}le sp\'ecifique, qu'une transition du second ordre d'un \'etat de
N\'eel vers un liquide de spins pourrait appara\^{i}tre en fonction de la
valeur des couplages. Ceci nous conduit au chapitre \ref{Chapter4} dans
lequel une \'etude de l'influence de la PFP est men\'ee sur diff\'erents 
\emph{ansatz} de champ moyen et pour des r\'eseaux bi-dimensionnels.

Le point originale de ce chapitre \ref{Chapter3} est l'utilisation de la
proc\'edure de Popov et Fedotov dans la calcul de propri\'et\'ees magn\'etique
\`{a} temp\'erature finie et pour un ordre \`{a} une boucle en perturbation
\cite{DRepjb-05}.

\section{Champ moyens pour le mod\`{e}le bidimensionnel de Heisenberg}

Nous r\'esumerons dans cette section le contenu du chapitre \ref{Chapter4}
o\`{u} nous consid\'erons la PFP appliqu\'ee \`{a} des syst\`{e}mes de spin
quantique \`{a} temp\'erature finie. Le but de ce chapitre est de confronter
l'approche utilisant la PFP \`{a} celle utilisant un multiplicateur de Lagrange
pour fixer la contrainte d'occupation par site du r\'eseau de spin.

La description des syst\`{e}mes quantiques de spins \`{a} temp\'erature
finie passe g\'en\'eralement par l'utilisation d'une proc\'edure de
\emph{point-selle} qui est une approximation d'ordre z\'ero de la function
de partition. La solution de champ moyen ainsi g\'en\'er\'ee peut fournir une
approximation r\'ealiste de la solution exacte.

Cependant, les solutions de champ moyen ainsi obtenue ne sont pas uniques.
Le choix d'un bon champ moyen repose essentiellement sur les propri\'et\'ees
du syst\`{e}me consid\'er\'e, en particulier sur ses sym\'etries.
Ceci g\'en\`{e}re des difficult\'es majeures. Une quantit\'e consid\'erable
de travail \`{a} \'et\'e fait sur ce point et il y a une litt\'erature 
importante sur le sujet. En particulier, les syst\`{e}mes qui sont bien
d\'ecrits par des mod\`{e}les de type Heisenberg sans frustration semblent
, selon leur dimension, bien compris par l'introduction de l'\'etat
ferromagn\'etique ou antiferromagn\'etique de N\'eel \`{a} temp\'erature nulle
$T = 0$ \cite{BernuLhuillier-92,BernuLhuillier-94}. 
Cependant il n'en n'est pas de m\^{e}me pour beaucoup de syst\`{e}mes de basse 
dimensionnalit\'e ($D \leq 2$) et/ou frustr\'es 
\cite{Zhang-88,MisguichLhuillier-03,Lecheminant-03}.
Ces syst\`{e}mes pr\'esentent des caract\'eristiques sp\'ecifiques.
Une analyse extensive et une discussion, pour des dimensions $D=2$, ont
\'et\'e pr\'esent\'ees par Wen \cite{Wen-02}. La comp\'etition entre
\'etat antiferromagn\'etique et liquide de spins ont fait l'objet de r\'ecentes
investigations dans le cadre de la th\'eorie quantique des champs \`{a} 
temp\'erature nulle \cite{TanakaHu-05,HermeleSenthilFisher-05}.

La raison de ces comportements sp\'ecifiques des syst\`{e}mes de basse
dimension peut \^{e}tre qualitativement reli\'e au fait que
les fluctuation quantiques et thermiques sont tr\`{e}s fortes,
et d\'etruisent ainsi l'ordre antiferromagn\'etique.
Ceci motive une transcription de l'Hamiltonian en termes d'op\'erateurs
compos\'es que nous appelons ``diffuson'' et ``coop\'eron''.

Le chapitre \ref{Chapter4} compare le traitement exact (obtenue par la
PFP) et moyen (par un multiplicateurs de Lagrange) de la contrainte 
d'occupation pour diff\'erentes approches de l'Hamiltonien de Heisenberg.

En r\'esum\'e, nous montrons au chapitre \ref{Chapter4} que la contrainte
stricte d'occupation induit une diff\'erence quantitative de la
temp\'erature critique en la comparant avec les r\'esultats obtenus
par une contrainte d'occupation moyenne.
En cons\'equence il apparait un effet mesurable sur le comportement
des param\`{e}tres d'ordre.
C'est l\`{a} encore un point original de cette th\`{e}se qui \`{a} conduit
\`{a} l'article donn\'e en r\'eference \cite{Dillen-05}.

Avec l'occupation exacte de spin, la temp\'erature de transition des \'etats
de liquides de spins et antiferromagn\'etiques de N\'eel sont le double
de la temp\'erature critique obtenue par la m\'ethode des multiplicateurs
de Lagrange.

En revanche, la PFP ne peut pas \^{e}tre employ\'ee dans une description en
terme de ``coop\'erons''. Les coop\'erons sont \`{a} consid\'erer comme des 
paires $BCS$ et d\'etruisent ainsi deux quasi-particules en faveur de la 
cr\'eation d'une nouvelle qui n'est autre qu'une paire du type $BCS$. 
Il en r\'esulte que dans ce cas
le nombre de particules n'est pas conserv\'ee au contraire de la contrainte
exacte. Or il se trouve que la PFP ne tol\`{e}re aucune fluctuation du 
nombre de particule.

Dans une description plus r\'ealiste nous devons tenir compte des
contributions des fluctuations quantiques et thermiques qui peuvent
\^{e}tre d'une importance cruciale en particulier au voisinage des points
critiques.
Le chapitre \ref{Chapter5} se consacre \`{a} l'implication des fluctuations
de phase autour du champ moyen dans l'\'etat $\pi$-flux.

\section{Du mod\`{e}le de Heisenberg \`{a} l'action $QED_3$ pour des 
temp\'erature finie}

Un ansatz de N\'eel n'est pas n\'ecessairement un bon candidat pour la
description des syst\`{e}mes de spin quantique a deux dimensions. Nous
avons montr\'e au chapitre \ref{Chapter3} qu'un tel ansatz brise la sym\'etrie
$SU(2)$ g\'en\'erant les bosons de Goldstone qui d\'etruisent l'ordre de 
N\'eel. Un meilleur candidat semble \^{e}tre le liquide de spins \'etant 
donn\'e qu'il conserve la sym\'etrie $SU(2)$ intacte. Pour cette raison, 
le chapitre \ref{Chapter5} se concentre sur la phase liquide de spins.

L'\'Electrodynamique Quantique \`{a} deux dimensions d'espace et une de temps
, la $QED_3$, est un cadre commun qui peut \^{e}tre utilis\'e pour d\'ecrire
les syst\`{e}mes fortement corr\'el\'es aussi bien que les ph\'enom\`{e}nes
sp\'ecifiques qui y sont reli\'es comme la supraconductivit\'e \`{a} haute 
temp\'erature 
\cite{Tesanovic-02,GhaemiSenthil-05,LeeNagaosaWen-04,Morinari-05}.
Une formulation en th\'eorie des champs du mod\`{e}le de Heisenberg
en $D = 2$ dimensions d'espace fait correspondre l'action initiale
\`{a} une action $QED_3$ pour les spinons \cite{GhaemiSenthil-05,Morinari-05}.
Avec cette description apparait le probl\`{e}me du champ moyen et les questions
corr\'el\'ees au confinement des charges tests qui peuvent conduire \`{a}
l'impossibilit\'e de d\'eterminer les contributions des fluctuations
quantiques au travers de d\'eveloppements en nombre de boucles
\cite{HandsKogutLucini, Herbut-02, NogueiraKleinert-05}.

Nous consid\'erons dans le chapitre \ref{Chapter5} l'\'etat $\pi$-flux
initialement introduit par Affleck et Marston 
\cite{AffleckMarston-88,MarstonAffleck-89}.
L'occupation des spins par site du r\'eseau est toujours fix\'ee par la PFP.
La PFP introduit un potentiel chimique imaginaire modifiant les fr\'equences
de Matsubara induisant en retour des modifications perceptibles au niveau
de la temp\'erature de restauration de la sym\'etrie chirale. La sym\'etrie
chirale est initialement bris\'ee par la g\'en\'eration dynamique de masse des
spinons.

Nous nous concentrons sur la g\'en\'eration et le comportement de la masse des
spinons due \`{a} la pr\'esence du champ de jauge $U(1)$.
Appelquist et cie. \cite{Appelquist1,Appelquist2} ont montr\'e qu'\`{a}
temp\'erature z\'ero les fermions initialement sans masse peuvent acqu\'erir
une masse g\'en\'er\'ee dynamiquement lorsque le nombre de saveur $N$
est inf\'erieur \`{a} la valeur critique $N_c = 32/\pi^2$. Plus tard
Maris \cite{Maris} a confirm\'e l'existence d'une valeur critique 
$N_c \simeq 3.3$ au-dessous de laquelle la masse dynamique peut-\^{e}tre 
g\'en\'er\'ee. \'Etant donn\'e que nous consid\'erons des spins $S=1/2$ nous 
avons $N=2$ et ainsi $N<N_c$.

\`{A} temp\'erature finie, Dorey et Mavromatos \cite{DoreyMavromatos} et Lee
\cite{Lee-98} ont montr\'{e} que la masse g\'en\'er\'ee dynamiquement s'annule
pour une temp\'erature $T$ plus grande que la valeur critique $T_c$.

Nous montrons que l'utilisation du potential chimique imaginaire introduit
par la PFP \cite{Popov-88} modifie notablement le potentiel effectif
entre deux particules charg\'ees et double la temp\'erature critique $T_c$
en accord avec les r\'esultats pr\'esent\'es pr\'ec\'edement \cite{Dillen-05}.

Le potentiel chimique imaginaire r\'eduit le ph\'enom\`{e}ne d'\'{e}crantage
du potentiel d'inter\-action entre des fermions test losrqu'on le compare
\`{a} celui obtenu par la m\'ethode des multiplicateurs de Lagrange.

Nous montrons \'egalement que la temp\'erature de transition de la restauration
de la sym\'etrie ``chirale'', o\`{u} la masse des spinons $m(\beta)$ s'annule,
est doubl\'ee par le potentiel chimique imaginaire de Popov-Fedotov.
Le param\`{e}tre d'ordre $r = \frac{2 m(0)}{k_B T_c}$ donn\'e par
Dorey et Mavromatos \cite{DoreyMavromatos} et Lee \cite{Lee-98} s'en trouve
r\'eduit de moiti\'e. La th\'eorie, dans l'\'etat dans lequelle elle est 
actuellement, ne peut pas rendre compte de la mesure exp\'erimentale de $r$ 
qui est environ de $8$ pour $Y Ba CuO$.

Tous ces r\'esultats ont donn\'e lieu \`{a} l'article cit\'e en r\'ef\'erence
\cite{DRcondmat-06}.

\section{Perspectives}

Marston \cite{Marston} a montr\'e que pour retirer les configurations
interdites de la jauge $U(1)$ dans le mod\`{e}le antiferromagn\'etique de 
Heisenberg, un terme de Chern-Simons apparait naturellement et doit \^{e}tre 
inclus dans l'action $QED_3$. Cette contrainte suppl\'ementaire conduit \`{a}
fixer \`{a} $\pi$ (modulo $2\pi$) le flux au travers de la plaquette form\'ee
par la maille \'el\'ementaire du r\'eseau de spins.
De nouveaux travaux, que nous n'avons pas abord\'es dans ce manuscrit, 
montrent que la temp\'erature de transition chirale ainsi que le rapport 
$r = m/T_c$ peuvent \^{e}tre control\'es par le coefficient de Chern-Simons 
\cite{DRcondmatChernSimons-06}.

D'autre points int\'eressants consernent la compactification du champ de jauge
$U(1)$ que nous avons utilis\'e pour obtenir la masse dynamique des spinons.
Dans le cas d'une th\'eorie de jauge compacte des instantons apparaissent
et interagissent avec la mati\`{e}re (ici les spinons) et peuvent changer
le comportement du syst\`{e}me de spin \cite{Polyakov-77,Polyakov-87}.
Le confinement des spinons par les instantons reste un probl\`{e}me ouvert.



\chapter{Introduction}

After the discovery of high temperature superconductivity 
\index{High-Tc superconductivity} by Bednorz and 
M\"{u}ller \cite{BednorzMuller-86} great theoretical efforts were devoted
to find an explanation of the mechanism underlying the very high critical
temperatures phenomenon. Many new superconductors were discovered amoung 
which one have cuprates with $La_{2-x} Sr_{x} Cu O_4$ and $Bi_2 Sr_2 Ca Cu_2 
O_8$ are examples of hole doped superconductors, ruthenates $Sr_2 Ru O_4$, 
heavy-fermion metals such as $U Pt_3$ and $Zr Zn_2$, and organic materials
like the well known $\kappa-(BEDT-TTF)_2-Cu[N(CN)_2]Br$.
\index{High-Tc superconductivity!Cuprate}
\index{High-Tc superconductivity!Ruthenate}
\index{High-Tc superconductivity!Heavy-fermion metals}
\index{High-Tc superconductivity!Organic}

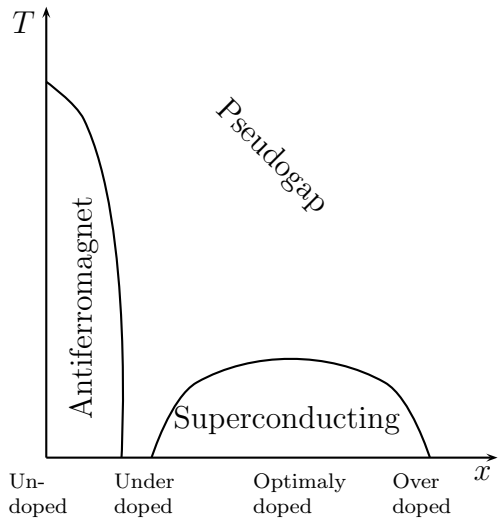
\begin{figure}[h]
\centering
\begin{pspicture}(6,6)(-1,-1)
\psline{->}(0,0)(6,0) \rput(5.8,-0.2){$x$}
\psline{->}(0,0)(0,6) \rput(-0.3,5.8){$T$}
\pscurve(0,5)(0.5,4.5)(1,0) \rput{90}(0.5,2){Antiferromagnet}
\pscurve(1.4,0)(2,1)(4.5,1)(5.1,0) \rput(3.2,0.5){Superconducting}
\rput{-45}(3,4){Pseudogap}
\rput(0,-0.5){\parbox{1cm}{\scriptsize Un\-doped}}
\rput(1.4,-0.5){\parbox{1cm}{\scriptsize Under doped}}
\rput(3.25,-0.5){\parbox{1cm}{\scriptsize Optimaly doped}}
\rput(5.1,-0.5){\parbox{1cm}{\scriptsize Over doped}}
\end{pspicture}
\caption{Schematic phase diagram of a cuprate as a function of hole 
doping $x$ and temperature $T$}
\index{High-Tc superconductivity!Un-doped}
\index{High-Tc superconductivity!Under-doped}
\index{High-Tc superconductivity!Optimaly-doped}
\index{High-Tc superconductivity!Over-doped}
\index{High-Tc superconductivity!Pseudogap}
\index{High-Tc superconductivity!Antiferromagnet}
\label{IntroFig1}
\end{figure}

In particular cuprates behave differently from
the conventional $BCS$ superconductors. Experimentalists observed
$d$-symmetry of the order parameter, strong electronic correlations and 
non-conventional but ``universal'' phase diagrams as shown in figure 
\ref{IntroFig1} \cite{BennemannKetterson-04}.
\index{High-Tc superconductivity!$d$-symmetry}
\index{High-Tc superconductivity!Conventional,$BCS$}
\index{High-Tc superconductivity!Non-conventional}
\index{High-Tc superconductivity!Phase diagram}

\noindent
All cuprates share the same kind of atomic structure which consists of a 
layered structure made of $Cu O_2$-layers which are considered as responsible
for Cooper-pairing and superconductivity.
\index{High-Tc superconductivity!$Cu O_2$ layer}
\index{High-Tc superconductivity!Cooper-pairing}

In the following we concentrate on the $Cu O_2$ layers and more 
precisely on the antiferromagnetic insulating phase of the high-$T_c$
superconductors which corresponds to the undoped regime as shown in figure 
\ref{IntroFig1}. The insulating phase of the cuprates like compound, the
so called parent compound, can be modelled by Heisenberg models.
\index{High-Tc superconductivity!Parent compound}
\index{High-Tc superconductivity!Insulating phase}

The study of the insulating phase is motivated by the fact
that under doping the parent compound should keep a \emph{memory} of the 
correlations underlying the antiferromagnet insulating phase. In other words
one believes that the superconducting phase must partially conserve some 
characters of the antiferromagnet insulating parent compound.

Our work is devoted to the study of the effects of a \emph{strict} 
site-occupation constraint caracterising the insulating phase in the 
description of quantum spin systems at finite temperature. This constraint 
consists in the enforcement of the occupation of each lattice site by exactly 
one $S=1/2$ spin. 
The implementation of such a constraint was introduced by Popov 
and Fedotov \cite{Popov-88}. Contrary to other methods which are based on the 
use of a Lagrange mutliplier 
\cite{ArovasAuerbach-88,Auerbach-94,ReadSachdev-91} 
leading to an average occupations the \textbf{P}opov and \textbf{F}edotov 
\textbf{p}rocedure (\textbf{PFP}) avoids this approximate treatment by means of
the introduction of an imaginary chemical potential as we will show in chapter 
\ref{chapterPathIntegralPFP}. We are interested to work out and analyse the 
application of the PFP onto antiferromagnet spin systems at finite temperature 
and to compare these results with the Lagrange multiplier method.
\index{Lagrange multiplier}
\index{Popov and Fedotov procedure!Imaginary chemical potential}

\vspace*{0.5cm}

\textbf{Chapter \ref{chapterPathIntegralPFP}}
introduces the mathematical tools found in the literature and which will
be used throughout this manuscript. We present the fermionization of spin 
Hamiltonian models and the PFP. We show that the PFP eliminates the 
\emph{unphysical} Fock states in the fermionization of spin models. 
Finally we construct a path integral formulation of the partition function in 
imaginary time and show that the PFP induces a modification of the Matsubara 
frequencies which characterizes the fermionic propagator.

\vspace*{0.5cm}

\textbf{In chapter \ref{Chapter3}} 
we concentrate on a N\'eel mean-field phase description of quantum spin 
systems. The possible competition or phase transition between spin liquid 
states and an antiferromagnetic N\'eel state which may be expected to 
describe Heisenberg type systems in two dimensions where discussed recently 
\cite{GhaemiSenthil-05,Morinari-05,SenthilFisher-05,SenthilFisher-04}. 
We aim to study the pertinence of the N\'eel state ansatz as a mean-field 
approximation for finite temperature using the PFP. Quantum and thermal 
fluctuations contributions are worked out at the one-loop level taking the 
Popov and Fedotov procedure into account.
The outcomes are compared with the spin-wave theory and extended to the
$XXZ$-model by working out the magnetization and the magnetic susceptibility 
for temperature below and up to critical point. We discuss the degree of 
realism of the mean-field N\'eel ansatz. Indeed Ghaemi and Senthil 
\cite{GhaemiSenthil-05} have recently shown, with the help of a specific model,
that a second order phase transition from a N\'eel mean-field to an ASL 
(algebraic spin liquid) may be at work depending on the strength of 
interaction parameter which enter the Hamiltonian of the system.

The originality of my work stays here on the fact that magnetic properties
have never been worked out using a strict site-occupation constraint. 
Chapter \ref{Chapter3} remedies to it and an article on this point can be 
found in \cite{DRepjb-05}.

\vspace*{0.5cm}

\textbf{In chapter \ref{Chapter4}} 
we consider different mean-fields possible choices of quantum spin systems 
at finite temperature in which each lattice site is occupied by exactly one 
spin imposed by the PFP. We also construct the formalism in which the 
occupation constraint is imposed on the average by means of the Lagrange 
mutliplier method with the aim of confronting these two approaches in the 
framework of Heisenberg-type models.

The description of strongly interacting quantum spin systems at finite 
temperature generally relies on a saddle point procedure which is a zeroth 
order approximation of the partition function. The so generated mean-field 
solution is aimed to provide a qualitatively realistic approximation of the 
exact solution.
The specific behaviour of low-dimensional systems is characterized by
the fact that low-dimensionality induces strong quantum and thermal 
fluctuations, hence disorder which destroys the antiferromagnet order. 
This motivates a transcription of the Hamiltonian in terms of composite 
operators which we call "diffusons" and "cooperons" which are the essence of 
the well known \textbf{R}esonant \textbf{V}alence \textbf{B}ond (\textbf{RVB}) 
spin liquid states proposed by Anderson \cite{Anderson-87}.
\index{Diffuson}
\index{Cooperon}

The original point concerns the confrontation of the magnetization obtained 
through the PFP with the result obtained by means of an average projection 
procedure in the framework of the mean-field approach characterized by a 
N\'eel state. 
The same confrontation is performed for the order parameter which characterizes
the system when its Hamiltonian is written in terms of so called Abrikosov 
fermions (also called pseudo-fermions or spinons) \cite{Dillen-05}.

\vspace*{0.5cm}

\textbf{Chapter \ref{Chapter5}} is devoted to a more realistic analysis in 
which we take care of the contributions of quantum fluctuations which may be 
of overwhelming importance particularly in the vicinity of critical points.

Quantum Electrodynamics $QED_{(2+1)}$ has attracted considerable interest in 
the last decade 
\cite{Aitchison-92,DoreyMavromatos,Herbut-02,Lee-98,Tesanovic-02} 
since it is a common framework which can be used to describe stron\-gly 
correlated systems such as quantum spin systems in $1$ time and $2$ space 
dimensions, as well as related specific phenomena like high-$T_c$ 
superconductivity 
\cite{Tesanovic-02,GhaemiSenthil-05,LeeNagaosaWen-04,Morinari-05}. 
A gauge field formulation of antiferromagnetic Heisenberg 
models in $d = 2$ space dimensions maps the initial action onto 
a $QED_3$ action for spinons \cite{GhaemiSenthil-05,Morinari-05}. 
We consider the $\pi$-flux state approach introduced by Affleck and 
Marston \cite{AffleckMarston-88,MarstonAffleck-89}. 
The strict site-occupation is introduced by the imaginary chemical potential 
proposed by Popov and Fedotov \cite{Popov-88} for $SU(2)$ spin symmetry 
which modifies the Matsubara frequencies as will be explained in chapter 
\ref{chapterPathIntegralPFP}.

We show that at zero temperature the ``chiral'' symmetry is broken by the
generation of a dynamical mass \cite{Appelquist1,Appelquist2} which vanishes
at finite temperature $T$ larger than the critical one $T_c$ 
\cite{DoreyMavromatos,Lee-98}.

We show that the imaginary chemical potential 
introduced by Popov and Fedotov \cite{Popov-88} modifies noticeably the 
effective potential between two charged particles, doubles the dynamical 
mass transition temperature $T_c$ and reduces the screening effect of this 
static potential between test fermions.
This last point led to two articles given in 
\cite{DRcondmatChernSimons-06,DRcondmat-06}.

\vspace*{0.5cm}

\textbf{Chapter \ref{Chapter6}} summarizes and comments our results and 
suggests further developments aimed to lead to clues to remaining open 
questions \cite{DRepjb-05,Dillen-05,DRcondmat-06}.



\chapter{Path integral formulation and the Popov-Fedotov procedure
\label{chapterPathIntegralPFP}}

\minitoc
\newpage

We present here a set of technical tools, extracted from literature and
mainly taken from Negele and Orland's book \cite{OrlandNegele}, which are 
necessary in order to construct the partition function in terms of a 
functional integral. The fermionization of a Heisenberg spin-$1/2$ model is 
presented. The spin operators are described in the fermionic Fock space by 
means of the introduction of anticommuting creation and annihilation operators.
In order to eliminate the \emph{unphysical} Fock states which correspond to the
presence of $0$ or $2$ particles on each site of the system, we introduce
the so-called Popov-Fedotov procedure (PFP). Finally we construct a path
integral formulation of the partition function in imaginary time and
show that the PFP induces a modification of the Matsubara frequencies
which characterize the fermionic propagator.

\newpage

\section{Fermionization of spin-1/2 Heisenberg models \label{Chapter2Section1}}

\begin{figure}[h]
\begin{tabular}{cc}
\epsfig{file=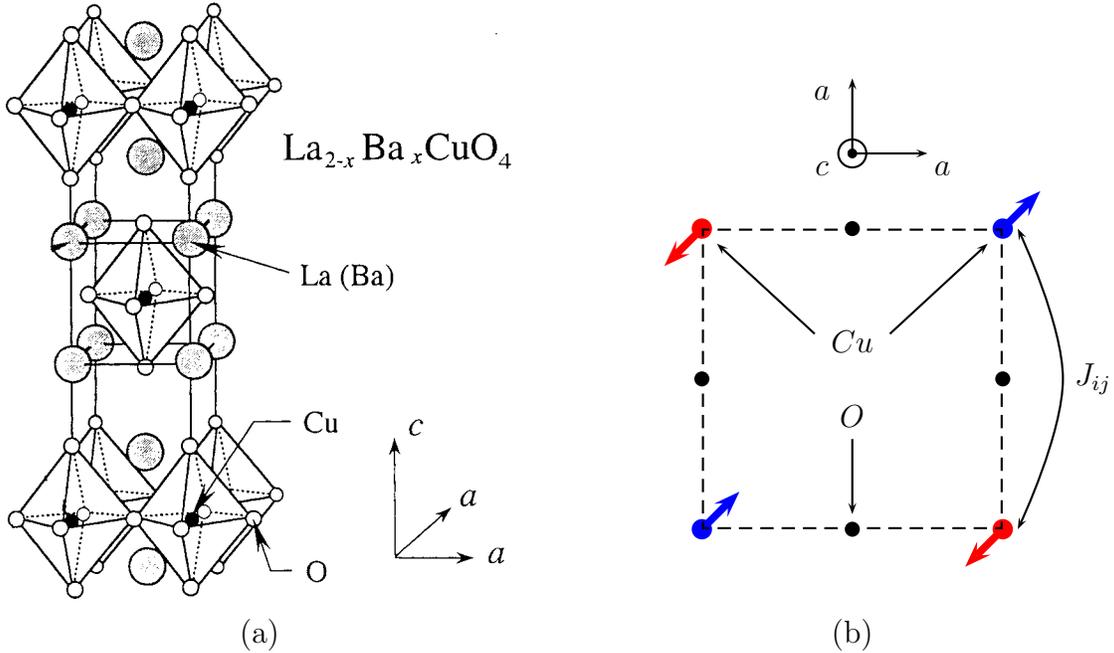}
 &
\begin{pspicture}(-4,-3)(4,4)
\psline[linewidth=1mm,linecolor=blue]{*->}(-2,-2)(-1.5,-1.5)
\psline[linewidth=1mm,linecolor=blue]{*->}(2,2)(2.5,2.5)
\psline[linewidth=1mm,linecolor=red]{*->}(-2,2)(-2.5,1.5)
\psline[linewidth=1mm,linecolor=red]{*->}(2,-2)(1.5,-2.5)
\pscircle*(-2,0){0.1}
\pscircle*(0,2){0.1}
\pscircle*(0,-2){0.1}
\pscircle*(2,0){0.1}
\psframe[linestyle=dashed](-2,-2)(2,2)
\psline{<-}(-1.8,1.8)(-0.5,0.6)
\psline{->}(0.5,0.6)(1.8,1.8)
\rput(0,0.5){$Cu$}
\psline{->}(0,-0.8)(0,-1.8)
\rput(0,-0.5){$O$}
\psline{*->}(0,3)(1,3) \rput(1.2,2.8){$a$}
\psline{*->}(0,3)(0,4) \rput(-0.4,3.8){$a$}
\pscircle(0,3){0.2} \rput(-0.4,2.8){$c$}
\pscurve{<->}(2.2,-2)(2.8,0)(2.2,2) \rput(3.2,0){$J_{ij}$}
\end{pspicture}
 \\
(a) & (b)
\end{tabular}
\caption{(a) Cristallographic structure of the cuprate superconductor 
$La_{2-x} Ba_x CuO_4$ and (b) a $CuO_2$ layer showing the (super)exchange
coupling between the copper atoms.}
\index{High-Tc superconductivity!Cristallographic structure}
\label{FigLaCuO1}
\end{figure}

Quantum antiferromagnet spin-$1/2$ models are of great interest for
theoretical studies. Indeed they have deep connections with 
high-temperature superconducting materials
such as $Nd_{2-x} Ce_x CuO_4$ and $La_{2-x} (Sr \text{ or } Ba)_x CuO_4$ 
for which the undoped region of the phase diagram shows an antiferromagnetic
phase described by a two dimensional square lattice of spin-$1/2$ particles, 
as explained in \cite{BennemannKetterson-04,LeeNagaosaWen-04}.
For example in the parent compound $La_2 CuO_4$ the copper is surrounded
by six oxygen atoms forming $CuO$ tetrahedron layers between $La$ layers.
Copper atoms interact with each other by means of a superexchange mechanism
forming an effective two dimensional square lattice of spin-$1/2$ as shown
in figures \ref{FigLaCuO1}.
\index{High-Tc superconductivity!Cuprate}
\index{High-Tc superconductivity!Parent compound}
\index{High-Tc superconductivity!Antiferromagnet}
\index{High-Tc superconductivity!Phase diagram}
\index{High-Tc superconductivity!$Cu O_2$ layer}
\index{Antiferromagnetic Heisenberg model}
\index{Antiferromagnetic Heisenberg model|see{High-Tc superconductivity}}
\index{Superexchange}

In the same direction Dagotto and Rice \cite{DagottoRice-96} show that 
the two-leg spin-$1/2$ ladder materials, the vanadyl pyrophosphate 
${( VO )}_2 P_2 O_7$ and the cuprate like $Sr Cu_2 O_3$,
can be modelised as a one-dimensional spin-$1/2$ Heisenberg model.
\index{Vanadyl pyrophosphate ${( VO )}_2 P_2 O_7$}
\index{High-Tc superconductivity!Cuprate}
\index{Spin ladder}

In order to study these antiferromagnet spin-$1/2$ systems we will
introduce the \\ \textbf{H}eisenberg \textbf{A}nti-\textbf{F}erromagnet 
\textbf{M}odel $\mathbf{(HAFM)}$ \index{Antiferromagnetic Heisenberg model}
with an external magnetic field $\vec{B}_i$
\index{Antiferromagnetic Heisenberg model}

\begin{equation}
H = -\frac{1}{2} \underset{i,j}{\sum} 
J_{ij} \vec{S}_i.\vec{S}_j + \underset{i}{\sum}
\vec{B}_i .\vec{S}_i
\label{GeneralH}
\end{equation}

\noindent
where the sums runs over the lattice sites $\vec{r}_i$ and $\vec{r}_j$.
The effective (super)exchange coupling $J_{ij}$ (see A.Auerbach 
\cite{Auerbach-94} chapter 1), acts between spins at position $\vec{r}_i$ 
and $\vec{r}_j$ on a $D$-dimensional hypercubic lattice. 
For antiferromagnet systems $J_{ij}$ is negative and positive in the
ferromagnetic case.
\index{Superexchange}
The second term of the HAFM is the coupling between the spins and the 
magnetic field $\vec{B}$ at each lattice site $\vec{r}_i$. The \emph{Land\'e
factor} and the \emph{Bohr magneton} coefficients are absorbed in 
the magnetic field $\vec{B}$.
\index{Bohr magneton}
\index{Land\'e factor}

The spins $\vec{S}$ are vector operators, their components obey a
$SU(2)$ Lie algebra $[S_x,S_y]= i S_z$. Here we adopt the 
convention $\hbar=1$. 
\index{Lie algebra}
The $S=1/2$ spin vector operators can be expressed using the \emph{Abrikosov}
fermionic creation and annihilation operators $f^{\dagger}_{i\sigma_1}$ 
and $f_{i\sigma_2}$ 
\index{Fermionization}
\index{Fermionization!Creation and annihilation operators}
\index{Abrikosov fermion}

\begin{eqnarray}
\vec{S}_i = \frac{1}{2} f^{\dagger}_{i\sigma_1} 
\vec{\sigma}_{\sigma_1\sigma_2} f_{i\sigma_2}
\label{FermionizedSpin}
\end{eqnarray}

\noindent
where $\sigma_1,\sigma_2=\uparrow,\downarrow$ and the 
$\vec{\sigma}_{\sigma_1 \sigma_2}$ vector components are Pauli matrices
\index{Pauli matrices}

\begin{eqnarray}
\sigma^x = \begin{pmatrix}
0 & 1 \\
1 & 0
\end{pmatrix}
, \quad
\sigma^y = \begin{pmatrix}
0 & -i \\
i & 0
\end{pmatrix}
, \quad
\sigma^z = \begin{pmatrix}
1 & 0 \\
0 & -1
\end{pmatrix}
\end{eqnarray}

\noindent
Explicitly spin operators at position $i$ read
\index{$SU(2)$ Spin operators}

\begin{eqnarray}
S^{+}_{i} &=& f^{\dagger}_{i \uparrow} f_{i \downarrow} \label{fSplus} \\
S^{-}_{i} &=& f^{\dagger}_{i \downarrow} f_{i \uparrow} \label{fSminus} \\
S^{z}_{i} &=& \frac{1}{2} (f^{\dagger}_{i \uparrow} f_{i \uparrow} - 
f^{\dagger}_{i \downarrow} f_{i \downarrow}) \label{fSz}
\end{eqnarray}

\noindent
The creation and annihilation operators 
$f_{i\sigma}$ and $f^{\dagger}_{i\sigma}$ verify the
anticommutation relations
\index{Fermionization!Creation and annihilation operators}

\begin{eqnarray}
\{ f_{i\sigma_1} , f^{\dagger}_{j\sigma_2} \} &=& \delta_{\sigma_1,\sigma_2}
\delta_{i,j} \\
\{ f_{i\sigma_1} , f_{j\sigma_2} \} &=& 0 \\
\{ f^{\dagger}_{i\sigma_1} , f^{\dagger}_{j\sigma_2} \} &=& 0
\end{eqnarray}

\noindent
Applying \eqref{fSplus}, \eqref{fSminus} and \eqref{fSz} onto the 
\emph{physical} Fock states $|1,0>$ and $|0,1>$ one gets
\index{Fermionization!Fock states}

\begin{eqnarray}
S^{+}_{i} |0,1> &=& |1,0> \equiv |\uparrow> \\
S^{-}_{i} |1,0> &=& |0,1> \equiv |\downarrow> \\
S^{z}_{i} |1,0> &=& -1/2 |\uparrow> \\
S^{z}_{i} |0,1> &=& 1/2 |\downarrow>
\end{eqnarray}

The insertion of \eqref{FermionizedSpin} into \eqref{GeneralH} generates the 
Fermionized Heisenberg Antiferromagnet Model which is quartic in the
fermion operators. Using this formulation it is possible to write 
the Hamiltonian in different forms. Indeed the spin interaction term 
$\vec{S}_i.\vec{S}_j$ can be expressed as 
\index{Diffuson} 
\index{Cooperon}
\index{Antiferromagnetic Heisenberg model}

\begin{eqnarray}
\vec{S}_i.\vec{S}_j
&=&
\vec{S}_i.\vec{S}_j
\notag \\
&=& \frac{1}{2} \mathcal{D}^{\dagger}_{ij} \mathcal{D}_{ij}
+ \frac{\tilde{n}_i  \tilde{n}_j}{4} - \frac{\tilde{n}_i}{2}
\notag \\
&=& \frac{1}{2} \mathcal{C}_{ij}^{\dagger} \mathcal{C}_{ij} 
+ \frac{\tilde{n}_i \tilde{n}_j}{4}
\label{TreeForms}
\end{eqnarray}

\noindent
where $\tilde{n}_i = \sum_\sigma f^\dagger_{i \sigma} f_{i \sigma}$,
$\mathcal{D}$ and $\mathcal{C}$ are quadratic in the fermionic
creation and annihilation operators

\begin{eqnarray}
\mathcal{D}_{ij} = f_{i \uparrow}^{\dagger} f_{j \uparrow} + 
f_{i \downarrow}^{\dagger} f_{j \downarrow}
\\
\mathcal{C}_{ij} =   f_{i \uparrow} f_{j \downarrow} - 
f_{i \downarrow} f_{j \uparrow}
\end{eqnarray}

\noindent
N\'eel states, Resonant Valence Bond (RVB) states
\cite{Anderson-87} as well as $\pi$-flux states
can be introduced by means of \eqref{TreeForms} in order to define different
mean-fields, as done by using the large-N method developed by 
Affleck \cite{Affleck-85},Affleck and Marston \cite{AffleckMarston-88}
, Read and Sachdev \cite{ReadSachdev-89,ReadSachdev-90,ReadSachdev-91}.
\index{N\'eel state}
\index{Resonant Valence Bond (RVB) states}
\index{$\pi$-flux state}
\index{large-N method}

From equation \eqref{fSz} we see that the \emph{physically} acceptable 
Fock states on a lattice site $i$ are those for which the expectation value 
of the number operator either $f_{i,\uparrow}^\dagger f_{i,\uparrow}$ or 
$f_{i,\downarrow}^\dagger f_{i,\downarrow}$ is equal to one an the other equal
to zero (corresponding to $<S_z>=\pm 1/2$).
The occupation by one fermion per lattice site is fulfilled if

\begin{eqnarray}
\sum_\sigma f^{\dagger}_{i\sigma} f_{i\sigma}=1
\label{Constraint}
\end{eqnarray}

\noindent
One way to implement this constraint consists of the introduction
of a projector onto the Fock space using a Lagrange multiplier $\lambda$
\index{Lagrange multiplier}
\index{Fermionization!Projector}

\begin{eqnarray}
\tilde{P}_{i} = \int \mathcal{D} \lambda_i e^{\lambda_i (
\sum_\sigma f^{\dagger}_{i\sigma} f_{i\sigma} - 1)}
\label{Chapter2LagrangeProjector}
\end{eqnarray}

\noindent
where $\lambda$ plays the role of a chemical
potential in the framework of the path integral formulation.
However such a propagator cannot impose a rigorous
constraint \eqref{Constraint}. The actual value of $\lambda$ is fixed
by a saddle point method, that is to say mean-field equations will fix
the mean value of the Lagrange parameter.
In order to implement the constraint \eqref{Constraint} in a \emph{strict} way
we shall introduce the Popov-Fedotov procedure.
\index{Popov and Fedotov procedure}

\section{The Popov-Fedotov procedure (PFP) \label{sectionPF}}

The introduction of this procedure is a key point in the present work.
In the following we aim to study the consequences of its use in the
description of spin systems at finite temperature. In the present section
we show how it is able to enforce the constraint \eqref{Constraint}.

The Fock space constructed with the fermionic operators 
$f,f^{\dagger}$ is not in bijective correspondence with the Hilbert space of 
spin states as shown in figure \ref{FigHilbertFockSpace}.
\index{Fermionization!Creation and annihilation operators}
Indeed, in Fock space and for spin-1/2 particles, the 
occupation of each site $i$ can be characterized by the states 
$|n_{i,\uparrow},n_{i,\downarrow}>$ with 
$n_{i,\sigma} \in \{0,1\}$ the eigenvalue of the occupation operator
$f^{\dagger}_{i \sigma}f_{i \sigma}$, that is the
states $|0,0>$, $|1,0>$, $|0,1>$ and $|1,1>$. Since the sites have to be
occupied by a single particle the \emph{unphysical} states $|0,0>$ and 
$|1,1>$ have to be eliminated. This is done by means of a projection
procedure proposed by Popov and Fedotov \cite{Popov-88} and generalized to 
$SU(N)$ symmetry by Kiselev \emph{et al.} \cite{KFO-01}.
\index{Popov and Fedotov procedure!Unphysical Fock states}
\index{Popov and Fedotov procedure!Hilbert space}
\index{Popov and Fedotov procedure!Fock space}

\begin{figure}[h]
\center
\begin{tabular}{ccc}
Hilbert space &  & Fock space $\{|n_{\uparrow},n_{\downarrow}> \}$ \\
$|\uparrow>$ & $\Longleftrightarrow$ & $|1,0>$ \\
$|\downarrow>$ & $\Longleftrightarrow$ & $|0,1>$ \\
- &  & $|0,0>$ \\
- &  & $|1,1>$
\end{tabular}
\caption{Correspondance between the Hilbert space and the $S=1/2$ Fock space}
\label{FigHilbertFockSpace}
\end{figure}

We introduce the projection operator 
$\tilde{P}=\frac{1}{i^{\tilde{N}}} e^{i \frac{\pi}{2} \tilde{N}}$
, where $\tilde{N}=\underset{i,\sigma}{\sum}f^{\dagger}_{i\sigma} f_{i\sigma}$
is the number operator, into the partition function $\cal{Z}$ which then reads
\index{Popov and Fedotov procedure!Projector}
\index{Popov and Fedotov procedure!Number operator}

\begin{eqnarray}
\mathcal{Z}=Tr \left[ e^{-\beta \tilde{H}}\tilde{P} \right]
\end{eqnarray}

\noindent
where $\tilde{H}$ is the fermionized Hamiltonian of the systems and 
$\beta=1/T$ is the inverse temperature, with the convention 
$k_B=1$ for the Boltzmann constant. Define the trace of the operator 
$\mathcal{O}$ in Fock space as
\index{Boltzmann constant}
\index{Trace in Fock space}
\index{Trace in Fock space|see{Path integrals}}

\begin{eqnarray}
Tr \left[ \mathcal{O}  \right] =
\underset{\{ n_{i,\alpha} \in \left[ 0,1 \right] \} }{\sum}
\left(\underset{i}{\bigotimes} <n_{i,\uparrow}, n_{i,\downarrow}|
\right) \mathcal{O} 
\left(\underset{i}{\bigotimes} |n_{i,\uparrow}, n_{i,\downarrow}>
\right)
\end{eqnarray}

\noindent 
where $\underset{i}{\bigotimes} |n_{i,\uparrow}, n_{i,\downarrow}>$ 
is the tensor product of the Fock states 
$|n_{i,\uparrow}, n_{i,\downarrow}>$ for different lattice site $i$.
The contributions of the Hamiltonian should be equal to zero when 
applied to the \emph{unphysical} states. 
Since $S_i^{+} |unphysical> = S_i^{-} |unphysical> = S_i^{z} |unphysical> = 0$
all Hamiltonians build with only spin operators $\left\{ S_i^{+}, S_i^{-}
, S_i^{z} \right\}$ give an energy equal to zero when applied to unphysical
states ($H |unphysical> = 0$).
The action of $\tilde{P}_j$ 
on each site $j$ is such that the contributions 
of states $|0,0>_j$ and $|1,1>_j$ to the partition function $\cal{Z}$ 
eliminate each other. Indeed 
\index{Popov and Fedotov procedure}
\index{Popov and Fedotov procedure!Unphysical Fock states}
\index{Popov and Fedotov procedure!Imaginary chemical potential}

\begin{gather*}
<0,0|_j \, e^{-\beta H}.e^{i\frac{\pi}{2}*0}  |0,0>_j +
<1,1|_j \, e^{-\beta H}.e^{i\frac{\pi}{2}*2} |1,1>_j \\
+ <1,0|_j \, e^{-\beta H}.e^{i\frac{\pi}{2}} |1,0>_j
+ <0,1|_j \, e^{-\beta H}.e^{i\frac{\pi}{2}} |0,1>_j \\
= i \left( <1,0|_j \, e^{-\beta H} |1,0>_j
+ <0,1|_j \, e^{-\beta H} |0,1>_j \right)
\end{gather*}

\noindent
Hence the partition function reads
 
\begin{eqnarray}
\mathcal{Z}=\frac{1}{i^{\mathcal{N}}}.
Tr \left[ e^{-\beta (\tilde{H}-\mu\tilde{N})} \right]
\label{Zmu}
\end{eqnarray} 

\noindent 
where $\mathcal{N}$ is the total number of spin sites in the 
considered lattice. Equation \eqref{Zmu} with the 
\emph{imaginary chemical potential} 
\index{Popov and Fedotov procedure!Imaginary chemical potential}
$\mu = i \frac{\pi}{2 \beta}$ describes a system with strictly one
 fermion (spin-$\uparrow$ or spin-$\downarrow$) per lattice site, 
in contrast with the usual method which introduces an average projection 
by means of a real Lagrange multiplier  \cite{Auerbach-94,ArovasAuerbach-88}.

In the following the Heisenberg model will be studied in the path integral
formulation using the Popov-Fedotov procedure.
The consequences of its application will be compared to those obtained
by means of Lagrange formulations.

\section{Path integral formulation of the partition function
\label{SectionPathIntegral}}

In the present section we construct the path integral formulation
of the partition function \eqref{Zmu}.
First we shall define the coherent states
of the fermionic Fock space. Then we derive from these
states the appropriate properties which lead from the trace of an 
operator over Fock state into the trace over coherent states
of the same Fock space. This leads to the functional integral
expression of the partition function \eqref{Zmu}.

\subsection{Fermionic coherent states}

The coherent states are an overcomplete linear combination of the set of
states in Fock space. They are eigenstates of the annihilation operator, 
bosonic as well as fermionic \cite{OrlandNegele}.

Define $|\xi >$ as the coherent state of the fermionic Fock space
\index{Path integrals!Coherent states}

\begin{equation}
|\xi > = \sum_{k=0}^\infty 
\sum_{ \{ n_\alpha \} =0}^1  \left(-1\right)^{\underset{\alpha}{\sum} n_\alpha}
\xi_1^{n_1} \dots \xi_\alpha^{n_\alpha} \dots \xi_k^{n_k} 
|{ n_1, \dots, n_\alpha, \dots, n_k }>
\end{equation}

\noindent where $|{ n_1, \dots, n_\alpha, \dots, n_k }>$ is a Fock state
constructed with the fermionic creation operator 
${\left( f^\dagger_1 \right)}^{n_1} 
\dots {\left( f^\dagger_\alpha \right)}^{n_\alpha} 
\dots {\left( f^\dagger_k \right)}^{n_k}$
 applied on the vacuum state $|0>$. 
The $\xi_\alpha$ are \emph{Grassmann} variables verifying the anticommutation
relations
\index{Path integrals!Fock space}
\index{Path integrals!Grassmann variables}

\begin{eqnarray}
\{ \xi_\alpha, \xi_\lambda \}= 
\xi_\alpha \xi_\lambda + \xi_\lambda \xi_\alpha &=& 0 \\
\{ \xi_\alpha, \xi_\lambda^{*} \} &=& 0 \\
\xi_\alpha^2 = \left( \xi_\alpha^{*} \right)^2 &=& 0
\end{eqnarray}

\noindent
More properties on the \emph{Grassmann} algebra are given in appendix
\ref{AppendixGrassmann} \index{Path integrals!Grassmann variables}. \\
Applying the annihilation operator $f_\alpha$ on the coherent state
$|\xi>$ we obtain

\begin{eqnarray}
f_\alpha |\xi > &=& \sum_{k=0}^\infty \sum_{ \{ n_\alpha \} =0}^1 
\xi_1^{n_1} \dots \xi_\alpha^{n_\alpha} \dots \xi_k^{n_k} 
f_\alpha |{ n_1, \dots, n_\alpha, \dots, n_k }>
\notag \\
&=& 
\sum_{k=0}^\infty
\sum_{
\begin{aligned}
\{ n_\lambda \} &= 0, \\
n_\alpha &= 1
\end{aligned}
}^1
\xi_1^{n_1} \dots \xi_\alpha^{n_\alpha} \dots \xi_k^{n_k} 
\left(-1\right)^{\overset{\alpha-1}{\underset{j = 1}{\sum}} n_j}
|{ n_1, \dots, n_\alpha-1, \dots, n_k }>
\notag \\
&=& \sum_{k=0}^\infty \sum_{ \{ n_\alpha \} =0}^1  
\left(-1\right)^{\overset{\alpha-1}{\underset{j = 1}{\sum}} n_j}
\xi_1^{n_1} \dots \xi_\alpha^{n_\alpha+1} \dots \xi_k^{n_k} 
|{ n_1, \dots, n_\alpha, \dots, n_k }>
\notag \\
&=& \xi_\alpha |\xi>
\end{eqnarray}

\noindent
Thus $\xi_\alpha$ and $|\xi>$ are the eigenvalue and the eigenvector of the 
fermionic annihilation operator $f_\alpha$. Since $f_\alpha^\dagger$ 
increases the number of particles in any Fock state by one, $|\xi>$
cannot be an eigenvector of $f_\alpha^\dagger$.

Coherent states are equivalently constructed combining the properties of 
the \emph{Grassmann} variables in exponential series such as
\index{Path integrals!Coherent states}

\begin{eqnarray}
|\xi> &=& e^{-\sum_\alpha \xi_\alpha f_\alpha^\dagger} |0> 
\,= \prod_\alpha (1-\xi_\alpha  f_\alpha^\dagger)|0>
\label{DefinitionXi}
\end{eqnarray} 

\noindent 
where $|0>$ is the vacuum state of the fermionic Fock space.
Similary the \emph{bras} of the coherent states $|\xi>$ verify  
\index{Path integrals!Vacuum state}

\begin{eqnarray}
<\xi| &=& <0|e^{-\sum_\alpha f_\alpha \xi_\alpha^*}=<0|e^{\sum_\alpha 
\xi_\alpha^* f_\alpha}
 \\
<\xi|f_\alpha^\dagger &=& <\xi|\xi_\alpha^*
\end{eqnarray}

\noindent
Useful properties can be extracted from these basic equalities.

\subsection{Properties of the coherent states}

In this subsection we review some properties of the coherent states 
$|\xi>$ which will be used later in the construction of path integrals.
The first one is the closure relation in the fermionic
Fock space and the second is the trace of a fermionized operator 
$\mathcal{O}$.

We recall some main points concerning the 
closure relation of coherent states which reads  
\index{Path integrals!Closure relation}

\begin{eqnarray}
\int \prod_\alpha d\xi_\alpha^* d\xi_\alpha 
e^{-\sum_\alpha \xi_\alpha^* \xi_\alpha} |\xi><\xi|= \Unitmatrix
\label{ClosureRelation}
\end{eqnarray}

\noindent
A detailed demonstration can be found in the book of Negele and Orland
\cite{OrlandNegele}.  
To proceed, we define the operator $\mathcal{A}$ as being the left hand side 
of equation \eqref{ClosureRelation}. Following Negele and Orland, the first 
step in the demonstration of the equality \eqref{ClosureRelation}
is to show that the operator $\mathcal{A}$ is proportional to the unit 
operator $\Unitmatrix$, the second step to show that the proportionality
factor $\kappa$ is equal to one.

\medskip
\noindent
Using the expression \eqref{DefinitionXi} and the definition of the 
derivation operator given in appendix \ref{AppendixGrassmann} 
the commutation relation of $f_\beta$ and the operator $|\xi><\xi|$ is given by

\begin{eqnarray}
\left[f_\beta,|\xi><\xi|\right] 
&=&
f_\beta |\xi><\xi|-|\xi><\xi|f_\beta
\notag \\
&=&
\xi_\beta |\xi><\xi| -|\xi> \frac{\partial}{\partial \xi_\beta^{*}} <\xi|
\notag \\
&=& 
\left[ \xi_\beta - \frac{\partial}{\partial \xi_\beta^{*}} \right] |\xi><\xi|
\end{eqnarray}

\noindent
Using the properties of the integration operator (see appendix 
\ref{AppendixGrassmann}) the commutation relation of the operator 
$\mathcal{A}$ and the annihilation operator $f_\beta$ is equal to 
zero by virtue of the \emph{Grassmann} algebra
\index{Path integrals!Grassmann algebra}

\begin{eqnarray}
<n_\gamma| \left[ f_\beta , \mathcal{A} \right] |n_\lambda> 
&=&
<n_\gamma|
\int \prod_\alpha d\xi_\alpha^* d\xi_\alpha 
e^{-\sum_\alpha \xi_\alpha^* \xi_\alpha} 
\left[\xi_\beta - \frac{\partial}{\partial \xi_\beta^{*}} \right]
|\xi><\xi||n_\lambda>
\notag \\
&=&
\int \prod_\alpha d\xi_\alpha^* d\xi_\alpha 
e^{-\sum_\alpha \xi_\alpha^* \xi_\alpha} 
\left[\xi_\beta - \frac{\partial}{\partial \xi_\beta^{*}} \right] 
\xi_\gamma \xi_\lambda^{*}
\notag \\
&=&
\int \prod_{\alpha = \left\{\beta, \gamma, \lambda \right\}} 
\left( d\xi_\alpha^* d\xi_\alpha 
\left( 1 - \xi^{*}_\alpha \xi_\alpha  \right) \right)
\left[ \xi_\beta \xi_\gamma \xi^{*}_\lambda + \xi_\gamma 
\delta_{\beta \lambda} \right]
\notag \\
&=& 0
\end{eqnarray}

\noindent
This equality can also be proven with the use of the creation operator 
$f_\beta^\dagger$.
Therefore the operator $\mathcal{A}$ commutes with any operator composed 
of operators $f_\alpha$ and $f_\alpha^\dagger$. 
Since the \emph{Schur} lemma stipulates that if an operator commutes with
any operator then it must be proportional to the unit operator.
$\mathcal{A}$ must be proportional to the unit operator $\Unitmatrix$,
$\mathcal{A}= \kappa \Unitmatrix$.
\index{Schur lemma}

\noindent
The matrix element of $\mathcal{A}$ between two vacuum states is given by

\begin{eqnarray}
<0| \mathcal{A} |0> 
&=& 
<0| \int \prod_\alpha d\xi_\alpha^* d\xi_\alpha 
e^{-\sum_\alpha \xi_\alpha^* \xi_\alpha} |\xi><\xi| |0>
\notag \\
&=& 
\int \prod_\alpha d\xi_\alpha^* d\xi_\alpha 
e^{-\sum_\alpha \xi_\alpha^* \xi_\alpha}
\notag \\
&=& 
\prod_\alpha 1 = <0| \kappa \Unitmatrix |0> = \kappa
\end{eqnarray}

\noindent
Hence $\kappa = 1$.

\bigskip
The closure relation is very useful in order to define the trace of 
a fermionic operator $\mathcal{O}$. Using the previously defined 
expression of the trace in the fermion Fock space given in Negele and Orland's
book \cite{OrlandNegele}
\index{Path integrals!Trace in Fock space}

\begin{eqnarray}
Tr \left[ \mathcal{O}  \right] &=&
\underset{\{ n_{i,\sigma} \in \left[ 0,1 \right] \} }{\sum}
\left(\underset{i}{\bigotimes} <n_{i,\uparrow}, n_{i,\downarrow}|
\right) \mathcal{O} 
\left(\underset{i}{\bigotimes} |n_{i,\uparrow}, n_{i,\downarrow}>
\right)
\notag \\
&\equiv& \underset{n}{\sum} <n| \mathcal{O} |n>
\end{eqnarray}

\noindent
and inserting relation \eqref{ClosureRelation} into the trace of the 
operator $\mathcal{O}$, we obtain :

\begin{eqnarray}
Tr \left[ \mathcal{O} \right] 
&=& \int \prod_\alpha d\xi_\alpha^* d\xi_\alpha 
e^{-\sum_\alpha \xi_\alpha^* \xi_\alpha} \sum_n <n|\xi><\xi|\mathcal{O}|n>
\notag \\
&=& \int \prod_\alpha d\xi_\alpha^* d\xi_\alpha 
e^{-\sum_\alpha \xi_\alpha^* \xi_\alpha}<-\xi|\mathcal{O} 
\sum_n |n><n| |\xi>
\notag \\
Tr \left[ \mathcal{O} \right] &=& \int \prod_\alpha d\xi_\alpha^* d\xi_\alpha 
e^{-\sum_\alpha \xi_\alpha^* \xi_\alpha} <-\xi|\mathcal{O}|\xi>
\label{TraceO}
\end{eqnarray}

\noindent
where $<-\xi|\mathcal{O}\left( \{ f_\alpha \}, \{f_\alpha^\dagger \} \right)
|\xi> 
= e^{-\sum_\alpha \xi^{*}_\alpha \xi_\alpha}
\mathcal{O}\left(\{-\xi_\alpha \}, \{\xi^{*}_\alpha \} \right)$.

\bigskip
These results complete the tools which enable the construction of
the path integral \eqref{Zmu}.

\subsection{Partition function of many-body systems 
\label{SubsectionPartitionFunction}}

In this subsection we construct the partition function
of a many-body Hamiltonian in a path integral formulation and we follow
the prescription given in \cite{OrlandNegele}. 
We start with the grand-canonical partition function of a general fermionic
Hamiltonian $\tilde{H}$ in Fock space 
\index{Path integrals!Partition function}

\begin{eqnarray}
\mathcal{Z} = Tr \left[ e^{-\beta \left(
\tilde{H}-\mu \tilde{N} \right)} \right]
\label{Zcanonic}
\end{eqnarray}

\noindent
Using the Lie-Trotter relation
$\underset{M \rightarrow \infty}{\lim} 
\left( e^{-\mathcal{A}/M} e^{-\mathcal{B}/M} \right)^M 
= e^{-\left(\mathcal{A}+\mathcal{B} \right)}$ equation \eqref{Zcanonic}
can be reexpressed as $\mathcal{Z} = \underset{M \rightarrow \infty}{\lim}
Tr \left[ \left( e^{- \varepsilon \left( \tilde{H}-\mu \tilde{N} \right) }
\right)^M \right]$, with $\varepsilon = \beta /M$. Inserting the closure
relation \eqref{ClosureRelation} between each operator
$e^{- \varepsilon \left( \tilde{H}-\mu \tilde{N} \right) }$ and using
the expression of the trace over coherent states equation \eqref{TraceO}, the
partition function takes the form
\index{Lie-Trotter relation}

\begin{eqnarray}
\mathcal{Z} 
&=&
\underset{M \rightarrow \infty}{lim} 
\int \prod_{k=1}^M \prod_\alpha d\xi_{k,\alpha}^{*} d\xi_{k,\alpha}
< - \xi_{M}|e^{- \varepsilon \left( \tilde{H}-\mu \tilde{N} \right) } 
|\xi_{M-1}><\xi_{M-1}| \dots |\xi_k><\xi_k|
\notag \\
&\,& 
\dots |\xi_2>
<\xi_2|e^{- \varepsilon \left( \tilde{H}-\mu \tilde{N} \right) } |\xi_1>
\notag \\
&=& \underset{M \rightarrow \infty}{lim} 
\int \prod_{k=1}^M \prod_\alpha d\xi_{k,\alpha}^{*} d\xi_{k,\alpha}
e^{-S(\xi^*,\xi)}
\notag \\
-S(\xi^*,\xi) &=& 
\varepsilon \sum_{k=2}^M \left[ 
\sum_\alpha \xi_{\alpha,k}^{*} \{\frac{\xi_{\alpha,k}-\xi_{\alpha,k-1}}
{\varepsilon}-\mu \xi_{\alpha,k-1} \} 
+ H(\xi_{\alpha,k}^{*},\xi_{\alpha,k-1}) \right]
\notag \\
&+& \varepsilon \left[ \sum_\alpha \xi_{\alpha,1}^{*} \{
\frac{\xi_{\alpha,1}+ \xi_{\alpha,M}}
{\varepsilon}+\mu \xi_{\alpha,M} \}
+ H(\xi_{\alpha,1}^{*},-\xi_{\alpha,M}) \right]
\notag \\
\end{eqnarray}

\noindent
where $\alpha$ refers to the particle position and its spin, and
$k$ to the slice in which the term 
$e^{- \varepsilon \left( \tilde{H}-\mu \tilde{N} \right) }$ appear in the 
Lie-Trotter formula.
It is convenient to introduce the continuum notation $\xi_\alpha(\tau)$ 
to represent the set 
$\{ \xi_{\alpha,1}, \dots, \xi_{\alpha,k}, \dots, \xi_{\alpha,M} \}$ 
and in the limit of $M \rightarrow \infty$ to define 

\begin{eqnarray}
\xi_{\alpha,k}^{*} 
\frac{\left( \xi_{\alpha,k} - \xi_{\alpha,k-1} \right)}{\epsilon} \equiv
\xi_\alpha^{*} \left( \tau \right) \frac{\partial}{\partial \tau} 
\xi_\alpha \left( \tau \right)
\end{eqnarray}

\noindent
The partition function $\mathcal{Z}$ of the many-body Hamiltonian 
$\tilde{H}$ with a chemical potential $\mu$ takes the functional integral form

\begin{eqnarray}
\mathcal{Z} = \int_{\xi_\alpha(\beta)=- \xi_\alpha(0)} 
\mathcal{D} \xi
e^{-\int_0^\beta d\tau \{ \sum_\alpha \xi_\alpha^*(\tau)
(\partial_\tau - \mu) \xi_\alpha(\tau) 
+ H(\xi_\alpha^*(\tau),\xi_\alpha(\tau)) \} } 
\notag \\
\label{PathIntegralZ}
\end{eqnarray}

\noindent
where $\mathcal{D} \xi \equiv \underset{M \rightarrow \infty}{\lim}
\overset{M}{\underset{k=1}{\prod}} \prod_\alpha d\xi_{k,\alpha}^{*} 
d\xi_{k,\alpha}$.
Expression \eqref{PathIntegralZ} is applicable for general fermionic 
Hamiltonian operators, in particular for the description of spin systems in
terms of path integrals. In the next subsection we introduce a Fourier 
transform with respect to the imaginary time $\tau$. This leads to
Matsubara frequencies which are shifted by the presence of the imaginary
chemical potential $\mu$.

\subsection{Modified Matsubara frequencies \label{SubsectionMatsubara}}

Here we show how the Matsubara frequencies are modified
by the introduction of an imaginary chemical potential in the partition
function of an spin-$1/2$ Heisenberg model. As we already saw above 
Popov and Fedotov introduced the imaginary chemical potential 
$\mu=i \frac{\pi}{2 \beta}$ in order to remove 
the \emph{unphysical} states of the Fock space as explained in 
section \ref{sectionPF} and in \cite{Popov-88}. 
Introducing this chemical potential in the term
$\int_0^\beta d\tau \{ \sum_\alpha \xi_\alpha^*(\tau)
(\partial_\tau - \mu) \xi_\alpha(\tau)$ of equation
\eqref{PathIntegralZ} the Fourier transform of $\xi_\alpha(\tau)$ reads

\begin{eqnarray}
\xi_\alpha (\tau) 
&=& 
\sum_{\omega_F} \xi_\alpha (\omega_F) e^{i \omega_F \tau}
\end{eqnarray}

\noindent
and its reverse

\begin{eqnarray}
\xi_\alpha (\omega_F) 
&=& 
\frac{1}{\beta} \int_0^\beta d\tau \xi_\alpha(\tau) e^{-i \omega_F \tau}
\end{eqnarray}

\noindent
Here $\omega_F = \frac{2 \pi}{\beta} (n+1/2)$ with $n \in \Z$  
are the well known fermionic Matsubara frequencies. This leads to
\index{Matsubara frequencies!Fermionic}

\begin{eqnarray}
\int_0^\beta d\tau \sum_\alpha \xi_\alpha^*(\tau)
(\partial_\tau - \mu) \xi_\alpha(\tau) =
\underset{\omega_F}{\sum} \sum_\alpha \xi_\alpha^*(\omega_F)
i (\omega_F - \frac{\pi}{2 \beta} ) \xi_\alpha(\omega_F)
\end{eqnarray}

\noindent
Since the chemical potential is imaginary we see that we can redefine
the fermionic Matsubara frequencies by the change of variable

\begin{eqnarray}
\tilde{\omega}_F 
= \omega_F - \frac{\pi}{2 \beta} 
= \frac{2 \pi}{\beta} (n+1/4)
\end{eqnarray}

\noindent
Redefining the Fourier transform of $\xi_\alpha(\tau)$ as

\begin{eqnarray}
\xi_\alpha(\tau)= \underset{\tilde{\omega}_F}{\sum} 
\xi_\alpha(\tilde{\omega}_F) e^{i \tilde{\omega}_F \tau}
\end{eqnarray}

\noindent
and shifting the partial derivative over $\tau$, 
$\frac{\partial}{\partial \tau} - \mu \rightarrow 
\frac{\partial}{\partial \tau}$, the partition function takes the form

\begin{eqnarray}
\mathcal{Z} &=& \int_{\xi_\gamma(\beta)=i\xi_\gamma(0)} 
\mathcal{D} \xi e^{-S \left( \xi^*, \xi \right)}
\notag \\
S (\xi_\alpha^{*}(\tau),\xi_\alpha(\tau)) &=&
\sum_{\alpha= \left(\vec{r}_i,\sigma \right)}
\left ( \xi_\alpha^{*}(\tau)\frac{\partial}{\partial \tau}\xi_\alpha(\tau)
\right)
+ H(\{\xi_\alpha^{*}(\tau)\},\{\xi_\alpha(\tau)\})
\label{PathIntegralZmu}
\end{eqnarray}

\noindent
The partition function itself does not change much by the 
introduction of the imaginary chemical potential $\mu=i \frac{\pi}{2 \beta}$.
However the antiperiodic integration condition 
$\xi_\gamma(\beta) = - \xi_\gamma(0)$ needs to be changed into 
$\xi_\gamma(\beta)=i\xi_\gamma(0)$ by modification of 
the Matsubara frequencies. 
\index{Matsubara frequencies!Modified Matsubara frequencies}
In the following chapters, we will show how these modified fermionic 
Matsubara frequencies $\tilde{\omega}_F=\frac{2 \pi}{\beta} (n+1/4)$ change 
the behaviour of the physical properties of spin-$1/2$ systems.


\chapter{Mean-field and fluctuation contributions 
to the magnetic properties of Heisenberg models \label{Chapter3}}

\minitoc
\newpage

This chapter intends to present and discuss applications of the
mean-field and loop expansion to the determination of physical
properties of antiferromagnetic Heisenberg-type systems in spatial
dimension D using the PFP. This original point of the present work was 
published in \cite{DRepjb-05}.
\index{Loop expansion}

Recent work on quantum spin systems discusses the possible existence of
spin liquid states and in two space dimensions
the competition or phase transition between spin liquid states and an 
antiferromagnetic N\'eel state which is naturally expected to describe 
Heisenberg type systems \cite{GhaemiSenthil-05,Morinari-05,
SenthilFisher-05,SenthilFisher-04}. It is also known that undoped 
superconducting systems show an antiferromagnetic phase 
\cite{LeeNagaosaWen-04}.
\index{Spin liquid states}
\index{Quantum fluctuations}

In the following we focus our attention on a mean-field N\'eel phase 
description of quantum spin systems described by Heisenberg models. 
More precisely we present below a detailed study of the magnetization and the 
parallel magnetic susceptibility of Heisenberg antiferromagnetic spin-1/2 
systems on $D$-dimensional lattices at finite temperature. The aim of the 
work is the study of the physical pertinence of the N\'eel state ansatz,
using the PFP, as a mean-field approximation in the temperature interval 
$0<T<T_c$ where $T_c$  is the critical temperature \cite{DRepjb-05}. In order 
to get a precise answer to this point we work out the quantum and thermal 
fluctuation contributions beyond the mean-field approximation under the 
constraint of \emph{strict} single site-occupancy 
\cite{Azakov-01,DRepjb-05,Dillen-05,KFO-01,Popov-88} 
which allows to avoid a Lagrange multiplier approximation 
\cite{ArovasAuerbach-88}. The results are also extended to anisotropic 
$XXZ$ systems and compared to those obtained in the framework of the 
spin-wave theory.
\index{Anisotropic spin model}

\newpage

\section{Nearest-Neighbour Heisenberg model \label{SectionNNH}}

In order to focus on the essential properties of a N\'eel state mean-field
we consider a Heisenberg model which present a bipartite spin lattice.
A bipartite lattice can be split into two disjoint sublattices $A$ and $B$, 
where $J_{ij}$ connects only $i \in A$ to $j \in B$ as defined in 
\cite{Auerbach-94}. The simplest isotropic Heisenberg model showing a 
bipartite lattice reads
\index{Bipartite lattice}
\index{Nearest-Neighbour model}
\index{N\'eel state}

\begin{eqnarray}
H = -\frac{1}{2} \underset{i,j}{\sum} 
J_{ij} \vec{S}_i.\vec{S}_j + \underset{i}{\sum}
\vec{B}_i .\vec{S}_i
\end{eqnarray}

\noindent
with 

\begin{eqnarray}
J_{ij} = J \underset{\vec{\eta} \in \{\vec{a}_1,
\dots, \vec{a}_D  \}}{\sum} \delta 
\left( \vec{r}_i - \vec{r}_j \pm \vec{\eta} \right)
\label{JNearestNeighbour}
\end{eqnarray}

\noindent
where $J$ is the negative antiferromagnet exchange coupling working between 
nearest neighbour sites $i$ and $j$
separated by the lattice vector $\vec{\eta} \in \{\vec{a}_1,
\dots, \vec{a}_D  \}$ on a D-dimensional lattice. Keeping the 
static external magnetic field always fixed in the $Oz$ direction
the Hamiltonian reads 

\begin{eqnarray}
H = - J \underset{<i,j>}{\sum} 
\vec{S}_i.\vec{S}_j + \underset{i}{\sum} B_i .S^{z}_i
\label{HNearestNeighbour}
\end{eqnarray}
where the sum $\underset{<i,j>}{\sum}$ runs over the nearest neighbour site 
$<i,j>$ at position $\vec{r}_i \in A$ and $\vec{r}_j \in B$. \\

\begin{figure}[h]
\center
\begin{pspicture}(3,3)
\psgrid[griddots=10,gridlabels=0,subgriddiv=1](3,3)
\multips(0,0)(0,2){2}{
\multips(0,0)(2,0){2}{
\psline[linewidth=1mm,linecolor=blue]{*->}(0,0)(0.5,0.5)}
\multips(1,1)(2,0){2}{
\psline[linewidth=1mm,linecolor=blue]{*->}(0,0)(0.5,0.5)}}
\multips(0,0)(0,2){2}{
\multips(0,1)(2,0){2}{
\psline[linewidth=1mm,linecolor=red]{*->}(0,0)(-0.5,-0.5)}
\multips(1,0)(2,0){2}{
\psline[linewidth=1mm,linecolor=red]{*->}(0,0)(-0.5,-0.5)}}
\end{pspicture}
\caption{A two dimensional bipartite lattice system in a N\'eel state.}
\label{Chapter3Fig1}
\end{figure}
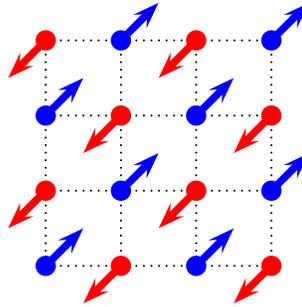

Figure \ref{Chapter3Fig1} is a two-dimensional representation of 
a bipartite lattice.
The spin sublattice $A$ and $B$ are represented by blue and red arrows,
dotted lines materialize the exchange interaction $J<0$ between spins 
(the blue and red arrows).
\index{Spin sublattice}
\index{Exchange interaction}

\section{Spin-wave theory \label{SectionSpinWave}}

A few years ago Coldea \emph{et al.} \cite{Coldea-01} 
measured the magnetic excitations of the square-lattice spin-$1/2$ 
antiferromgnet and high-$T_c$ parent compound $La_2 Cu O_4$. They showed
that the inclusion of some interactions beyond the nearest-neighbour
Heisenberg term \eqref{JNearestNeighbour} leads to a good 
description of the dispersion relation observed by high-resolution 
inelastic neutron scattering in the framework of spin-wave theory.
\index{Inelastic neutron scattering}
\index{Spin-Wave}
\index{Spin-Wave!Dispersion relation}

A modified spin-wave theory developed
by Takahashi \cite{Takahashi-89} led to the Auerbach and Arovas equations 
\cite{ArovasAuerbach-88} which were obtained by Schwinger-boson formulation.
\index{Schwinger boson}

In this section we calculate the sublattice magnetization as well as
the magnetic susceptibility in the framework of spin-wave theory applied on the
nearest neighbour antiferromagnet Heisenberg model \eqref{HNearestNeighbour}.
Later these results will be compared to those obtained using 
the Popov-Fedotov procedure with the Hamiltonian \eqref{HNearestNeighbour}.

\subsection{The Holstein-Primakoff approach}

In a broken symmetry phase such as the N\'eel state at least
one spin component shows a non-zero expectation value. For small temperatures
the fluctuations about this expectation value can be studied by means of the
Holstein and Primakoff (H-P) spin-deviation creation and annihilation boson 
operators $a_i$ and $a_i^\dagger$.
\index{N\'eel state!Broken symmetry}

Since the external magnetic field is applied in the $Oz$ direction the 
non-zero spin component is $S^{z}$. Following H-P we express the spin 
operators in the form \cite{HolsteinPrimakoff-40}
\index{Spin-Wave!Holstein-Primakoff transformation}

\begin{eqnarray}
S_i^{+} &=& \sqrt{2S - n_i^a} \, a_i
\notag \\
S_i^{-} &=& a_i^\dagger \, \sqrt{2S - n_i^a} 
\notag \\
S_i^{z} &=& S - n_i^a
\label{SHP}
\end{eqnarray}

\noindent
where $n_i^a = a_i^\dagger a_i = S - S_i^{z}$ is the so called 
spin-deviation operator, $a_i$ and $a_i^\dagger$ are boson operators.
Here $S$ is defined from the relation $\vec{S}^2 = S\left(S+1\right)$
of the quantum spin vector $\vec{S}$.
For small temperatures and/or for large-$S$ the expectation value of the 
spin-deviation operator $<n_i^a>$ is small compared to $2S$ and leads to
\index{Spin-Wave!Deviation operator}

\begin{eqnarray}
\sqrt{2S - n_i^a} = \left( 2S \right) \left( 1 - \frac{n_i^a}{4S}
- \frac{(n_i^a)^2}{32S^2} + \dots \right)
\label{sqrtHP}
\end{eqnarray}

\noindent
Since the Hamiltonian \eqref{HNearestNeighbour} shows a bipartite structure
on the lattice the spin expectation values are of opposite sign from
sublattice $A$ to sublattice $B$ as depicted in figure \ref{Chapter3Fig1}.
We can define two type of spin operators depending to which sublattice 
($A$ or $B$) they belong. Introducing \eqref{sqrtHP} into \eqref{SHP} 
and neglecting all terms $n_i^a / S$ the spin components for sublattice $A$
read

\begin{eqnarray}
S_{A,i}^{+} &=& \sqrt{2S} \, a_i
\notag \\
S_{A,i}^{-} &=& a_i^\dagger \, \sqrt{2S}
\notag \\
S_{A,i}^{z} &=& S - n_i^a
\label{SpinSubA}
\end{eqnarray}

\noindent
and for the sublattice $B$

\begin{eqnarray}
S_{B,j}^{+} &=& \sqrt{2S} \, a_j^\dagger 
\notag \\
S_{B,j}^{-} &=& a_j \, \sqrt{2S}
\notag \\
S_{B,j}^{z} &=& -S + n_j^a
\label{SpinSubB}
\end{eqnarray}

\noindent
Following Igarashi \cite{Igarashi-92}, Kubo \cite{Kubo-52}, Oguchi
\cite{Oguchi-59} and Takahashi \cite{Takahashi-89} the use of
\eqref{SpinSubA} and \eqref{SpinSubB} in \eqref{HNearestNeighbour} leads to

\begin{eqnarray}
H_{SW} &=& |J| \underset{i \in A}{\sum} 
\underset{
\begin{aligned}
j \in B,& \\
\vec{r}_j = \vec{r}_i + \vec{\eta}&
\end{aligned}}{\sum}
\left( S_{A,i}^z S_{B,j}^z + \frac{1}{2} \left[ S_{A,i}^{+} S_{B,j}^{-} +
S_{A,i}^{-} S_{B,j}^{-} \right]  \right)
\notag \\
&+& \sum_{i \in A} B_i S_{A,i}^z + \sum_{j \in B} B_j S_{B,j}^z
\notag \\
&=& - \frac{\mathcal{N}}{2} z |J| S^2
+ z |J| S \left[ \underset{i \in A}{\sum} \left(1 - \mathcal{B}_i \right) 
a_i^\dagger a_i
+ \underset{j \in B}{\sum} \left(1 + \mathcal{B}_j \right) 
a_j^\dagger a_j \right]
\notag \\
&&+ |J| S \underset{i \in A}{\sum} \underset{j \in B}{\sum}
\left( a_i a_j + a_i^\dagger a_j^\dagger \right)
\label{Hspinwave}
\end{eqnarray}

\noindent
where $\mathcal{N}$ is the total number of spin-$S$ in the system and 
$z=2D$ is the coordination of a spin in a D-dimensional 
hypercubical lattice. The magnetic field $\mathcal{B}_i$ is redefined as 
$\mathcal{B}_i = \frac{B_i}{z |J| S}$.
Since we admit that the expectation value of the spin-derivation 
operator $<n^a>$ is small only quadratic terms in the boson operator $a$ 
appear in the Hamiltonian \eqref{Hspinwave} which amounts to neglect 
the interaction between spin-waves.
\index{Coordination}
\index{Hypercubical lattice}

The spin-wave partition function $\mathcal{Z_{SW}}$ at finite 
temperature reads

\begin{eqnarray}
\mathcal{Z_{SW}} = 
Tr \left[ e^{\beta H_{SW}\left( \{ B_i \} \right)} \right]
\label{SWPartition}
\end{eqnarray}

\noindent
Magnetization and susceptibility of the spin system 
will be extracted from \eqref{SWPartition} taking derivatives with respect
to the magnetic field $B_i$.

\subsection{N\'eel Spin-wave magnetization \label{SubsectionSWM}}

In order to derive the sublattice magnetization $m_{A(B)}$ 
from the spin-wave free energy $\mathcal{F}_{SW}=-\frac{1}{\beta} \ln 
\mathcal{Z}_{SW}$ we set $B_i = B$ if $i$ belongs to sublattice $A$ sites 
and $B_i = -B$ if it belongs to sublattice $B$. Taking the derivative with 
respect to $B$ in \eqref{SWPartition} we obtain

\begin{eqnarray}
m_A &=& \frac{1}{\mathcal{N}} 
\left( \underset{i \in A}{\sum} < S_{A,i}^z >_{SW} 
- \underset{j \in B}{\sum} < S_{B,j}^z >_{SW} \right)
\notag \\
&=& - m_B = - \frac{1}{\mathcal{N} \beta} \frac{\partial}{\partial B} 
\ln {\mathcal{Z}_{SW}}{\Big\vert_{B=0}}
= \frac{1}{\mathcal{N}} \frac{\partial}{\partial B} 
\mathcal{F}_{SW}{\Big\vert_{B=0}}
\end{eqnarray}

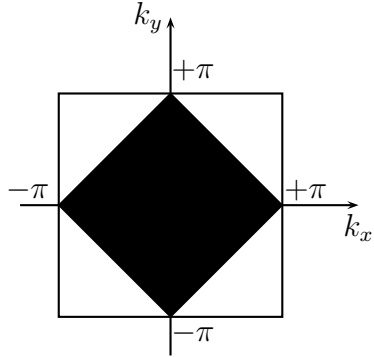
\begin{figure}[h]
\center
\begin{pspicture}(4,4)
\psframe(0,0)(3,3)
\psline{->}(-0.5,1.5)(4,1.5)
\rput(4,1.2){\text{$k_x$}}
\psline{->}(1.5,-0.5)(1.5,4)
\rput(1.2,4){\text{$k_y$}}
\pspolygon*(0,1.5)(1.5,3)(3,1.5)(1.5,0)
\rput(3.3,1.7){\text{$+\pi$}}
\rput(1.8,3.3){\text{$+\pi$}}
\rput(1.8,-0.2){\text{$-\pi$}}
\rput(-0.4,1.7){\text{$-\pi$}}
\end{pspicture}
\caption{Two dimensional Spin Brillouin Zone (shaded area)}
\label{Chapter3Fig2}
\end{figure}

\noindent
Define the Fourier transform of the boson operator $a$

\begin{eqnarray}
a_i &=& \frac{1}{\sqrt{\mathcal{N}_A/2}} \sum_{\vec{k} \in SBZ}
b_{\vec{k}}^{(1)} e^{i \vec{k}.\vec{r}_i}
\label{FourierA1} \\
a_j &=& \frac{1}{\sqrt{\mathcal{N}_B/2}} \sum_{\vec{k} \in SBZ}
b_{\vec{k}}^{(2)} e^{i \vec{k}.\vec{r}_j}
\label{FourierA2}
\end{eqnarray}

\noindent
where $SBZ$ is the Spin Brillouin Zone which is shown in Figure 
\ref{Chapter3Fig2} (shaded area) inside the lattice Brillouin Zone of a 
two dimensional bipartite spin system (large square).
More details on the construction of the Spin Brillouin Zone
are given in appendix \ref{AppendixSpinBrillouinZone}.
\index{Spin Brillouin Zone}
\index{Bipartite lattice}

\noindent
Fourier transforming \eqref{Hspinwave} with the definitions \eqref{FourierA1}
and \eqref{FourierA2} the Hamiltonian goes over to

\begin{eqnarray}
H_{SW} &=&
- \frac{\mathcal{N}}{2} z |J| S^2 + \mathcal{N} z |J| \mathcal{B} S^2
+ z |J| S \underset{\vec{k} \in SBZ}{\sum} \left[\left(1-\mathcal{B}\right)
b_{\vec{k}}^{(1)\dagger} b_{\vec{k}}^{(1)} + \left( 1 - \mathcal{B}\right) 
b_{\vec{k}}^{(2)\dagger} b_{\vec{k}}^{(2)}\right]
\notag \\
&&+ z |J| S \underset{\vec{k} \in SBZ}{\sum} \gamma_{\vec{k}}
\left[b_{\vec{k}}^{(1)} b_{-\vec{k}}^{(2)} 
+ b_{\vec{k}}^{(1)\dagger} b_{-\vec{k}}^{(2)\dagger} \right]
\label{HspinwaveB}
\end{eqnarray}

\noindent
where $\gamma_{\vec{k}}$ is the Fourier transform of 
$\sum_{\vec{\eta}} \delta \left( \vec{r}_i - \vec{r_j} + \vec{\eta} \right)$ 
and $\vec{\eta}$ was defined in equation \eqref{JNearestNeighbour}, section
\ref{SectionNNH}. $\gamma$ is related to $\vec{\eta}$ by

\begin{eqnarray}
\gamma_{\vec{k}} = \frac{1}{z} \underset{\vec{\eta} \in \{\pm \vec{a}_1,
\dots, \pm \vec{a}_D  \} }{\sum} e^{i \vec{k}.\vec{\eta}}
\label{GammaK}
\end{eqnarray}

\noindent
Notice that the coordination $z$ is simply related to the lattice vectors
by $z = \underset{\vec{\eta} \in \{\pm \vec{a}_1,
\dots, \pm \vec{a}_D  \} }{\sum} 1$. The Hamiltonian \eqref{HspinwaveB} can
be diagonalized by means of a Bogoliubov transformation applied to
the boson operators $b,b^{\dagger}$ \cite{Bogoliubov-47,Bogoliubov-58}
\index{Bogoliubov transformation}

\begin{eqnarray}
\left(
\begin{array}{c}
b_{\vec{k}}^{(1)} \\
b_{-\vec{k}}^{(2)\dagger}
\end{array}
\right)
=
\left[
\begin{array}{cc}
u_{\vec{k}} & v_{\vec{k}} \\
v_{\vec{k}} & u_{\vec{k}}
\end{array}
\right]
\left(
\begin{array}{c}
\beta_{(+),\vec{k}} \\
\beta_{(-),-\vec{k}}^\dagger
\end{array}
\right)
\label{SWBogoliubov}
\end{eqnarray}

\noindent
where $u_{\vec{k}}$ and $v_{\vec{k}}$ are real coefficients verifying
$u_{\vec{k}}^2 - v_{\vec{k}}^2 =1$ as a consequence of the commutation 
relations obeyed by the boson operators 
$\beta_{(+),\vec{k}}$ and $\beta_{(-),\vec{k}}$.
We set $u_{\vec{k}} = \cosh \theta_{\vec{k}}$ and 
$v_{\vec{k}} = \sinh \theta_{\vec{k}}$. Using \eqref{SWBogoliubov} 
leads to the diagonalized Hamiltonian \eqref{HspinwaveB}

\begin{eqnarray}
H_{SW} &=& - \frac{\mathcal{N}}{2} z |J| S^2 
+ z |J| \mathcal{N} \mathcal{B} S^2 
- \frac{\mathcal{N}}{2} z |J| S \left( 1 - \mathcal{B} \right)
\notag \\
&&+ \underset{\vec{k} \in SBZ}{\sum} \omega_{\vec{k}}
\left( \beta_{(+),\vec{k}}^\dagger \beta_{(+),\vec{k}} 
+ \beta_{(-),\vec{k}}^\dagger \beta_{(-),\vec{k}}
+ 1 \right)
\label{HspinwaveDiag}
\end{eqnarray}

\noindent
with

\begin{eqnarray}
\tanh 2 \theta_{\vec{k}} &=& - \frac{\gamma_{\vec{k}}}{\left(1 -
\mathcal{B} \right)}
 \\
\omega_{\vec{k}} &=& z |J| S 
\sqrt{\left(1-\mathcal{B} \right) - \gamma_{\vec{k}}^2 }
\end{eqnarray}

\noindent
Here $\omega_{\vec{k}}$ is the spin-wave spectrum and 
$\beta_{(\pm),\vec{k}}^\dagger \beta_{(\pm),\vec{k}}$ is the number operator 
of \emph{magnons} occupying the energy "level" $\omega_{\vec{k}}$.
In the absence of the magnetic field $\mathcal{B}$ and for
$\vec{k}$ close to $\vec{Q}=0$ or $\vec{\pi}$ 
($\vec{\pi} = \left( \pi, \dots, \pi \right)$)
the dispersion relation shows a relativistic spectrum 
$\omega_{\vec{k}} \sim c |\vec{k} - \vec{Q}|$
where $c$ is the spin-wave velocity 
for a D-dimensional hypercubical lattice, $c = \sqrt{D} |J| S $.
\index{Magnons}

The sublattice magnetization $m_A = - m_B$ is derived from the spin-wave
free energy 
$\mathcal{F}_{SW} = - \frac{1}{\beta} \ln Tr \left[ e^{-\beta H_{SW}}\right]$
and reads
\index{Spin-Wave!Free energy}
\index{Spin-Wave!SW Magnetization}

\begin{eqnarray}
m_A &=& 1 - \frac{1}{\mathcal{N}} \underset{\vec{k} \in SBZ}{\sum}
\frac{1}{\tanh \frac{\beta}{2} \omega_{\vec{k}}}.
\frac{1}{\sqrt{1 - \gamma_{\vec{k}}^2}}
\notag \\
&=& 1 - \frac{1}{\mathcal{N}} \underset{\vec{k} \in BZ}{\sum}
\left( n_{\vec{k}} + \frac{1}{2} \right).
\frac{1}{\sqrt{1 - \gamma_{\vec{k}}}}
\label{SpinWaveMagnetization}
\end{eqnarray}

\noindent
$n_{\vec{k}} = \frac{1}{e^{\beta \omega_{\vec{k}}}-1}$ is the boson 
occupation number. Notice that in the second equation of $m_A$ the sum
runs over the whole Brillouin Zone. Later the magnetization
\eqref{SpinWaveMagnetization} will be used as a reference for comparaison
with the magnetization obtained by means of our method using the PFP.

\subsection{N\'eel Spin-wave susceptibility}

With a uniform magnetic field $B_i = B$ applied
on the spin system the spin-wave susceptibility is obtained from
the free energy by

\begin{eqnarray}
\chi_\parallel = - \frac{\partial^2 \mathcal{F}_{SW}}{\partial B^2}
\Big{\vert}_{B=0}
\end{eqnarray}

\noindent
where $\mathcal{F}_{SW} = - \frac{1}{\beta} \ln \mathcal{Z}_{SW} =
- \frac{1}{\beta} \ln Tr \left[ e^{-\beta H_{SW}}\right]$ and the 
spin-wave Hamiltonian $H_{SW}$

\begin{eqnarray}
H_{SW} 
&=& - \frac{\mathcal{N}}{2} z |J| S^2
+ z |J| S \left[ \underset{i \in A}{\sum} \left(1 - \mathcal{B} \right) 
a_i^\dagger a_i
+ \underset{j \in B}{\sum} \left(1 + \mathcal{B} \right) 
a_j^\dagger a_j \right]
\notag \\
&&+ |J| S \underset{i \in A}{\sum} \underset{j \in B}{\sum}
\left( a_i a_j + a_i^\dagger a_j^\dagger \right)
\end{eqnarray}

\noindent
A Bogoliubov transformation leads to the diagonalized spin-wave Hamiltonian

\begin{eqnarray}
H_{SW} &=& - \frac{N}{2} z |J| S^2 + z |J| S \underset{\vec{k} \in SBZ}{\sum}
\Big[ \omega_{\vec{k}} + \left(\omega_{\vec{k}} + \mathcal{B} \right)
\beta_{(+),\vec{k}}^\dagger \beta_{(+),\vec{k}}
\notag \\
&&+ \left(\omega_{\vec{k}} - \mathcal{B} \right)
\beta_{(-),\vec{k}}^\dagger \beta_{(-),\vec{k}}  \Big]
 \\
\tanh \left( 2 \theta_{\vec{k}}\right) &=& - \gamma_{\vec{k}}
 \\
\omega_{\vec{k}} &=& z |J| S \sqrt{1 - \gamma_{\vec{k}}^2}
\end{eqnarray}

\noindent
$\mathcal{B}= B/{z |J| S}$. The free energy is then given by
\index{Spin-Wave!Free energy}

\begin{eqnarray}
\mathcal{F}_{SW} &=& - \frac{1}{\beta} \ln \mathcal{Z}_{SW}
\notag \\
&=& - \frac{1}{\beta} \underset{\vec{k} \in SBZ}{\sum}
\Bigg[ - \beta \omega_{\vec{k}} - \ln \left[ 
1 - e^{- \beta \left(\omega_{\vec{k}} + \mathcal{B} \right)} \right]
\left[ 1 - e^{- \beta \left(\omega_{\vec{k}} - \mathcal{B} \right)} \right]
\Bigg]
\end{eqnarray}

\noindent
Finally the spin-wave susceptibility reads
\index{Spin-Wave!SW Susceptibility}

\begin{eqnarray}
\chi_\parallel &=& - \frac{\partial^2 \mathcal{F}_{SW}}{\partial B^2}
\Big{\vert}_{B=0}
\notag \\
&=& 2 \beta \underset{\vec{k} \in SBZ}{\sum} n_{\vec{k}} \left(n_{\vec{k}}
+1 \right)
\label{SpinWaveChi}
\end{eqnarray}

\noindent
where $n_{\vec{k}} = \frac{1}{e^{\beta \omega_{\vec{k}}}-1}$ is the 
boson occupation number of the magnons.
\index{Magnons!Occupation number}

\section{Effective action \label{SectionEffectiveAction}}

After this review on spin-wave theory we return to the study of 
the Heisenberg model by functional integrals.
The aim of this work is to extract the magnetization and 
the spin susceptibility from the path integral formulation 
(see section \ref{chapterPathIntegralPFP}, equation \eqref{PathIntegralZmu}) 
for the Hamiltonian \eqref{HNearestNeighbour} of a bipartite spin 
system in a N\'eel state.
As we shall see later the analytic development of the effective action
necessitates some approximations. 
In order to be able to appreciate the
pertinence of these approximations, that is the N\'eel mean-field with 
the one-loop corrections for example, we define a ``Ginzburg-Landau 
parameter'' which evaluates the relative importance of the quantum and thermal 
fluctuations on the mean-field N\'eel state. We compare the magnetization 
and the susceptibility worked out from functional integral formulation 
with those obtained from the spin-wave theory for the Hamiltonian 
\eqref{HNearestNeighbour}.
\index{Ginzburg-Landau parameter}

\subsection{The Hubbard-Stratonovich transform 
\label{Chapter3Section3Subsection1} }

As was shown in chapter \ref{chapterPathIntegralPFP} the Hamiltonian
\eqref{HNearestNeighbour} can be expressed in terms of creation and 
annihilation fermion operators $f^\dagger$ and $f$ by replacing the 
spin operator $\vec{S}_i$ by \eqref{FermionizedSpin}. 
Performing this transformation and injecting the resulting
fermionic Hamiltonian \eqref{HNearestNeighbour} into the path integral 
\eqref{PathIntegralZmu} we see that the term $\vec{S}_i.\vec{S}_j$ 
in the exponent leads to a quartic expression in the \emph{Grassmann} variables

\begin{eqnarray}
S\left( \xi^{*},\xi \right) &=& \underset{\alpha = \vec{r}_i,\sigma}{\sum}
\left( \xi^\dagger_\alpha \left(\tau\right) \frac{\partial}{\partial \tau}
\xi_\alpha \left(\tau\right) \right)
+ H\left( \xi^{*},\xi \right)
\notag \\
H\left( \xi^{*},\xi \right) &=&
\underset{i}{\sum} B_i.\frac{1}{2}\left(
\xi^{*}_{i,\uparrow}(\tau) \xi_{i,\uparrow}(\tau) - 
\xi^{*}_{i,\downarrow}(\tau) \xi_{i,\downarrow}(\tau) \right)
\notag \\
&-& \underset{<i,j>}{\sum} J \left( \xi^{*}_{i,\sigma_1}(\tau) 
\vec{\sigma}_{\sigma_1 \sigma_2} \xi_{i,\sigma_2}(\tau) \right).
\left( \xi^{*}_{j,\sigma_3}(\tau) \vec{\sigma}_{\sigma_3 \sigma_4}
\xi_{j,\sigma_4}(\tau) \right)
\label{GrassmannAction}
\end{eqnarray}

\noindent
where repeated indices mean summations over them. 
Recall that $\vec{\sigma}$ are 
the $SU(2)$ Pauli matrices defined in \ref{FermionizedSpin}. Since integration 
over the \emph{Grassmann} variables of the path integral 
\eqref{PathIntegralZmu} is not possible as it stands we have to reduce the 
action \eqref{GrassmannAction} to a quadratic expression by means of a 
Hubbard-Stratonovich (HS) transformation \cite{Hubbard-59,Stratonovich-58}. 
We define the Hubbard-Stratonovich auxiliary field action $S_0$ as
\index{Hubbard-Stratonovich auxiliary field}

\begin{eqnarray}
S_0\left[ \varphi^{'}(\tau) \right] =
\frac{1}{2} \underset{i,j}{\sum} \left( J^{-1} \right)_{ij} 
\vec{\varphi^{'}}_i(\tau).
\vec{\varphi^{'}}_j(\tau)
\end{eqnarray}

\noindent
where $\left( J^{-1}\right)_{ij}$ is the inverse of the coupling matrix 
$J_{ij}$ and $\vec{\varphi}$ is the Hubbard-Stratonovich auxiliary field. 
The matrix $\left(J^{-1}\right)_{ij}$ always exists if one
considers periodic bondary conditions on the spin system.
Adding the HS action $S_0$ to the fermionic Hamiltonian 
$H\left( \xi^{*},\xi \right)$ and performing the change of variable 
\index{Periodic bondaries}

\begin{eqnarray}
\vec{\varphi^{'}}_i(\tau) = \vec{\varphi}_i(\tau) 
+ \underset{j}{\sum} J_{ij} \vec{S}_j(\tau)
\label{VariableChangeVarphi}
\end{eqnarray}

\noindent 
where $\vec{S}_j(\tau) = \xi^{*}_{j,\sigma_1}(\tau)
 \vec{\sigma}_{\sigma_1 \sigma_2} \xi_{j,\sigma_2}(\tau)$ leads to
the Hubbard-Stratonovich transformed Hamiltonian which reads
\index{Hubbard-Stratonovich transformation}

\begin{eqnarray}
S_0\left[ \varphi^{'}(\tau) \right] + H\left( \xi^{*},\xi \right)
\rightarrow S_0\left[ \varphi(\tau) \right] + \underset{i}{\sum} 
\left(\vec{\varphi}_i(\tau) + \vec{B}_i \right).\vec{S}_i(\tau)
\end{eqnarray}

\noindent
and the partition function

\begin{eqnarray}
\mathcal{Z} &=& \frac{1}{\mathcal{Z}_0}
\int_{\varphi(\beta)=\varphi(0)} \mathcal{D}  \vec{\varphi}
\int_{\xi_{i\sigma}(\beta)=i\xi_{i\sigma}(0)} 
\mathcal{D} \xi
e^{-\int_0^\beta d\tau S\left[\varphi,\xi^{*},\xi \right] }
\label{PartitionFunctionSVarphiXi}
 \\
S\left[\varphi,\xi^{*},\xi\right] &=&
 \underset{i,\sigma}{\sum}
\xi_{i\sigma}^{*}(\tau) \frac{\partial}{\partial \tau} \xi_{i\sigma} (\tau)
+ S_0 \left[ \varphi(\tau) \right] +
\underset{i}{\sum} \left(\vec{\varphi}_i(\tau) 
+ \vec{B}_i\right).\vec{S}_i(\tau)
\label{SVarphiXi}
\end{eqnarray}

\noindent
Here $\mathcal{Z}_0$ stands for the partition function of the HS auxiliary
field and reads

\begin{eqnarray}
\mathcal{Z}_0
&=& \int_{\varphi(\beta)=\varphi(0)} \mathcal{D} \vec{\varphi}
e^{-\int_0^\beta d\tau S_0 \left[ \vec\varphi(\tau) \right]}
\end{eqnarray}

\noindent
and $\mathcal{D} \vec{\varphi} \equiv 
\underset{M \rightarrow \infty}{\lim} \overset{M}{\underset{k=1}{\prod}}
\underset{i}{\prod}
d\varphi^{x}_{i,k} d\varphi^{y}_{i,k} d\varphi^{z}_{i,k}$ as explained for
the \emph{Grassmann} variables in subsection \ref{SubsectionPartitionFunction}.

\subsection{Integration over the \emph{Grassmann} variables}

The action \eqref{SVarphiXi} can also be written as

\begin{eqnarray}
S\left[\varphi,\xi^{*},\xi\right] = S_0 \left[ \varphi(\tau) - B \right] +
 \underset{i}{\sum}
\left( 
\begin{array}{cc}
\xi^{*}_{i,\uparrow}(\tau) &
\xi^{*}_{i,\downarrow}(\tau)
\end{array}
\right)
M_i(\tau)
\left(
\begin{array}{c}
\xi_{i,\uparrow}(\tau) \\
\xi_{i,\downarrow}(\tau)
\end{array}
\right)
\label{SVarphiXiwithMatrixM}
\end{eqnarray}

\noindent
after the variable shift
$\vec{\varphi}_i(\tau) \rightarrow \vec{\varphi}_i(\tau) - \vec{B}_i$.
The matrix $M_i(\tau)$ contains the factor of the quadratic
terms in the \emph{Grassmann} variables

\begin{eqnarray}
M_i(\tau) =
\left[
\begin{array}{cc}
\frac{\partial}{\partial \tau} + \frac{1}{2} \varphi_i^{z}(\tau) &
\frac{1}{2} \varphi_i^{-}(\tau) \\
\frac{1}{2} \varphi_i^{+}(\tau) &
\frac{\partial}{\partial \tau} - \frac{1}{2} \varphi_i^{z}(\tau)
\end{array}
\right]
\end{eqnarray}

\noindent
Fourier transforming \eqref{SVarphiXiwithMatrixM} with the definitions
of subsection \ref{SubsectionMatsubara} one gets

\begin{eqnarray}
\int_0^{\beta} d\tau S\left[\varphi,\xi^{*},\xi\right] &=&
\frac{\beta}{2} \underset{\omega_{B,n}}{\sum} \underset{i,j}{\sum} 
\left( J^{-1}\right)_{ij}
\left[\vec{\varphi}_i(-\omega_{B,n})-\vec{B}_i \right]
\left[\vec{\varphi}_j(\omega_{B,n}) -\vec{B}_j \right]
\notag \\
&+& \beta
\underset{i}{\sum} \underset{\tilde{\omega}_{F,p},\tilde{\omega}_{F,q}}{\sum}
\left(
\begin{array}{cc}
\xi^{*}_{i,\uparrow}(\tilde{\omega}_{F,p}) &
\xi^{*}_{i,\downarrow}(\tilde{\omega}_{F,q})
\end{array}
\right)
M_i(p-q)
\left(
\begin{array}{c}
\xi_{i,\uparrow}(\tilde{\omega}_{F,q}) \\
\xi_{i,\downarrow}(\tilde{\omega}_{F,q})
\end{array}
\right)
\notag \\
\end{eqnarray}

\noindent
where $\omega_{B,n} \equiv \frac{2 \pi}{\beta} n$ are boson Matsubara 
frequencies since from equation \eqref{VariableChangeVarphi} the auxiliary 
field $\varphi$ needs to be periodic in $\tau$ so that 
$\vec{\varphi}_i(\beta)=\vec{\varphi}_i(0)$. The modified fermionic Matsubara
frequencies $\tilde{\omega}_{F,p} = \frac{2 \pi}{\beta} (p+1/4)$ are defined
in subsection \ref{SubsectionMatsubara}. After the Fourier transform the
matrix $M_i$ reads
\index{Matsubara frequencies!Modified Matsubara frequencies}

\begin{eqnarray}
M_i(p-q) =
\left[
\begin{array}{cc}
i \tilde{\omega}_{F,p} \delta_{p,q} + 
\frac{1}{2} \varphi_i^{z}(\tilde{\omega}_{F,p}-\tilde{\omega}_{F,q}) &
\frac{1}{2} \varphi_i^{-}(\tilde{\omega}_{F,p}-\tilde{\omega}_{F,q}) \\
\frac{1}{2} \varphi_i^{+}(\tilde{\omega}_{F,p}-\tilde{\omega}_{F,q}) &
i \tilde{\omega}_{F,p} \delta_{p,q} -
\frac{1}{2} \varphi_i^{z}(\tilde{\omega}_{F,p}-\tilde{\omega}_{F,q})
\end{array}
\right]
\end{eqnarray}

\noindent
Then after integration over the \emph{Grassmann} variables using equation
\eqref{AppendixGaussianGrassmannIntegral} one obtains

\begin{eqnarray}
\mathcal{Z} 
= \frac{1}{\mathcal{Z}_0} \int_{\varphi(\beta)=\varphi(0)} \mathcal{D} 
\vec\varphi e^{-S_{eff}\left[ \vec\varphi 
\right]}
\label{ZEffectiveAction}
\end{eqnarray}

\noindent
where the effective action $S_{eff}$ reads
\index{N\'eel state!Effective action}

\begin{eqnarray}
S_{eff} = \int_0^{\beta} S_0 \left[ \varphi(\tau) - B \right]
- \underset{i}{\sum} \ln \det \beta M_i
\label{EffectiveAction}
\end{eqnarray}

\section{Mean-field equation and One-loop contributions 
\label{Chapter3Section4} }
 
The mean-field equations are obtained from 
$\frac{\delta S_{eff}}{\delta \varphi}=0$ 
which is the stationnarity condition in the application of the
least action principle.
Assuming that the mean-field and fluctuations contributions of the
effective action $S_{eff}$ can be identified and separated,
the matrix $M$ can be decomposed into a mean-field part $G_0^{-1}$ and
a fluctuation contribution $M_1$ and thus reads
\index{Quantum fluctuations}

\begin{eqnarray}
M &=& -G_0^{-1} + M_1
\end{eqnarray}

\noindent
where the mean-field matrix $G_0^{-1}$ depend on the choice of the
mean-field Hubbard-Stratonovich auxiliary field $\vec{\bar{\varphi}}$ and
$M_1$ is composed of the fluctuations $\vec{\delta \varphi} = \vec{\varphi} -
\vec{\bar{\varphi}}$.
At first glance it does not seem obvious to choose the mean-field 
components of the HS auxiliary field $\vec{\bar{\varphi}}$.
From \eqref{VariableChangeVarphi} it is clear that
the mean-field of $\vec{\varphi}$ is related to the mean-field of the
spins and thus $\vec{\bar{\varphi}}$ should show the same symmetries as the 
spin mean-field.
When the temperature is increased it is expected that thermal fluctuations 
$\vec{\delta \varphi}$ become more and more important and the auxiliary field 
$\vec{\varphi}$ may move away from the mean-field $\vec{\bar{\varphi}}$. 
Considering the temperature limit $T \rightarrow 0$ the boson 
(and also fermion) Matsubara frequencies $\omega_{B,n}$ go to zero for 
any value of $n$. Then the relevant mean-field Fourier components of
$\vec{\bar{\varphi}}(\omega_{B,n})$ are those for which $\omega_{B,n}=0$.
Fourier transforming $M$ with respect to the imaginary 
time $\tau$ and extracting the mean-field part $\vec{\bar{\varphi}}$
from $M$ one obtains

\begin{eqnarray}
{G_0}_{p,q} = 
\begin{bmatrix}
-\frac{1}{\det G_p}
 \left[ i \omega_{F,p} - \frac{1}{2} \bar{\varphi}_i^z
(\omega_{F,p}-\omega_{F,q}) \right] \delta_{p,q}
 &
\frac{1}{\det G_p} \frac{1}{2} \bar{\varphi}_i^{-}
(\omega_{F,p}-\omega_{F,q}) \delta_{p,q}
 \\
\frac{1}{\det G_p} \frac{1}{2} \bar{\varphi}_i^{+}
(\omega_{F,p}-\omega_{F,q})\delta_{p,q}
 &
-\frac{1}{\det G_p} \left[ i \omega_{F,p} + \frac{1}{2} \bar{\varphi}_i^z
(\omega_{F,p}-\omega_{F,q}) \right] \delta_{p,q}
\end{bmatrix}
\notag \\
\end{eqnarray}

\noindent
and the fluctuating part $\vec{\delta \varphi}$ are given by

\begin{eqnarray}
{M_1}_{p,q} = 
\begin{bmatrix}
\frac{1}{2} \delta \varphi_i^z(\omega_{F,p}-\omega_{F,q})
 &
\frac{1}{2} \delta \varphi_i^{-}(\omega_{F,p}-\omega_{F,q})
 \\
\frac{1}{2} \delta \varphi_i^{+}(\omega_{F,p}-\omega_{F,q})
 &
-\frac{1}{2} \delta \varphi_i^z(\omega_{F,p}-\omega_{F,q})
\end{bmatrix}
\end{eqnarray}

\noindent
with $\delta \vec{\varphi}_i(\omega_{F,p}- \omega_{F,q})=
\vec{\varphi}_i(\omega_{F,p}- \omega_{F,q})- 
\vec{\bar{\varphi}}_i(\omega_{F,p}- \omega_{F,q})\delta_{p,q}$ and 
$\det G_p =$\\$ -\left[\omega_{F,p}^2 + \left(\frac{\vec{\bar{\varphi}}_\alpha
(\omega_{F,p}- \omega_{F,q}=0)}{2}\right)^2\right]$. 
The expression $\ln \det (\beta M)$ in the effective action 
\eqref{EffectiveAction} can now be developed into a series 

\begin{eqnarray*}
\ln \det (\beta M) &=& \ln \det \beta \left[-G_0^{-1}(1-G_0 M_1)\right]\\
&=& \ln \det (-\beta G_0^{-1}) + Tr\,\ln (1-G_0 M_1) \\
&=& \ln \det (-\beta G_0^{-1}) - 
Tr\{ \sum_{n=1}^{\infty} \frac{1}{n} (G_0 M_1)^n \}
\end{eqnarray*}
The first term $\ln \det (-\beta G_0^{-1})$ leads to the expression 
$\sum_i \ln 2 \cosh \frac{\beta}{2} \| \vec{\varphi}_i(\omega_B=0) \|$ and
the effective action over the auxiliary field $\vec{\varphi}$ reads
\index{Loop expansion}

\begin{eqnarray}
S_{eff}\left[ \vec\varphi \right] 
= \int_0^\beta d\tau S_0 \left[ \vec\varphi(\tau) \right]
-\sum_i \ln 2 \cosh \frac{\beta}{2} \| \vec{\bar{\varphi}}_i(\omega_B=0) \|
+ Tr \left[ \sum_{n=1}^{\infty} \frac{1}{n} (G_0 M_1)^n \right]
\notag \\
\label{EffectiveAction2}
\end{eqnarray}

\noindent
The first term $n=1$ in the sum over $n$ gives the contributions at the 
first order in the fluctuations $\delta \varphi$, the second one
the one-loop correction to the mean-field for $n=2$. It is quadratic
in $\delta \varphi$. Hence in a loop expansion beyond the mean-field 
approximation $\vec{\bar{\varphi}}$ the effective action given by 
(\ref{EffectiveAction2}) is a Taylor series expansion in powers of 
$\vec{\delta \varphi}$. To second order (one-loop contribution) in the 
fluctuations $\vec{\delta \varphi}^2$ 
of $\vec{\varphi} = \vec{\bar{\varphi}} + \delta \vec{\varphi}$.
 
\begin{eqnarray}
S_{eff}\left[ \vec{\varphi} \right] =
{S_{eff}} {\Big\vert}_{\left[\vec{\bar{\varphi}}\right]}
+ \frac{\delta S_{eff}} {\delta \vec{\varphi}}
 {\Big\vert}_{\left[\vec{\bar{\varphi}}\right]} \vec{\delta \varphi}
+ \frac{1}{2} \frac{\delta^2 S_{eff}} {{\vec{\delta \varphi}}^2}
{\Big\vert}_{\left[\bar{\vec{\varphi}}\right]}
 {\delta \vec{\varphi}}^2 + \mathcal{O} ({\vec{\delta \varphi}}^3)
\end{eqnarray}

We now give a more precise definition of our mean-field $\vec{\bar{\varphi}}$.
The partition function \eqref{ZEffectiveAction} can be worked out by
means of a \emph{saddle-point} method in which 
$\frac{\partial S_{eff}}{\delta \vec{\varphi}}
{\Big\vert}_{\left[\bar{\vec{\varphi}}\right]} \delta \vec{\varphi} = 0$
and leads to the equation
\index{Saddle-point method}

\begin{eqnarray}
\mathcal{Z} = 
e^{-{S_{eff}} {\Big\vert}_{\left[\vec{\bar{\varphi}}\right]}}
.\int \mathcal{D} \delta \varphi 
e^{-\left( 
\frac{1}{2} \frac{\delta^2 S_{eff}} {\delta \vec{\varphi}^2}
{\Big\vert}_{\left[\bar{\vec{\varphi}}\right]}
 {\delta \vec{\varphi}}^2 + \mathcal{O} (\delta \vec{\varphi}^3)
\right)}
\end{eqnarray}

\noindent
where the mean-field solutions  verify the self-consistent set of equations

\begin{eqnarray}
\sum_j \left( J^{-1}\right)_{ij} 
\left[\vec{\bar{\varphi}}_j-\vec{B}_j \right]
=\frac{1}{2} \frac{\vec{\bar{\varphi}}_i}{\bar{\varphi}_i}
\tanh \left[ \frac{\beta\bar{\varphi}_i}{2} \right]
\label{MeanFieldEquation}
\end{eqnarray}

\noindent
These equations lead directly to the cancellation of the first order term in
$\vec{\delta \varphi}$ as worked out in appendix 
\ref{AppendixOneLoopCorrections}.

In the following we consider a N\'eel mean-field order
$\vec{\bar{\varphi}}_i(\tau) = (-1)^{\vec{\pi}.\vec{r}_{i}} 
\bar{\varphi}^z \vec{e}_z = \bar{\varphi}^z_i \vec{e}_z$ where 
$\vec{\pi}$ is the Brillouin spin sublattice vector as defined
in section \ref{SectionSpinWave}.
A magnetic field aligned along the direction $\vec{e}_z$ is applied to the 
system. The partition function can be decomposed into a product of three terms

\begin{eqnarray}
\mathcal{Z} = {\mathcal{Z}_{MF}}.{\mathcal{Z}_{zz}}
.{\mathcal{Z}_{+-}}
\end{eqnarray}

\noindent
where $\mathcal{Z}_{MF}$, $\mathcal{Z}_{zz}$ and
$\mathcal{Z}_{+-}$ are given by

\begin{eqnarray}
\mathcal{Z}_{MF} &=& e^{-{S_{eff}} {\Big\vert}_{\left[\bar
{\varphi}\right]} }
\label{PartitionFunctionMF}
 \\
\mathcal{Z}_{zz} &=& \frac{1}{\mathcal{Z}^{zz}_{0}}
\int \mathcal{D}\delta \varphi^z e^{-
\frac{1}{2} \frac{\partial^2 S_{eff}}{{\partial \varphi^z}^2}
{\Big\vert}_{\left[\bar{\varphi}\right]}
 {\delta \varphi^z}^2 }
\label{PartitionFunctionZZ}
 \\
\mathcal{Z}_{+-} &=& \frac{1}{\mathcal{Z}_0^{+-}}
\int \mathcal{D}(\delta \varphi^{+}, \delta \varphi^{-}) e^{-
\frac{1}{2} \frac{\partial^2 S_{eff}}{{\partial \varphi^{+}
\partial \varphi^{-}}}
{\Big\vert}_{\left[\bar{\varphi}\right]}
 {\delta \varphi^{+}}.{\delta \varphi^{-}} }
\label{PartitionFunctionPM}
\end{eqnarray}

\noindent
with

\begin{equation}
{S_{eff}} {\Big\vert}_{\left[\bar{\varphi}\right]}
 =
\frac{\beta}{2} \sum_{i,j} \left( J^{-1}\right)_{ij} 
\left[(\bar{\varphi}^z_i-B^z_i)
.(\bar{\varphi}^z_j-B^z_j) \right]
-\sum_i \ln 2 \cosh \frac{\beta}{2} 
\| \bar{\varphi}^z_i \|
\end{equation}

\noindent
The one-loop corrections $\left({\delta \varphi^z}\right)^2$ and 
$\delta \varphi^{+} \delta \varphi^{-}$ terms are worked out in details 
in appendix \ref{AppendixOneLoopCorrections} and read

\begin{align}
\frac{1}{2} \frac{\delta^2 S_{eff}}{{\delta \varphi^z}^2}
{\Big\vert}_{\left[\bar{\varphi}\right]}
 {\delta \varphi^z}^2
= &
\underset{\omega_B}{\sum} \sum_{i,j} \frac{\beta}{2}
\Bigg[
 \left( J^{-1}\right)_{ij} 
\notag \\
&-
\left(
\frac{\beta}{4} 
\tanh^{'} \left( \frac{\beta}{2} \bar{\varphi}_i^z \right)\right)
\delta_{ij} \delta(\omega_B=0) \Bigg]
\delta \varphi_i^z(-\omega_B)\delta \varphi_j^z(\omega_B)
\notag \\
\frac{1}{2} \frac{\partial^2 S_{eff}}{{\delta \varphi^{+}
\delta \varphi^{-}}}
{\Big\vert}_{\left[\bar{\varphi}\right]}
 {\delta \varphi^{+}} {\delta \varphi^{-}}
= &
\underset{\omega_B}{\sum} \sum_{i,j} \frac{\beta}{2}
\left[
\frac{1}{2} \left( J^{-1}\right)_{ij} - 
\left( \frac{1}{2} 
\frac{\tanh \left( \frac{\beta}{2} \bar{\varphi}_i^z \right) }
{\bar{\varphi}_i^z - i \omega_B } \right)
\delta_{ij} \right]
\delta \varphi_i^{+}(-\omega_B)\delta \varphi_j^{-}(\omega_B)
\notag \\
& +
\underset{\omega_B}{\sum} \sum_{i,j} \frac{\beta}{2}
\left[ \frac{1}{2} \left( J^{-1}\right)_{ij} \right]
\delta \varphi_i^{+}(\omega_B) \delta \varphi_j^{-}(-\omega_B)
\notag \\
\label{delta2Seff}
\end{align}

\noindent
$\mathcal{Z}_{MF}$ is the mean-field contribution, $\mathcal{Z}_{zz}$
and $\mathcal{Z}_{+-}$ are the one-loop contributions 
respectively for the longitudinal part $\delta \varphi^z$ and
the transverse parts of $\vec{\varphi}$, $\delta \varphi^{+-}$,
which take account of the fluctuations around the mean-field value 
$\bar{\varphi}^z$. 
 
The contributions $\mathcal{Z}_{zz}$ and $\mathcal{Z}_{+-}$ 
are quadratic in the 
field variables $\delta \varphi^z, \delta \varphi^{+-}$ and can
be worked out in the presence of a staggered magnetic field $B_i^z$. 
Studies involving a uniform magnetic field acting on 
antiferromagnet quantum spin systems can also be found in ref.\cite{KFO-01}.
\index{Staggered magnetic field}

\section{Magnetization and susceptibility for D-dimension\-al systems}

\subsection{Relation between the Hubbard-Stratonovich auxiliary field
and the magnetization \label{Chapter3Section5Subsection1}}

We define by $\mathcal{F}_{MF} \equiv -\frac{1}{\beta} \ln \mathcal{Z}_{MF}$ 
the mean-field free energy where $\mathcal{Z}_{MF}$ is given by equation
\eqref{PartitionFunctionMF}. The local fields $\{\vec{\bar{\varphi}}_i\}$ 
can be related to the local magnetizations $\{\vec{\bar{m}}_i\}$. Using  
$\vec{\bar{m}}_i = - \frac{\partial \mathcal{F}_{MF}}{\partial \vec{B}_i}$ 
one gets

\begin{eqnarray}
\vec{\bar{m}}_i &=& - \frac{\partial \mathcal{F}_{MF}}{\partial \vec{B}_i}
= \frac{1}{\beta} \frac{\partial }{\partial \vec{B}_i}
{S_{eff}} {\Big\vert}_{\left[\bar
{\varphi}\right]}
\notag \\
&=& \underset{j}{\sum} \left( J^{-1} \right)_{ij} 
\left[\vec{\bar{\varphi}}_j - \vec{B}_j \right]
\end{eqnarray}

\noindent
and from the mean-field relation \eqref{MeanFieldEquation} one deduces also
the relation

\begin{eqnarray}
\vec{\bar{m}}_i = \frac{1}{2} \frac{\vec{\bar{\varphi}}_i}{\bar{\varphi}_i}
\tanh \frac{\beta}{2} \bar{\varphi}_i
\end{eqnarray}

\noindent
where $\bar{\varphi}_i = \| \vec{\bar{\varphi}}_i \|$.
Considering the N\'eel state $\vec{\bar{\varphi}}_i 
= \bar{\varphi}_i^z \vec{e}_z$ and keeping the external magnetic field 
$\vec{B}_i$ applied in the direction $Oz$ one gets

\begin{eqnarray}
\bar{\varphi}_j^z &=& \frac{2}{\beta} {\tanh}^{-1} 2 \bar{m}_i 
\label{Relation1}
 \\
\bar{\varphi}_j^z - B_j &=& \sum_j J_{i,j} \bar{m}_j
\label{Relation2}
\end{eqnarray}

\noindent
Combining \eqref{Relation1} and \eqref{Relation2} the self-consistent 
mean-field equation of the magnetization $\bar{m}_i$ is obtained by means of 
the set of equations

\begin{eqnarray}
\frac{2}{\beta} {\tanh}^{-1} 2 \bar{m}_i &=&
B_i + \sum_j J_{i,j} \bar{m}_j
\notag \\
\bar{m}_i &=& \frac{1}{2} \tanh \left[ 
\frac{\beta}{2} \left( B_i + \sum_j J_{i,j} \bar{m}_j \right) \right]
\label{MagnetizationMeanFieldEquation}
\end{eqnarray}
\index{N\'eel state!Self-consistent equation}

\subsection{Linear response theory}

The magnetization mean-field equation \eqref{MagnetizationMeanFieldEquation}
can be solved in the linear response theory. If the applied magnetic 
field is weak enough the mean-field magnetization $\bar{m}_i$ can be 
developed linearly with respect to the magnetic field

\begin{eqnarray}
\bar{m}_i &=& (-1)^{\vec{r}_i.\vec{\pi}} \bar{m} + \Delta m B_i 
+ \mathcal{O}(B^2)
\label{LinearMeanFieldMagnetization}
\end{eqnarray}

\noindent
where $\vec{\pi}$ is the Brillouin vector coming from the existence of
the sublattices in the N\'eel state defined below \eqref{MeanFieldEquation}, 
$\vec{r}_i$ the lattice position,
$\bar{m}$ the sublattice magnetization and $\Delta m$ the linear
coefficient in the magnetic field $B_i$.
According to the dependence of the magnetic field on the site
$i$ of the lattice and by inspection of 
\eqref{LinearMeanFieldMagnetization} and \eqref{MagnetizationMeanFieldEquation}
the linear coefficient $\Delta m$ reads
\index{Brillouin vector}

\begin{eqnarray}
\Delta m =
\begin{cases}
\Delta \tilde{m}_0 = \frac{\frac{\beta}{4}\left(1-4\bar{m}^2 \right)}
{1-\frac{\beta}{2}D|J|\left(1-4\bar{m}^2 \right)}
\quad \text{, when $B_i = (-1)^{\vec{r}_i.\vec{\pi}} B$}
\\
\Delta m_{\chi 0} = \frac{ \frac{\beta}{4}\left( 1 - 4\bar{m}^2 
\right) }{1+\frac{\beta}{2}D|J|\left(1-4\bar{m}^2 \right)}
\quad \text{, when $B_i = B$.}
\end{cases}
\label{DeltaM}
\end{eqnarray}

\noindent
where $\bar{m}$ is the mean-field sublattice magnetization without 
external magnetic field which verifies

\begin{eqnarray}
\bar{m} = \frac{1}{2} th \frac{\beta}{2}D|J|\bar{m}
\label{MeanFieldEquation2}
\end{eqnarray}

\noindent
The solution depends on the dimension $D$ and the coupling
constant $J$ of the nearest-neighbour Heisenberg model 
\eqref{HNearestNeighbour}. The self-consistent equation 
\eqref{MeanFieldEquation2} is easily solved numerically using the 
Newton method \cite{NumericalRecipies}. As can be seen on 
figure \ref{fig:MagP3D100} the mean-field magnetization saturates at
$1/2$ for the temperature $T=0$ and vanishes at the critical temperature
$T_c \equiv {D |J|}/2$.
Depending on the configuration of the magnetic field one can either
compute the magnetization with $B_i = (-1)^{\vec{r}_i.\vec{\pi}} B$ or
the susceptibility with $B_i = B$ as was previoulsy explained in
section \ref{SectionSpinWave} concerning spin-wave theory.

\subsection{N\'eel magnetization with fluctuation corrections : results}

Substituting equation \eqref{LinearMeanFieldMagnetization} with
the magnetic field configuration $B_i = (-1)^{\vec{r}_i.\vec{\pi}} B$ 
and relations \eqref{Relation1},\eqref{Relation2} 
into equations \eqref{PartitionFunctionMF}, \eqref{PartitionFunctionZZ}
and \eqref{PartitionFunctionPM}
integrating over the HS auxiliary fluctuation field $\vec{\delta \varphi}$
one obtains the free energy $\mathcal{F}$. The derivation of the free energy 
is given in details in appendix \ref{AppendixOneLoopCorrections}.
The components $\mathcal{F}_{MF}, \mathcal{F}_{zz}$ and 
$\mathcal{F}_{+-}$ of the free energy for a linear approximation in the
magnetic field read
\index{N\'eel state!Free energy}

\begin{eqnarray}
\mathcal{F}_{MF} =
\mathcal{N} D |J| \left( \bar{m}+{\Delta \tilde{m}_0}.B \right)^2
- \frac{\mathcal{N}}{\beta} \ln \cosh \left( \frac{\beta}{2} 
\left[ B+2 D |J|(\bar{m}+{\Delta \tilde{m}_0.B})\right] \right)
\notag \\
\end{eqnarray}

\begin{eqnarray}
\delta\mathcal{F}_{zz} =
\frac{1}{2\beta} \underset{\vec{k} \in SBZ}{\sum}
\ln \left[ 1 - \left( \frac{\beta D |J| \gamma_{\vec{k}} }{2} \right)^2
\left[ 1- 4(\bar{m}+{\Delta \tilde{m}_0.B})^2 \right]^2 \right]
\end{eqnarray}

\begin{align}
\delta\mathcal{F}_{+-} = &
\frac{2}{\beta} \underset{\vec{k} \in SBZ}{\sum}
\ln \Bigg(
\notag \\
&
\frac{\sinh \left(\frac{\beta}{2}
\left(\left[B + 2 D |J|(\bar{m}+{\Delta \tilde{m}_0.B})\right]^2
- \left[2 D |J| \gamma_{\vec{k}} (\bar{m}+{\Delta \tilde{m}_0.B})\right]^2 
\right)^{1/2} \right)}
{\sinh \left(\frac{\beta}{2} 
\left[ B+2 D |J|(\bar{m}+{\Delta \tilde{m}_0.B})\right] \right)}
\Bigg)
\notag \\
\end{align}

\noindent
The magnetization $m$ on site $i$ is 
the sum of a mean-field contribution $\bar{m} = -\frac{1}{\beta} 
\frac{\partial ln \mathcal{Z}_{MF}}{\partial B^z}$, a transverse contribution 
$\delta m_{+-} = -\frac{1}{\beta} \frac{\partial ln \mathcal{Z}_{+-}}
{\partial B^z}$ and a longitudinal contribution $\delta m_{zz} = -\frac{1}
{\beta} \frac{\partial ln \mathcal{Z}_{zz}}{\partial B^z}$. 
For a small magnetic field $\vec B$ a linear approximation leads to 
$m = \bar{m} + \delta m_{zz} + \delta m_{+-}$ where
\index{N\'eel state!PFP magnetization}

\begin{align}
\bar{m} =& \frac{1}{2} \tanh \frac{\beta}{2}D|J|\bar{m}
 \\
\delta m_{zz} =&
 -\frac{1}{\mathcal{N} \beta} \underset{\vec{k} \in SBZ}{\sum}
\frac{8\bar{m} \Delta \tilde{m}_0 \left(1-4\bar{m}^2\right)
\left(\frac{\beta D |J| \gamma_{\vec{k}}}{2}\right)^2}
{\left[1 -\left(\frac{\beta D |J| \gamma_{\vec{k}}}{2}\right)^2 
 \left(1-4\bar{m}^2\right)^2\right]}
 \\
\delta m_{+-} =&
\frac{\left(1 + 2 D |J| \Delta \tilde{m}_0 \right)}{4\bar{m}}
\notag \\
& - \frac{1}{\mathcal{N}} \underset{\vec{k} \in SBZ}{\sum}
\frac{\left(1 + 2 D |J| \Delta \tilde{m}_0 (1 - \gamma_{\vec{k}}^2)\right)}
{\sqrt{1 - \gamma_{\vec{k}}^2}}
\frac{1}{\left[\tanh \left(\beta D |J| \bar{m}
\sqrt{1 - \gamma_{\vec{k}}^2} \right) \right]}
\notag \\
\end{align}

\noindent
$\mathcal{N}$ is the number of spin-1/2 sites, 
$\Delta \tilde{m}_{0} = \frac{\frac{\beta}{4}\left(1-4.\bar{m}^2 \right)}
{1-\frac{\beta}{2}D|J|\left(1-4.\bar{m}^2 \right)}$ and 
$\gamma_{\vec{k}} = \frac{1}{D} \underset{\vec{\eta}}{\sum} 
\cos (\vec{k}.\vec{\eta}) $ as defined in section \ref{SectionSpinWave}.

At low temperature $(T \rightarrow 0)$ it is seen that the magnetization 
goes over to the corresponding spin-wave expression 
\cite{Igarashi-92,Manousakis-91,HolsteinPrimakoff-40,Azakov-01} and reads

\begin{eqnarray}
m = 1 - \frac{1}{\mathcal{N}} \underset{\vec{k} \in SBZ}{\sum}
\frac{1}{\tanh
\left( \frac{\beta D |J|}{2}\sqrt{1-\gamma_{\vec{k}}^2} \right) }
.\frac{1}{\sqrt{1-\gamma_{\vec{k}}^2}}=\bar{m}+\delta m
\end{eqnarray}

\noindent 
where $\bar{m}=1/2$ is the mean-field contribution and 
$\delta m$ is generated by thermal and quantum fluctuations.

\begin{figure}
\centering
\epsfig{file=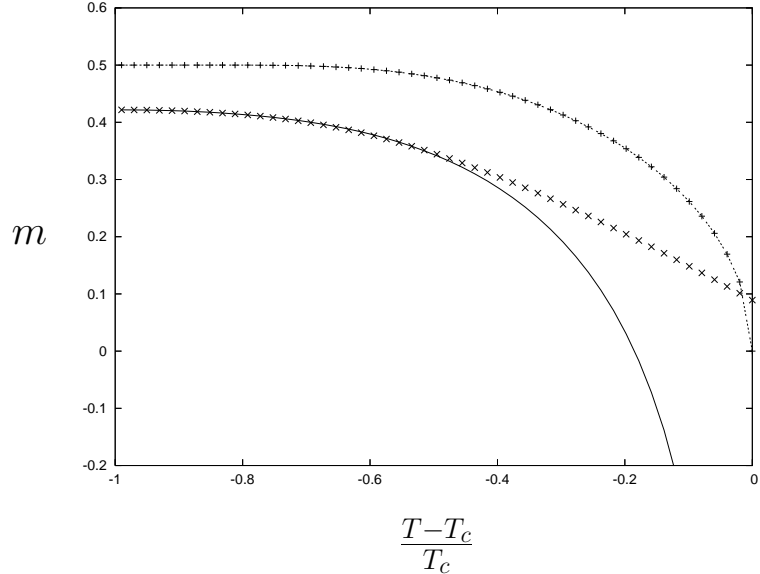,width=10cm}
\caption{Magnetization in a 3D Heisenberg antiferromagnet cubic 
lattice. 
Dotted line~: Mean-field magnetization, 
Dots : Spin-wave magnetization, 
Full line : One-loop corrected magnetization.}
\label{fig:MagP3D100}
\end{figure}

Figure \ref{fig:MagP3D100} shows the magnetization $m$ in the mean-field,
the one-loop and the spin-wave approach (as given in section 
\ref{SectionSpinWave}) for temperatures $T \leq T_c$
where $T_c = D|J|/2$ corresponds to a critical point. 
One observes a sizable
contribution of the quantum and thermal fluctuations generated by the 
loop contribution over the whole range of temperatures as well as an 
excellent and expected agreement between the quantum corrected and the 
spin-wave result at very low temperatures. 
\index{Spin-Wave}
 
The magnetization shows a singularity in the neighbourhood of the critical
point. This behaviour can be read from the analytical expressions of 
$\delta m_{+-}$ and $\delta m_{zz}$ and is generated by the $|\vec{k}| = 0$  
mode which leads to $\gamma_{\vec{k}} = 1$ and by cancellation of $\bar{m}$. 
The N\'eel state mean-field 
approximation is a realistic description at very low $T$. With increasing
temperature this is no longer the case. The chosen ansatz breaks a symmetry
whose effect is amplified as the temperature increases and leads to the
well-known divergence disease observed close to $T_c$. Hence if higher order
contributions in the loop expansion cannot cure the singularity the N\'eel 
state antiferromagnetic ansatz does not describe the physical symmetries of the
system at the mean-field level at temperatures in the neighbourhood of the
critical point. Consequently it is not a pertinent mean-field approximation for
the description of the system.

\begin{figure}
\centering
\epsfig{file=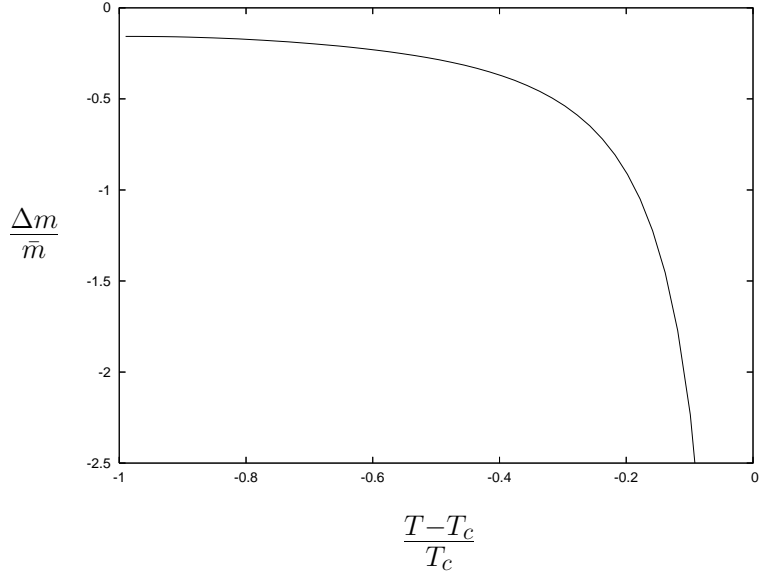,width=10cm}
\caption{Ginzburg criterion $\frac{\Delta m}{\bar{m}}$ 
for the 3D Heisenberg model.}
\label{fig:GinzP3D100}
\end{figure}

\begin{figure}
\centering
\epsfig{file=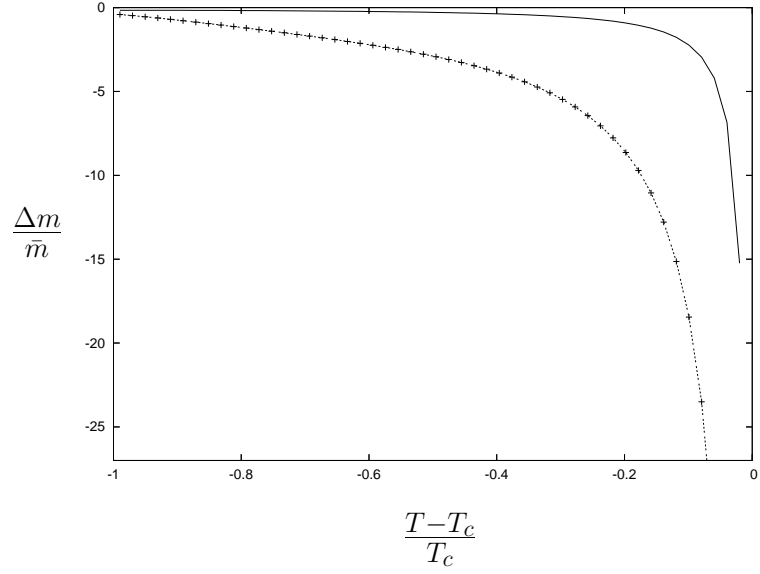,width=10cm}
\caption{Comparison of the Ginzburg criterion applied to a
2D (dotted line) and 3D (full line) Heisenberg model.}
\label{fig:GinzDim}
\end{figure}

The discrepancy can be quantified by means of the quantity $\frac{|\Delta m|}
{\bar{m}}$ where $\Delta m = m - \bar{m}=\delta m_{zz} + \delta m_{+-}$. 
Figure \ref{fig:GinzP3D100}  shows the result. The relation 
$\frac{|\Delta m|}{\bar{m}} < 1$ (Ginzburg criterion) fixes a limit 
temperature $T_{lim}$ above which the quantum and thermal fluctuations 
generate larger contributions than the mean-field. For 3$D$ systems 
this leads to $T_{lim} \simeq 0.8 T_c$, for $2D$ systems the criterion
is never satisfied except maybe for very low temperature
, see figure \ref{fig:GinzDim}.
\index{Ginzburg-Landau parameter}
\index{Quantum fluctuations}

The pathology is the stronger the smaller the space dimensionality.
It is also easy to see on the expression of the magnetization that, 
as expected,
the contributions of the quantum fluctuations decrease with increasing $D$. 
As can be seen in the figure \ref{fig:GinzDim}, the saddle point breaks down
earlier in two than in three dimensions.
  
In fact, the Heisenberg model spin-wave spectrum shows a Goldstone mode
as a consequence of the symmetry breaking by the N\'eel state. When   
$|\vec{k}|$ goes to zero 
\index{Goldstone modes}

\begin{eqnarray}
\omega_{\vec{k}} &=& Z D S \sqrt{1 - \gamma_{\vec{k}}^2 } \\
\underset{\vec{k} \rightarrow \vec{0}}{\text{lim }}\omega_{\vec{k}} &\sim& 
 |\vec{k}|
\end{eqnarray}
The zero mode destroys the long range order in 1D and 2D as expected from
the Mermin-Wagner theorem \cite{MerminWagner-66}.
\index{Mermin-Wagner theorem}

\subsection{The susceptibility : results}

We consider the parallel susceptibility $\chi_{\parallel}$ which characterizes 
a magnetic system on which a uniform magnetic field is applied in the $Oz$ 
direction. The expression of $\chi_{\parallel}$ decomposes again into three 
contributions
\index{N\'eel state!PFP parallel susceptibility}

\begin{eqnarray}
\chi_\parallel = -\frac{1}{\mathcal{N}} 
\frac{\partial^2 \mathcal{F}}{\partial B^2} 
\Bigg{\vert}_{B=0} =\chi_{MF} + \chi_{zz} + \chi_{+-} 
\end{eqnarray}

\noindent
with

\begin{align}
{\chi_\parallel}_{MF} 
=& \Delta m_{\chi 0} = \frac{ \frac{\beta}{4}\left( 1 - 4\bar{m}^2 
\right) }{1+\frac{\beta}{2}D|J|\left(1-4\bar{m}^2 \right)}
 \\
{\chi_\parallel}_{zz}
=& -\frac{1}{\mathcal{N} \beta} \underset{\vec{k} \in SBZ}{\sum}
\frac{8\left( \frac{\beta D |J| \gamma_{\vec{k}} }{2} \right)^2
\Delta m_{\chi 0}^2 \left(1 + 4 \bar{m}^2 \right)  }
{\left[ 1 - \left( \frac{\beta D |J| \gamma_{\vec{k}} }{2} \right)^2 
\left( 1 - 4 \bar{m}^2 \right)^2 \right]}
 \\
{\chi_\parallel}_{+-}
=&
 \frac{1}{\mathcal{N}} \underset{\vec{k} \in SBZ}{\sum} \Bigg{\{}
- \frac{1}{2} \frac{\beta \left(1 - 2 D |J| \Delta m_{\chi 0} \right)^2}
{\sinh^2 \left(\beta D |J| \bar{m} \right)}
 \\
& 
+ \frac{1}{\sinh^2 \left(\beta D |J| \bar{m} \sqrt{1-\gamma_{\vec{k}}^2} 
\right)} 
 \\
& \Bigg{[}
\frac{\beta}{2} \left(1 - 2 D |J| \Delta m_{\chi 0} \right)^2
- \beta \left(D |J| \Delta m_{\chi 0} \gamma_{\vec{k}} \right)^2 
\frac{\sinh 2\beta D |J| \bar{m}\sqrt{1-\gamma_{\vec{k}}^2} }
{\beta D |J| \bar{m}\sqrt{1-\gamma_{\vec{k}}^2} }
\Bigg{]}
\Bigg{\}}
\end{align}

\begin{figure}
\centering
\epsfig{file=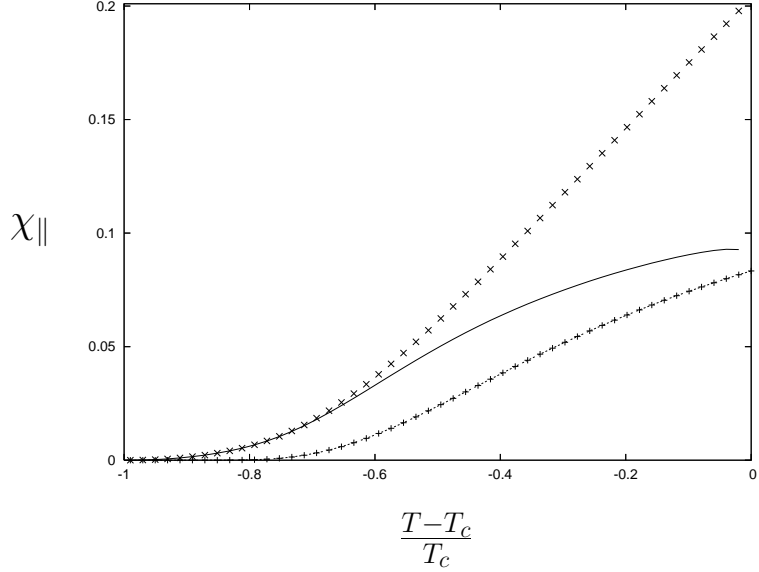,width=10cm}
\caption{Parallel magnetic susceptibility at 3D for the Heisenberg model.
 Dots : Spin-wave susceptibility. 
Dotted line : Mean-field susceptibility ${\chi_\parallel}_{MF}$.
Full line : total susceptibility (${\chi_\parallel}_{MF}
+\delta{\chi_\parallel}$).}
\label{Susceptibility}
\end{figure}

The behaviour of $\chi_{\parallel}$ is shown in Figure \ref{Susceptibility}
where we compare the mean-field, spin-wave and the one-loop corrected
contributions for a system on a 3 $D$ cubic lattice. 
One observes again a good agreement between the quantum corrected and the spin 
wave expressions at low temperatures. For higher temperatures the curves depart
from each other as expected. The mean-field  contribution remains in 
qualitative agreement with the total contribution.
\index{Spin-Wave}
 
\section{The XXZ-model}

To shed light on the fluctuations created by the symmetry breaking 
on an Heisenberg model when a N\'eel state is used as a mean-field state 
we add an anisotropic term $-\frac{J}{2} (1+\delta) S_i^z S_j^z$.
In this case we consider a so called XXZ-model. The corresponding Hamiltonian 
of the system can be written
\index{Anisotropic spin model}

\begin{eqnarray}
H^{XXZ} = -\frac{J}{2} \sum_{<ij>} \left( S_i^x S_j^x+ S_i^y S_j^y 
+ (1+\delta)S_i^z S_j^z \right)
\end{eqnarray}

\noindent
where $\delta$ governs the anisotropy. The self-consistent
mean-field magnetization of the XXZ-model reads

\begin{eqnarray}
\bar{m} = \frac{1}{2} \tanh \frac{\beta}{2}D|J(1+\delta)|\bar{m}
\end{eqnarray}

\noindent
Following the same steps as for the Heisenberg model we can compute the linear
response of the system to the application of a magnetic field $B_i$ in the
direction of the $Oz$ axis. The corresponding response to the mean-field
magnetization $\bar{m}_i = (-1)^{i} \bar{m}+\Delta \tilde{m}_{XXZ}.B_i$ reads

\begin{eqnarray}
\Delta \tilde{m}_{XXZ}
= \frac{\frac{\beta}{4}\left(1-4.\bar{m}^2 \right)}
{1-\frac{\beta}{2}D|J(1+\delta)|\left(1-4.\bar{m}^2 \right)}
\end{eqnarray}

\noindent
The magnetization is derived from the free energy given in the appendix
\ref{AppendixOneLoopCorrections} and reads

\begin{eqnarray}
m = \bar{m} + \delta m_{zz} + \delta m_{+-}
\end{eqnarray}

\noindent
where

\begin{eqnarray}
\bar{m} = -\frac{1}{\mathcal{N}} \frac{\partial \mathcal{F}_{MF}}{\partial B}
\Bigg{\vert}_{B=0} = \frac{1}{2} \tanh \frac{\beta}{2}D|J(1+\delta)|\bar{m}
\end{eqnarray}

\begin{eqnarray}
\delta m_{zz} &=& -\frac{1}{\mathcal{N}} 
\frac{\partial \mathcal{F}_{zz}}{\partial B} \Bigg{\vert}_{B=0} 
\notag \\
&=& -\frac{1}{\mathcal{N} \beta} \underset{\vec{k} \in SBZ}{\sum}
\frac{8.\bar{m} \Delta \tilde{m}_{XXZ} \left(1-4\bar{m}^2\right)
\left(\frac{\beta D |J(1+\delta)| \gamma_{\vec{k}}}{2}\right)^2}
{\left[1 -\left(\frac{\beta D |J(1+\delta)| \gamma_{\vec{k}}}{2}\right)^2 
 \left(1-4\bar{m}^2\right)\right]}
\end{eqnarray}

\begin{eqnarray}
\delta m_{+-} &=& -\frac{1}{\mathcal{N}} 
\frac{\partial \mathcal{F}_{+-}}{\partial B} \Bigg{\vert}_{B=0}
\notag  \\
&=&
\frac{\left(1 + 2 D |J(1+\delta)| \Delta \tilde{m}_{XXZ} \right)}{4\bar{m}}
 \\
&\phantom{=}& 
\begin{aligned}
- \frac{1}{\mathcal{N}} \underset{\vec{k} \in SBZ}{\sum}
 &
\frac{\left(1 + 2 D |J(1+\delta)| \Delta \tilde{m}_{XXZ} 
(1 -\left(\frac{J}{J(1+\delta)} \gamma_{\vec{k}}\right)^2)\right)}
{\sqrt{1 - \left(\frac{1}{(1+\delta)} \gamma_{\vec{k}}\right)^2}}
 \\
 &
\frac{1}{\left[\tanh \left(\beta D |J| \bar{m}
\sqrt{1 - \left(\frac{1}{1+\delta} \gamma_{\vec{k}}\right)^2} \right) \right]}
\end{aligned}
\notag \\
\end{eqnarray}

\noindent
The critical temperature $T_c^{XXZ}$ of the mean-field magnetization
for the XXZ-model reads

\begin{gather}
T_c^{XXZ} = \frac{D |J(1+\delta)|}{2}
\end{gather}

\noindent
For the zero temperature the excitation spectrum of the spin-wave obtained
from the previous magnetization expressions leads now to a finite 
$|\vec{k}|=0$ energy $\omega_{\vec{k}}$

\begin{eqnarray}
\omega_{\vec{k}} &=& Z D S \sqrt{1 - 
\left(\frac{1}{1+\delta}\gamma_{\vec{k}}\right)^2 } \\
\underset{\vec{k} \rightarrow \vec{0}}{\text{lim }}\omega_{\vec{k}} &\sim& 
\sqrt{1 - \left(\frac{1}{1+\delta}\right)^2 \left(1 - \frac{\vec{k}^2}{2 D}
\right)}
\end{eqnarray}

By examination the expressions show that the
zero momentum mode is no longer responsible for a breakdown of the saddle point
procedure near $T_c^{XXZ} = \frac{D |J(1+\delta)|}{2}$. 
However the magnetization 
of the XXZ-model remains infinite near $T_c^{XXZ}$. This is due to
the common disease shared with the Heisenberg model that the mean-field 
magnetization appearing in the denominator of $\delta m_{+-}$ goes to zero near
the critical temperature. One sees that the mean-field N\'eel state 
solution makes only sense at low temperatures, that is  for 
$T \lesssim T_{lim}$, whatever the degree of symmetry breaking induced by the 
mean-field ansatz.

One notices that the spectrum $\omega_{\vec{k}}$ no longer vanishes in the 
limit $\vec{k} \rightarrow 0$ when the anisotropic coupling $\delta \neq 0$. 
Since this is the case the transverse modes need a finite amount of energy 
to get excited. Thus they are not Goldstone modes 
\index{Goldstone modes} by virtue of the 
Goldstone theorem \cite{Auerbach-94,Goldenfeld}. 
The N\'eel state does not break the $O(2)$ symmetry
but only the discrete one $Z_2$ (in the anisotropy direction $Oz$) 
of the $XXZ$-model.
\index{N\'eel state!Broken symmetry}

\newpage

\section{Summary and conclusions}

In the present chapter we aimed to work out the expression of physical 
observables (magnetization and susceptibility) starting from a specific
mean-field ansatz and including contribution up to first
order in a loop expansion in order to investigate the effect of fluctuation
corrections to mean-field contributions at gaussian approximation.
The mean-field was chosen as a N\'eel state which is an \emph{a priori}
reasonable choice for spin systems described in terms of unfrustrated bipartite
Heisenberg model. The results were compared to those obtained in the 
framework of spin-wave theory. 
The PFP does not qualitatively change the well known magnetic properties of 
the Heisenberg model. However it must be pointed out that strict 
site-occupation had never been taken into account. This work remedies to this 
lack and was published in \cite{DRepjb-05}.

The number of particles per site was fixed by means of an strict constraint 
implemented in the partition function. 
It has been shown elsewhere \cite{Azakov-01,Dillen-05} that this fact 
introduces a large shift of the critical temperature compared to the case 
where the constraint is generated through an ordinary Lagrange multiplier term.

At low temperature the magnetization and the magnetic susceptibility are close 
to the spin-wave value as expected, also in agreement with former work 
\cite{Azakov-01}. Quantum corrections are sizable at low 
temperatures. With increasing temperature increasing thermal fluctuations 
add up to the quantum fluctuations.

At higher temperature the fluctuation contributions of quantum and thermal 
nature grow to a singularity 
in the neighbourhood of the critical temperature. The assumption that the 
N\'eel mean-field contributes for a major part to the magnetization and the
susceptibility is no longer valid. 
Approaching the critical temperature $T_c$ 
the mean-field contribution to the magnetization goes to 
zero and strong diverging fluctuations are generated at the one-loop order.
This behaviour is common to the Heisenberg and XXZ magnetization. In addition 
the N\'eel order breaks $SU(2)$ symmetry of the Heisenberg Hamiltonian inducing
low momentum fluctuations near $T_c$ which is not the case in the XXZ-model.

The influence of fluctuations decreases with the increase of the dimension $D$ 
of the system due to the expected fact that the mean-field contribution 
increases relatively to the loop contribution. 

In dimension $D=2$ the magnetization verifies the Mermin and Wagner theorem 
\cite{MerminWagner-66} for $T \neq 0$, the fluctuations are larger than the 
mean-field contribution for any temperature. In a more realistic description
another mean-field ansatz may be necessary in order to describe the correct 
physics. Indeed Ghaemi and Senthil \cite{GhaemiSenthil-05} have shown 
with the help of a specific model that a second order phase transition
from a N\'eel mean-field to an ASL (algebraic spin liquid) may be at work
depending on the strength of interaction parameter which enter the Hamiltonian
of the system. This confirms that another mean-field solution like ASL 
may be a better starting point than a N\'eel state when the temperature
$T$ increases.


\chapter{Mean-field ansatz for the $2d$ Heisenberg model \label{Chapter4}}

\minitoc
\newpage

We consider ordered quantum spin systems at finite temperature in which each 
lattice site is occupied by one electron with a given spin. Such a 
configuration can be constructed by means of constraints imposed 
through the specific projection operation \cite{Popov-88} which fixes the 
occupation in a strict sense. The constraint can also be implemented on the 
average by means of a Lagrange multiplier procedure \cite{Auerbach-94}.
It is the aim of the present chapter to confront these two approaches in the 
framework of Heisenberg-type models.

The description of strongly interacting quantum spin systems at finite 
temperature generally goes through a saddle point procedure which is a zeroth 
order approximation of the partition function. The so generated mean-field 
solution is aimed to provide a qualitatively realistic approximation of the 
exact solution.

However mean-field solutions are not unique. The implementation of a 
mean-field structure is for a large part subject to an educated guess which 
should rest on essential properties of the considered system, in particular 
its symmetries. This generates a major difficulty. A considerable amount of 
work on this point has been made and a huge litterature on the subject is 
available. In particular, systems which are described by Heisenberg-type 
models without frustration are seemingly well described by ferromagnetic
or antiferromagnetic (AF) N\'eel states at temperature $T = 0$ 
\cite{BernuLhuillier-92,BernuLhuillier-94}. 
It may however no longer be the case for many systems 
which are of low-dimensionality ($d \leq 2$) and (or) frustrated 
\cite{Zhang-88,MisguichLhuillier-03,Lecheminant-03}. 
These systems show specific features. An extensive 
analysis and discussion in space dimension $d = 2$ has recently been 
presented by Wen \cite{Wen-02}. The competition between AF and chiral spin 
state order has been the object of very recent investigations in the framework
of continuum quantum field approaches at $T = 0$ temperature, see 
\cite{TanakaHu-05,HermeleSenthilFisher-05}.

The reason for the specific behaviour of low-dimensional systems may 
qualitatively be related to the fact that low-dimensionality induces strong 
quantum fluctuations, hence disorder which destroys the AF 
order. This motivates a transcription of the Hamiltonian in terms of 
composite operators which we call "diffusons" and "cooperons" below, with 
the hope that the actual symmetries different from those which 
are generated by AF order are better taken into account at the mean-field 
level \cite{Auerbach-94}.

In the present work we aim to work out a strict versus average treatment 
of the site-occupation constraint on systems governed by 
Heisenberg-type Hamiltonians at the mean-field level, for different 
types of order. This original work was published in \cite{Dillen-05}.

The outline of the chapter is the following. 
Section \ref{Chapter4Section1} is 
devoted to the confrontation of the magnetization obtained through this 
procedure with the result obtained by means of an average projection
procedure in the framework of the mean-field approach characterized by a 
N\'eel state. 
The same confrontation is done in section \ref{Chapter4Section2} for the order 
parameter which characterizes the system when its Hamiltonian is written in 
terms of so called Abrikosov fermions \cite{MisguichLhuillier-03,Auerbach-94} 
in $d = 2$ space dimensions. 
In section \ref{Chapter4Section3} we show that the rigorous 
projection \cite{Popov-88} 
is no longer applicable when the Hamiltonian is written in terms of composite 
"cooperon" operators.
\index{Abrikosov fermion}

\newpage

\section{Antiferromagnetic mean-field ansatz \label{Chapter4Section1}}

\subsection{Exact occupation procedure \label{Chapter4Section1Subsection1}}

In this subsection we repeat the main steps of the partition function 
derivation given in section \eqref{SectionEffectiveAction} with a slight 
modification. Instead of integrating over the Grassmann variables we construct 
a mean-field Hamiltonian $\mathcal{H}_{MF}$ expressed in terms of the creation 
and annihilation fermion operators ($f^\dagger$ and $f$) and giving the same 
mean-field partition function $\mathcal{Z}_{MF}$ as if one integrates over 
the Grassmann variables. Another point is the fact that instead of using the 
modified Matsubara frequencies as shown in subsection 
\ref{SubsectionMatsubara} we keep the imaginary chemical potential as an 
explicit term outside of the time derivation $\partial_\tau$.
\index{Popov and Fedotov procedure!Imaginary chemical potential}

Starting with the nearest-neighbour Heisenberg Hamiltonian defined in 
equation \eqref{HNearestNeighbour}

\begin{eqnarray}
H = - J \underset{<i,j>}{\sum} 
\vec{S}_i.\vec{S}_j + \underset{i}{\sum} \vec{B}_i .\vec{S}_i
\end{eqnarray}

\noindent
the partition function $\mathcal{Z}$ can be written in the form 
\eqref{PathIntegralZ}

\begin{eqnarray}
\mathcal{Z}^{(PFP)} &=& \int_{\xi_{i,\sigma}(\beta)=- \xi_{i,\sigma}(0)} 
\mathcal{D} \xi
e^{-\int_0^\beta d\tau \{ \underset{i,\sigma}{\sum} 
\xi_{i,\sigma}^*(\tau)
(\partial_\tau - \mu) \xi_{i,\sigma}(\tau) 
+ H(\{ \xi_{i,\sigma}^*(\tau),\xi_{i,\sigma}(\tau)\}) \} }
\notag \\
&=& \int \mathcal{D} \xi e^{ -S(\{ \xi^{*}_{i,\sigma},\xi_{i,\sigma} \}) }
\label{Chapter4eq2}
\end{eqnarray} 

\noindent
where the $\{\xi^{*}_{i,\sigma},\xi_{i,\sigma}\}$ are Grassmann variables 
corresponding to the operators $\{f^{\dagger}_{i \sigma}, f_{i \sigma}\}$ 
defined in section \ref{Chapter2Section1}. 
These \emph{Grassmann} variables depend on the imaginary time $\tau$ in the 
interval $[0,\beta]$. The action $S$ is given by

\begin{eqnarray}
S(\{\xi^{*}_{i,\sigma},\xi_{i,\sigma}\}) = 
\int_{0}^{\beta} d\tau \sum_{i,\sigma} (\xi^{*}_{i,\sigma} 
(\tau) \partial_{\tau} \xi_{i,\sigma} (\tau) 
+ \mathcal{H}^{(PFP)}(\{\xi^{*}_{i,\sigma} (\tau),\xi_{i,\sigma} (\tau)\}))
\label{Chapter4eq3}
\end{eqnarray}

\noindent
where

\begin{eqnarray}
\mathcal{H}^{(PFP)} (\tau) = H(\tau) - \mu N(\tau)
\label{Chapter4eq4}
\end{eqnarray}

\noindent
and $N(\tau) \equiv \underset{i,\sigma}{\sum} 
\xi_{i,\sigma}^{*}(\tau) \xi_{i,\sigma}(\tau)$ is the particle number operator.
$\mu$ is the imaginary chemical potential introduced in section \ref{sectionPF}
describing the Popov and Fedotov procedure.
The Hubbard-Stratonovich (HS) transformation defined in subsection 
\ref{Chapter3Section3Subsection1} which generates the vector fields 
$\{\vec {\varphi}_i\}$ leads to the partition function $\mathcal{Z}$ which 
can be written in the form

\begin{eqnarray}
\mathcal{Z}^{(PFP)} = \int \mathcal{D} (\{\xi^{*}_{i, \sigma},\xi_{i,\sigma},
\vec {\varphi}_i\})
e^{- \int_{0}^{\beta} d\tau  [\sum_{i,\sigma} \xi^{*}_{i,\sigma} 
(\tau) \partial_{\tau} \xi_{i,\sigma} (\tau) + \mathcal{H}^{(PFP)}
(\{\xi^{*}_{i,\sigma},\xi_{i,\sigma}, \vec {\varphi}_i\})]}
\label{Chapter4eq5}
\end{eqnarray}

\noindent
In equation \eqref{Chapter4eq5} the expression of $\mathcal{Z}$ is quadratic 
in the Grassmann variables $\{\xi^{*}_{i,\sigma},\xi_{i,\sigma}\}$ over which 
the expression can be integrated. The remaining expression depends on the 
fields $\{\vec {\varphi}_i(\tau)\}$. A saddle point procedure decomposes 
$\vec{\varphi}_i(\tau)$ into a mean-field contribution and a fluctuating term

\begin{eqnarray}
\vec {\varphi}_i(\tau) = 
\left( \vec{\bar{\varphi}}_i - \vec{B}_i \right) + \delta \vec 
{\varphi}_i(\tau)
\end{eqnarray}

\noindent
where $\vec{\bar{\varphi}}_i$ are the constant solutions of the 
self-consistent equation

\begin{eqnarray} 
\vec{\bar{\varphi}}_i - \vec{B}_i = \frac{1}{2} \underset{j}{\sum} J_{ij}
\frac{\vec{\bar{\varphi}}_j }{\| \vec{\bar{\varphi}}_j \|} \tanh
\left( \frac{\beta \|\vec{\bar{\varphi}}_j \|}{2} \right) 
\end{eqnarray}

\noindent
as was shown in section \ref{Chapter3Section4} for equation 
\eqref{MeanFieldEquation}.

The partition function takes the form

\begin{eqnarray} 
\mathcal{Z}^{(PFP)} = \mathcal{Z}_{MF}^{(PFP)} 
\int {\mathcal{D}} (\{\delta \vec {\varphi}_i\})
e^{- S_{eff}(\{\delta \vec {\varphi}_i\})}
\end{eqnarray}

\noindent
where the first term on the right hand side corresponds to the mean-field 
contribution and the second term describes the contributions of the 
quantum and thermal fluctuations.

In the following we focus our attention only to the mean-field part of
the partition function

\begin{eqnarray}
\mathcal{Z}_{MF}^{(PFP)} = 
e^{-{S_{eff}} {\Big\vert}_{\left[\vec{\bar{\varphi}}\right]}}
= Tr \left[ e^{- \beta \mathcal{H}_{MF}^{(PFP)}} \right]
\end{eqnarray}

\noindent
where $\mathcal{H}_{MF}^{(PFP)}$ is the mean-field part of the Hamiltonian 
\eqref{Chapter4eq4} and reads

\begin{eqnarray}
\mathcal{H}_{MF}^{(PFP)} = -\frac{1}{2} \underset{i,j}{\sum} 
\left( J^{-1} \right)_{ij} \left(\vec{\bar{\varphi}}_i - \vec{B}_i \right)
.\left(\vec{\bar{\varphi}}_j - \vec{B}_j \right)
+ \underset{i}{\sum} \vec{\bar{\varphi}}_i.\vec{S}_i
- \mu \tilde{N}
\label{Chapter4MeanFieldHamiltonian}
\end{eqnarray}

\noindent
where spin operators $\vec{S}_i$ are given by \eqref{FermionizedSpin},
$\left( J^{-1} \right)_{ij}$ is the inverse of the coupling matrix $J_{ij}$ 
defined by \eqref{JNearestNeighbour} and $\tilde{N} = \underset{i,\sigma}
{\sum} f^{\dagger}_{i\sigma} f_{i\sigma}$ is the number operator. 
The mean-field Hamiltonian \eqref{Chapter4MeanFieldHamiltonian} expressed
in terms of the creation and annihilation operators 
$\{ f^{\dagger}_{i,\sigma},f_{i,\sigma} \}$ reads

\begin{eqnarray}
\mathcal{H}_{MF}^{(PFP)} &=& -\frac{1}{2} \underset{i,j}{\sum} 
\left( J^{-1} \right)_{ij} \left( \vec{\bar{\varphi}}_i - \vec{B}_i \right)
.\left(\vec{\bar{\varphi}}_j - \vec{B}_j \right)
\notag \\
&+& \underset{i}{\sum} \left( 
\begin{array}{cc}
f^{\dagger}_{i,\uparrow} & f^{\dagger}_{i,\downarrow}
\end{array}
 \right) 
\left[
\begin{array}{cc}
\left(- \mu + \frac{\bar{\varphi}_{i}^z}{2}\right) 
& \frac{\bar{\varphi}^{-}_i}{2} \\
\frac{\bar{\varphi}^{+}_i}{2} 
& -\left(\mu + \frac{\bar{\varphi}^z_i}{2} \right)
\end{array}
\right]
\left(
\begin{array}{c}
f_{i,\uparrow} \\
f_{i,\downarrow}
\end{array}
\right)
\notag \\
\label{Chapter4MeanFieldHamiltonian2}
\end{eqnarray}

The mean-field Hamiltonian \eqref{Chapter4MeanFieldHamiltonian2} can
be diagonalized by means of a Bogoliubov transformation as shown in
appendix \ref{AppendixDiagonalization} and leads to
\index{Bogoliubov transformation}

\begin{eqnarray}
\mathcal{H}_{MF}^{(PFP)} &=& -\frac{1}{2} \underset{i,j}{\sum} 
\left( J^{-1} \right)_{ij} \left( \vec{\bar{\varphi}}_i - \vec{B}_i \right)
.\left(\vec{\bar{\varphi}}_j - \vec{B}_j \right)
\notag \\
&+& \underset{i}{\sum} \left\{
\omega^{(PFP)}_{i,(+)} \beta^{\dagger}_{i,(+)} \beta_{i,(+)}
+ \omega^{(PFP)}_{i,(-)} \beta^{\dagger}_{i,(-)} \beta_{i,(-)}
\right\}
\end{eqnarray}

\noindent
Fermion creation and annihilation operators 
$\{\beta^{\dagger}_{i,(\pm)},\beta_{i,(\pm)} \}$ are linear combinations
of operators $\{f^{\dagger}_{i,\sigma},f_{i,\sigma} \}$ and the corresponding
excitation energies $\omega^{(PFP)}_{i,(\pm)}$ reads

\begin{eqnarray}
\omega^{(PFP)}_{i,(\pm)} = \mu \pm \frac{\| \vec{\bar{\varphi}}_i \|}{2}
\end{eqnarray}

\noindent
The partition is then easily worked out and reads

\begin{eqnarray}
Z_{MF}^{(PFP)} &=& 
i^{-\mathcal{N}} e^{\frac{1}{2} \beta \underset{ij}{\sum} J^{-1}_{ij} 
\left( \vec{\bar{\varphi}}_i - \vec{B}_i \right)
.\left(\vec{\bar{\varphi}}_j - \vec{B}_j \right) }
\underset{i}{\prod} 
(1 + e^{-\beta \omega^{(PFP)}_{i,(+)} })
(1 + e^{-\beta \omega^{(PFP)}_{i,(-)} })
\notag \\
&=&  e^{\frac{1}{2} \beta \underset{ij}{\sum} J^{-1}_{ij} 
\left( \vec{\bar{\varphi}}_i - \vec{B}_i \right)
.\left(\vec{\bar{\varphi}}_j - \vec{B}_j \right) } 
\underset{i}{\prod} \cosh 
\left(\frac{\beta \| \vec{\bar{\varphi}}_i \|}{2}\right)
\label{Chapter4eq13}
\end{eqnarray}

\noindent
and the free energy is given by the expression

\begin{eqnarray} 
\mathcal{F}_{MF}^{(PFP)} = -\frac{1}{\beta} \ln \mathcal{Z}_{MF} 
= - \frac{1}{2} \underset{ij}{\sum} J^{-1}_{ij} 
\left( \vec{\bar{\varphi}}_i - \vec{B}_i \right)
.\left(\vec{\bar{\varphi}}_j - \vec{B}_j \right)
- \frac{1}{\beta} \underset{i}{\sum} \ln 2 
\cosh \left(\frac{\beta \| \vec{\bar{\varphi}}_i \|}{2}\right)
\notag \\
\label{Chapter4eq14}
\end{eqnarray}

Going through the same steps as in section \ref{Chapter3Section5Subsection1} 
the local mean-field magnetization $\left\{ \vec{\bar{m}}_i \right\}$ is 
obtained from

\begin{eqnarray}
\vec{\bar{m}}_i = - \frac{\partial \mathcal{F}_{MF}^{(PFP)}} 
{\partial \vec{B}_i} {\Big\vert}_{\left\{ \vec{B}_i = \vec{0} \right\} }
\end{eqnarray}

\noindent
and is related to the $\{ \vec{\bar{\varphi}}_i \}$'s by 

\begin{eqnarray}
\vec{\bar{m}}_{i}
= \frac{1}{2} \frac{\vec{\bar{\varphi}}_i} {\bar{\varphi}_i }
\tanh \left[ \frac{\beta\bar{\varphi}_i} {2} \right]
\end{eqnarray}

\noindent
and by virtue of the relations \eqref{Relation1} and \eqref{Relation2} one 
gets the self-consistent equation for the $\{ \vec{\bar{m}}_i \}$
 
\begin{eqnarray}
\vec{\bar{m}}_i = \frac{2}{\beta} \underset{j}{\sum} 
J^{-1}_{ij} \tanh^{-1} \left( 2 \bar{m}_j \right) 
\frac{\vec{\bar{m}}_j}{\| \vec{\bar{m}}_j \|}
\end{eqnarray}

\noindent
If the local fields $\{\vec{B}_{i}\}$ are oriented along a fixed direction 
$\vec {e}_{z}$, $\vec{\bar{m}}_i =  \bar{m}_i \vec {e}_{z}$, the 
magnetizations are the solutions of the self-consistent equations

\begin{eqnarray}
\bar{m}_i = \frac{1}{2} \tanh \left( \frac{\beta \underset{j}{\sum} J_{ij} 
\bar{m}_j}{2}\right)
\label{Chapter4eq17}
\end{eqnarray}

\subsection{Lagrange multiplier approximation 
\label{Chapter4Section1Subsection2}}

In section \ref{Chapter2Section1} we introduced the projector
\eqref{Chapter2LagrangeProjector} 
\index{Lagrange multiplier}

\begin{eqnarray}
\tilde{P}_{i} = \int \mathcal{D} \lambda_i e^{\lambda_i (
\underset{\sigma}{\sum} f^{\dagger}_{i\sigma} f_{i\sigma} - 1)}
\label{Chapter4eq20}
\end{eqnarray}

\noindent
which allows to fix to one the number of spin-$1/2$ per lattice site.
Similarly to the preceding case in which the Popov and Fedotov imaginary
chemical potential was used one can introduce the one-particle site 
occupation by means of a Lagrange procedure. In order to do that one has
to replace the Popov and Fedotov projector 
$\tilde{P}=\frac{1}{i^{\tilde{N}}} e^{i \frac{\pi}{2} \tilde{N}}$ by 
\eqref{Chapter2LagrangeProjector}. 
The Hamiltonian $\mathcal{H}$ then reads

\begin{eqnarray}
\mathcal{H}^{(\lambda)} 
=  \frac{1}{2} \underset{i,j}{\sum} J_{ij} \vec S_{i}.\vec S_{j} 
+ \underset{i}{\sum} \vec{B}_i.\vec{S}_i + \sum_{i} \lambda_i (n_{i} - 1)
\label{Chapter4eq21}
\end{eqnarray}

\noindent
where $\lambda_i$ is a variational parameter and 
$\{ {n_i = \underset{\sigma}{\sum} f^{\dagger}_{i\sigma} f_{i\sigma}} \}$ 
are particle number operators.

Following the same lines as in section \ref{Chapter4Section1Subsection1} 
with the help of a Hubbard-Stratonovich transformation and a Bogoliubov
transformation as shown in appendix \ref{AppendixDiagonalization} the 
mean-field partition function $\mathcal{Z}_{MF}^{(\lambda)}$ can be worked 
out
\index{Hubbard-Stratonovich transformation}
\index{Bogoliubov transformation}

\begin{eqnarray}
Z_{MF}^{(\lambda)} = e^{-\frac{1}{2} \beta \underset{i,j}{\sum} J^{-1}_{ij} 
\left( \vec{\bar{\varphi}}_i - \vec{B}_i \right)
\left( \vec{\bar{\varphi}}_j - \vec{B}_j \right) 
- \mathcal{N} \lambda }
\underset{i}{\prod} 
(1 + e^{-\beta \omega^{\lambda}_{i,(+)} })
(1 + e^{-\beta \omega^{\lambda}_{i,(-)} })
\label{Chapter4eq22}
\end{eqnarray}

\noindent
with

\begin{eqnarray}
\omega^{\lambda}_{i,(+)} = \lambda 
+ \frac{\| \vec{\bar{\varphi}}_j \|}{2}
\end{eqnarray}

\begin{eqnarray}
\omega^{\lambda}_{i,(-)} = \lambda 
- \frac{\| \vec{\bar{\varphi}}_j \|}{2}
\end{eqnarray}

The parameter $\lambda$ is fixed through a minimization of the free energy
with respect to $\lambda$

\begin{eqnarray}
\frac{\partial \mathcal{F}_{MF}^{(\lambda)} }{\partial \lambda_i}
{\Big\vert}_{\left\{ \lambda_i = \lambda \right\} } = 0
\end{eqnarray}

\noindent
The minimization shows that the extremum solution is obtained for 
$\lambda = 0$ and

\begin{eqnarray}
\mathcal{F}_{MF}^{(\lambda)} 
= - \frac {1}{2} \underset{i,j}{\sum} J^{-1}_{ij} 
\left( \vec{\bar{\varphi}}_i - \vec{B}_i \right) 
.\left( \vec{\bar{\varphi}}_j - \vec{B}_j \right) 
- \frac{2}{\beta} \underset{i}{\sum}
\ln 2 \cosh \left(\frac{\beta \| \vec{\bar{\varphi}}_i \|}{4}\right) 
\label{Chapter4eq26}
\end{eqnarray}

\noindent
which is different from the expression of equation \eqref{Chapter4eq14} by
a factor $1/2$ in the argument of the $\cosh$ term.

The magnetization can be obtained in the same way as done in subsection
\ref{Chapter4Section1Subsection1}. One obtains

\begin{eqnarray}
\bar{m}_i^{(\lambda = 0)} = \frac{1}{2} \tanh \left(
\frac{ \beta \underset{j}{\sum} J_{ij} 
\bar{m}_j^{(\lambda = 0)} }{4}\right)
\label{Chapter4eq27}
\end{eqnarray}

\noindent
which is again different from the expression obtained in the case of a 
rigorous projection, see equation \eqref{Chapter4eq17}.

\begin{figure}
\centering
\epsfig{file=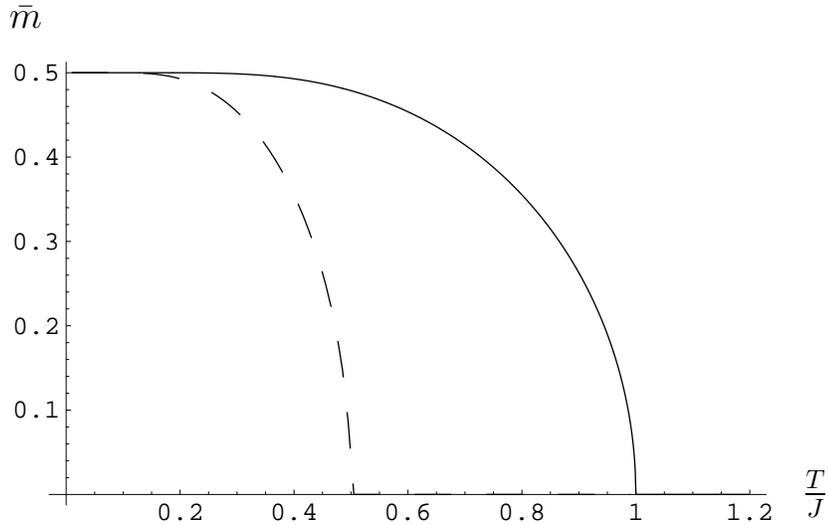}
\caption{Magnetization vs. reduced temperature $t = T / |J|$.
Full line: exact site-occupation. Dashed line: average site-occupation.}
\label{Chapter4Fig1}
\end{figure}

The uniform solutions $\bar{m}_i^{(PFP)} = (-1)^i \bar{m}^{(PFP)}$ and 
$\bar{m}_i^{(\lambda = 0)} =(-1)^i \bar{m}^{(\lambda = 0)}$ 
for $J_{ij} = J \underset{\vec{\eta} \in \{\vec{a}_1,
\dots, \vec{a}_D  \}}{\sum} \delta 
\left( \vec{r}_i - \vec{r}_j \pm \vec{\eta} \right)$
have been calculated by solving the selfconsistent equations
\eqref{Chapter4eq17} and \eqref{Chapter4eq27}.
The results are shown in figure (\ref{Chapter4Fig1}).
It is seen that the treatment of 
the site-occupation affects sizably the quantitive behaviour of observables. 
In particular it shifts the location of the critical temperature $T_c$ by a 
factor 2. Such a strong effect has already been observed on the behaviour of 
the specific heat, see refs. \cite{Azakov-01,KFO-01}.

\section{Spin state mean-field ansatz in 2d \label{Chapter4Section2}}

In 2$d$ space the Heisenberg Hamiltonian given by 

\begin{eqnarray}
H = - J \underset{<i,j>}{\sum} 
\vec{S}_i.\vec{S}_j
\end{eqnarray}

\noindent
can be written in terms of composite non-local operators 
$\{ \mathcal{D}_{ij} \}$ ("diffusons") \cite{Auerbach-94} defined as 
\index{Diffuson}

\begin{eqnarray}
\mathcal{D}_{ij} = f_{i, \uparrow}^{\dagger} f_{j, \uparrow} + 
f_{i, \downarrow}^{\dagger} f_{j, \downarrow}
\label{Chapter4eq29}
\end{eqnarray}

\noindent
If the coupling strengths are fixed as 

\begin{eqnarray}
J_{ij} = J 
\underset{\vec{\eta}}{\sum} \delta \left(\vec{r}_i - \vec{r}_j \pm \vec{\eta} 
\right)
\end{eqnarray}

\noindent
where $\vec{\eta}$ is a lattice vector $\{a_1,a_2\}$ in the $\vec {Ox}$  
and $\vec {Oy} $ directions the Hamiltonian takes the form

\begin{eqnarray}
H = - J \sum_{<ij>} (\frac{1}{2} {\cal D}^{\dagger}_{ij} {\cal D}_{ij} - 
\frac{n_i}{2} + \frac{n_i  n_j}{4})  
\label{Chapter4eq31}
\end{eqnarray}

\noindent
where $i$ and $j$  are nearest neighbour sites.

The number operator products $\{n_i  n_j\}$ in Eq.\eqref{Chapter4eq31} 
are quartic in terms of creation and annihilation operators in Fock space. 
In principle the formal treatment of these terms requires the introduction
of a mean-field procedure.
One can however show that the presence of this term has no
influence on the results obtained from the partition function. 
Indeed these terms lead to a constant quantity under the exact site-occupation
constraint and hence are of no importance for the physics described by the 
Hamiltonian \eqref{Chapter4eq31}.
As a consequence we leave it out from the beginning as well as the 
contribution corresponding to the $\{n_i\}$ terms.

\subsection{Exact occupation procedure}

Starting with the Hamiltonian

\begin{eqnarray}
\mathcal{H}^{(PFP)}
= - \frac{J}{2} \sum_{<ij>}   {\mathcal D}^{\dagger}_{ij} {\cal D}_{ij} 
- \mu N
\label{Chapter4eq32}
\end{eqnarray}

\noindent
the partition function $\mathcal{Z}$ can be written in the form 
\eqref{Chapter4eq2}
and the Hamiltonian in the form \eqref{Chapter4eq4}.
A Hubbard-Stratonovich transformation on the corresponding functional integral 
partition function in which the action contains the occupation number 
operator as seen in equation \eqref{Chapter4eq3} eliminates the quartic 
contributions generated by equation \eqref{Chapter4eq29}
and introduces the mean-fields $\{\Delta_{ij}\}$.
The Hamiltonian takes then the form

\begin{eqnarray}
\mathcal{H}^{(PFP)}
= \frac{2}{|J|}\underset{<ij>}{\sum} \bar{\Delta}_{ij} \Delta_{ij} 
+\underset{<ij>}{\sum} \left[ \bar{\Delta}_{ij} \mathcal{D}_{ij} +
\Delta_{ij} \mathcal{D}^{\dagger}_{ij} \right]  - \mu N
\end{eqnarray}

\noindent
The fields $\left\{ \Delta_{ij} \right\}$ and their complex conjugates 
$\left\{ \bar{\Delta}_{ij} \right\}$ 
can be decomposed into a mean-field contribution and a fluctuation term
 
\begin{eqnarray}
\Delta_{ij} = \Delta_{ij}^{MF} + \delta \Delta_{ij}
\end{eqnarray}

\begin{figure}
\center
\begin{pspicture}(6,6)
\psset{unit=1.5cm}
\psgrid[griddots=10,gridlabels=0,subgriddiv=1](3,3)
\multips(0,0)(0,2){2}{
\multips(0,0)(2,0){2}{
\psdots[dotscale=2,dotstyle=*](0,0)
\psdots[dotscale=2,dotstyle=*](1,1)
\psdots[dotscale=2,dotstyle=*](1,0)
\psdots[dotscale=2,dotstyle=*](0,1)}}
\psline[linewidth=0.05]{-}(1,1)(1.5,1) \psline[linewidth=0.05]{>-}(1.5,1)(2,1)
\psline[linewidth=0.05]{-}(2,1)(2,1.5) \psline[linewidth=0.05]{>-}(2,1.5)(2,2)
\psline[linewidth=0.05]{-}(2,2)(1.5,2) \psline[linewidth=0.05]{>-}(1.5,2)(1,2)
\psline[linewidth=0.05]{-}(1,2)(1,1.5) \psline[linewidth=0.05]{>-}(1,1.5)(1,1)
\rput(0.8,0.7){\text{$\vec{i}$}}
\rput(2.4,0.7){\text{$\vec{i}+\vec{e}_x$}}
\rput(2.4,2.4){\text{$\vec{i}+\vec{e}_x + \vec{e}_y$}}
\rput(0.8,2.4){\text{$\vec{i}+\vec{e}_y$}}
\end{pspicture}
\caption{Plaquette ($\Box$) on a two dimensionnal spin lattice}
\index{Plaquette}
\label{Chapter4Fig2}
\end{figure}
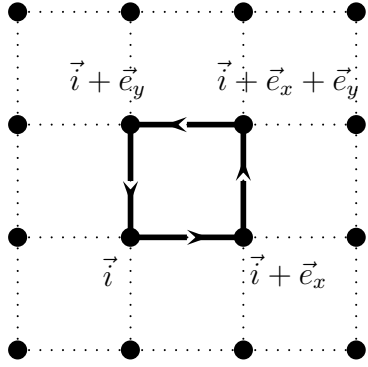

\noindent
The field $\Delta_{ij}^{MF}$ can be chosen as a complex quantity 
$\Delta_{ij}^{MF}=|\Delta_{ij}^{MF}|e^{i \phi_{ij}^{MF}}$. 

The phase $\phi_{ij}^{MF}$ is fixed in the following way. Consider a 
square plaquette 
$\Box \equiv \left( \vec{i}, \vec{i} + \vec{e}_{x}, \vec{i} + \vec{e}_{x} 
+ \vec {e}_{y}, \vec{i} + \vec{e}_{y} \right)$ 
where $\vec{e}_{x}$ and $\vec{e}_{y}$ 
are the unit vectors along the directions $\vec{Ox}$ and $\vec{Oy}$ 
starting from site $\vec i$ on the lattice as shown in figure 
(\ref{Chapter4Fig2}).
On this plaquette we define

\begin{eqnarray}
\phi = \underset{(ij) \in \Box}{\sum} \phi_{ij}^{MF}
\end{eqnarray}

\noindent
which is taken to be constant. If the gauge phase $\phi_{ij}^{MF}$ fluctuates 
in such a way that $\phi$ stays constant the average of $\Delta_{ij}^{MF}$ 
will be equal to zero in agreement with Elitzur's theorem \cite{Elitzur}. 
\index{Elitzur theorem}
In order to guarantee the $SU(2)$ invariance of the mean-field Hamiltonian
along the plaquette we follow \cite{AffleckMarston-88,ArovasAuerbach-88,LeeNagaosa-02,MarstonAffleck-89,Wen-02}
and introduce

\begin{eqnarray}
\phi_{ij}=
\begin{cases}
e^{i.\frac{\pi}{4}(-1)^i}, \text{if } \vec{r}_j=\vec{r}_i+\vec{e}_x \\
e^{-i.\frac{\pi}{4}(-1)^i}, \text{if } \vec{r}_j=\vec{r}_i+\vec{e}_y \\
\end{cases}
\end{eqnarray}

\noindent
where $\vec{e}_x$ and $\vec{e}_y$ join the site $i$ to its nearest neighbours
$j$. Then the total flux through the fundamental plaquette is such that 
$\phi = \pi$ \index{$\pi$-flux state} 
which guarantees that the $SU(2)$ symmetry of the plaquette is
respected \cite{Marston}.

At the mean-field level the partition function reads 

\begin{eqnarray}
\mathcal{Z}_{MF}^{(PFP)} = e^{- \beta \mathcal{H}_{MF}^{(PFP)}}
\end{eqnarray}

\noindent
where

\begin{eqnarray}
\mathcal{H}_{MF}^{(PFP)} &=& \frac{2}{|J|}\underset{<ij>}{\sum} 
\bar{\Delta}_{ij}^{MF}.\Delta_{ij}^{MF}
+ \underset{<ij>}{\sum} \left[
\bar{\Delta}_{ij}^{MF} \mathcal{D}_{ij} + 
\Delta_{ij}^{MF} \mathcal{D}^{\dagger}_{ij} \right]  - \mu N
\label{Chapter4eq38}
\end{eqnarray}

\noindent
After a Fourier transformation the Hamiltonian \eqref{Chapter4eq38} 
takes the form

\begin{eqnarray}
\mathcal{H}_{MF}^{(PFP)} = \mathcal{N} z \frac{\Delta^2}{|J|}
+ \underset{\vec{k} \in SBZ}{\sum} \underset{\sigma}{\sum}
\left(
f^\dagger_{\vec{k},\sigma} \,
f^\dagger_{\vec{k}+\vec{\pi},\sigma}
\right)
\left[
\widetilde{\mathcal{H}}^{(PFP)}
\right]
\left(
\begin{array}{c}
f_{\vec{k},\sigma} \\
f_{\vec{k}+\vec{\pi},\sigma}
\end{array}
\right)
\label{Chapter4eq39}
\end{eqnarray}
with

\begin{eqnarray}
\left[ \widetilde{\mathcal{H}}^{(PFP)} \right] =
\left[
\begin{array}{cc}
-\mu + \Delta \cos \frac{\pi}{4} z \gamma_{k_x,k_y} &
-i \Delta \sin \frac{\pi}{4} z \gamma_{k_x,k_y+\pi} \\
+i \Delta \sin \frac{\pi}{4} z \gamma_{k_x,k_y+\pi} &
-\mu - \Delta \cos \frac{\pi}{4} z \gamma_{k_x,k_y} 
\end{array}
\right]
\label{Chapter4eq40}
\end{eqnarray}

\noindent
where $\Delta \equiv |\Delta^{MF}|$. The Spin Brillouin Zone (SBZ) covers
half of the Brillouin Zone (see figure \ref{Chapter3Fig2} in subsection
\ref{SubsectionSWM}). The $\gamma_{\vec{k}}$'s are defined by
\index{Spin Brillouin Zone}

\begin{eqnarray}
\gamma_{\vec{k}} = \frac{1}{z} \underset{\vec{\eta}}{\sum} 
e^{i \vec{k}.\vec{\eta}} = \frac{1}{2} \left( \cos k_x a_1  + \cos k_y a_2 
\right)
\end{eqnarray}

\noindent
where $z=4$ is the coordination and $\mathcal{N}$ the number of sites.
$a_1$ and $a_2$ are the lattice parameters in direction $\vec{Ox}$ and
$\vec{Oy}$. The lattice parameters are not important for our study. We 
renormalize the wave vector $\vec{k}$ by the relations $k_x = k_x . a_1$
and $k_y = k_y . a_2$ as shown in appendix \ref{AppendixSpinBrillouinZone}.
The momenta $\{\vec k\}$ act in the first half Brillouin zone (spin Brillouin 
zone). 
\index{Spin Brillouin Zone}

Performing a Bogolioubov transformation which diagonalizes the remaining 
expression \eqref{Chapter4eq40} in Fourier space leads to 
\index{Bogoliubov transformation}

\begin{eqnarray}
\mathcal{H}_{MF}^{(PFP)} = \frac{\mathcal{N} z \Delta^{2}} {|J|} 
+ \underset{\vec k, \sigma}{\sum}
\left[ \omega^{(PFP)}_{(+),\vec k, \sigma}
\beta^{\dagger}_{(+), \vec k, \sigma} \beta_{(+), \vec k, \sigma}
+ \omega^{(PFP)}_{(-),\vec k, \sigma} 
\beta^{\dagger}_{(-),\vec k, \sigma} \beta_{(-), \vec k, \sigma} \right]
\label{Chapter4eq42}
\end{eqnarray}

\noindent
The transformation is worked out in appendix \ref{AppendixDDiffuson}.
The eigenenergies $\omega^{(PFP)}_{(+),\vec k, \sigma}$ 
and $\omega^{(PFP)}_{(-),\vec k, \sigma}$ are given by
 
\begin{eqnarray}
\omega^{(PFP)}_{(+),\vec k, \sigma} = -\mu + 2 \Delta [cos^{2}(k_{x}) 
+ cos^{2}(k_{y})]^{1/2}  
\end{eqnarray}

\noindent
and similarly

\begin{eqnarray}
\omega^{(PFP)}_{(-),\vec k, \sigma} = -\mu - 2 \Delta [cos^{2}(k_{x}) 
+ cos^{2}(k_{y})]^{1/2}
\end{eqnarray}

The partition function $Z_{MF}$ has the same structure as the corresponding 
partition function in equation \eqref{Chapter4eq13} and the free energy is 
given by 

\begin{eqnarray}
\mathcal{F}_{MF}^{(PFP)} = \frac{\mathcal{N} z \Delta^{2}} {|J|} - 
\frac{1} {\beta} \underset{\vec k, \sigma}{\sum}
\ln \left( 2 \cosh \beta \Delta \varepsilon_{\vec{k}} \right)
\label{Chapter4eq45}
\end{eqnarray}

\noindent
with

\begin{eqnarray}
\varepsilon_{\vec{k}} = 2 \left[cos^{2}(k_{x}) + cos^{2}(k_{y}) \right]^{1/2}
\label{Chapter4eq46}
\end{eqnarray}

Finally the variation of $\mathcal{F}_{MF}$ with respect to $\Delta$ leads to 
the self-consistent mean-field equation

\begin{eqnarray}
\tilde{\Delta}^{(PFP)} = \frac{1} {2 \mathcal{N}} 
\underset{\vec k, \sigma}{\sum} \varepsilon_{\vec{k}} 
\tanh \left( \frac{\beta|J|\varepsilon_{\vec{k}} \tilde{\Delta}^{(PFP)}} 
{z}\right)
\label{Chapter4eq47}
\end{eqnarray}

\noindent
with $\tilde{\Delta}^{(PFP)} = z \Delta/|J|$ which fixes $\Delta$.

\subsection{Lagrange multiplier approximation}

Similarly to equation \eqref{Chapter4eq21} one may introduce a Lagrange 
constraint and write \index{Lagrange multiplier}

\begin{eqnarray}
\mathcal{H}^{(\lambda)} = \frac{2}{|J|} \underset{<ij>}{\sum} 
\bar{\Delta}_{ij} \Delta_{ij} 
+ \underset{<ij>}{\sum} \left( \bar{\Delta}_{ij} \mathcal{D}_{ij} 
+ \Delta_{ij} \mathcal{D}^{\dagger}_{ij} \right) + \underset{i}{\sum}
\lambda_i \left( n_{i} - 1 \right)
\end{eqnarray}

\noindent
In the mean-field approximation one have

\begin{eqnarray}
\lambda_i = \lambda
\end{eqnarray}

\noindent
and after a Bogoliubov transformation as for equation \eqref{Chapter4eq42}
the mean-field Hamiltonian reads

\begin{eqnarray}
H_{MF}^{(\lambda)} = \frac{\mathcal{N} z \Delta^{2}} {|J|} 
+ \underset{\vec k, \sigma}{\sum} 
\left( \omega^{(\lambda)}_{(+),\vec k, \sigma}
\beta^{\dagger}_{(+) \vec k, \sigma} \beta_{(+) \vec k, \sigma}
+ \omega^{(\lambda)}_{(-),\vec k, \sigma}
\beta^{\dagger}_{(-) \vec k, \sigma} \beta_{(-) \vec k, \sigma}\right) 
\end{eqnarray}

\noindent
with the eigenenergies

\begin{eqnarray}
\omega^{(\lambda)}_{(+),\vec k, \sigma} = \lambda + 2 \Delta [cos^{2}(k_{x})  
+  cos^{2}(k_{y})]^{1/2}
\end{eqnarray}

\noindent
and similarly

\begin{eqnarray}
\omega^{(\lambda)}_{(-),\vec k, \sigma} = \lambda - 2 \Delta [cos^{2}(k_{x})  
+  cos^{2}(k_{y})]^{1/2}
\end{eqnarray}

\noindent
The expression of the free energy is now given by

\begin{eqnarray}
\mathcal{F}_{MF}^{(\lambda)} = -\mathcal{N} \lambda 
+ \frac{\mathcal{N} z \Delta^{2}} {|J|} 
- \frac{1}{\beta} \underset{\vec k, \sigma}{\sum}
\ln \left[ 1 + e^ {- \beta  \omega^{(\lambda)}_{(+),\vec k, \sigma} }\right]
\left[ 1 + e^ {- \beta \omega^{(\lambda)}_{(-),\vec k, \sigma} }\right]
\end{eqnarray}

As was shown in subsection \ref{Chapter4Section1Subsection2}
the minimization of this expression in terms of $\lambda$ delivers the 
solution $\lambda = 0$ and 

\begin{eqnarray}
\mathcal{F}_{MF}^{(\lambda)} = 
+ \frac{\mathcal{N} z \Delta^{2}} {|J|} -
\frac{1} {\beta} \underset{\vec k, \sigma}{\sum} 2 
\ln \left( 2 \cosh \frac {\beta \Delta \varepsilon_{\vec{k}}} {2} \right)
\label{Chapter4eq54}
\end{eqnarray}

\noindent
where $\varepsilon_{\vec{k}}$ is given by equation \eqref{Chapter4eq46}.

The variation of $\mathcal{F}_{MF}^{(\lambda)}$ with respect to $\Delta$ leads 
to the self-consistent mean-field equation

\begin{eqnarray}
\tilde{\Delta}^{(\lambda)}= \frac{1} {\mathcal{N}} 
\underset{\vec k, \sigma}{\sum}
\varepsilon_{\vec{k}}
\tanh \left( \frac{\beta |J| \varepsilon_{\vec{k}} \tilde{\Delta}^{(\lambda)} }
{2 z}\right)
\label{Chapter4eq55}
\end{eqnarray}

\noindent
with $\tilde{\Delta}^{(\lambda)} = z \Delta/|J|$.

\begin{figure}
\centering
\epsfig{file=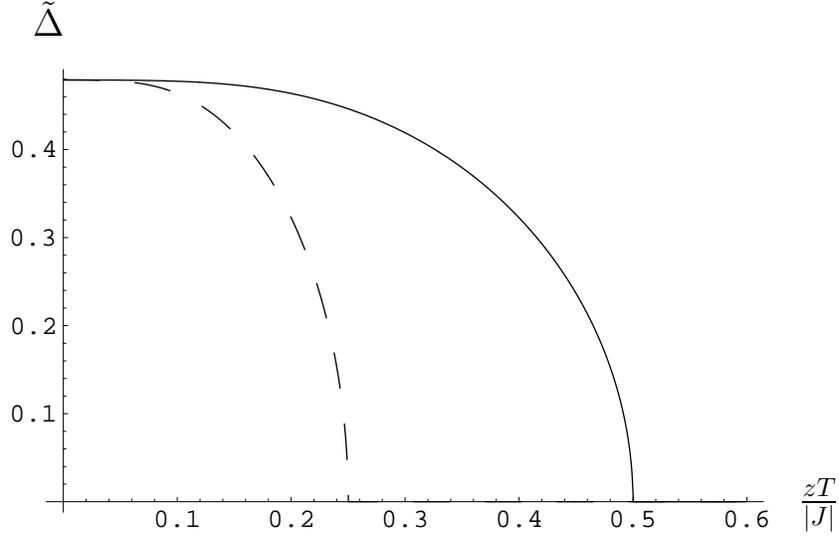}
\caption{$\tilde{\Delta}$ vs. reduced temperature $\tilde{t}=zT/|J|$.
Full line: exact site-occupation $\tilde{\Delta}^{(PFP)}$. 
Dashed line: average site ocupation $\tilde{\Delta}^{(\lambda)}$.}
\label{Chapter4Fig3}
\end{figure}

Expressions in equation \eqref{Chapter4eq54} and  
equation \eqref{Chapter4eq55} should be 
compared to the expressions obtained in equation \eqref{Chapter4eq45} and 
equation \eqref{Chapter4eq47}.
Figure \ref{Chapter4Fig3} shows the behaviour of 
$\tilde{\Delta}$ for the two different treatments of site-occupation on the 
lattice. The exact occupation procedure compared to
Lagrange multiplier approximation
doubles the critical temperature of the order parameter $\tilde{\Delta}$ as
was shown in section \ref{Chapter4Section1} and in \cite{Azakov-01} for the
N\'eel state.

\section{Cooperon mean-field ansatz \label{Chapter4Section3}}

Starting from  the Hamiltonian 
\begin{eqnarray}
H =   - |J| \sum_{<i,j>} \vec S_{i} \vec S_{j} 
\end{eqnarray}

\noindent
one can introduce a further set of non-local composite operators 
$\left\{ \mathcal{C}_{ij} \right\}$ ("cooperons") \index{Cooperon}

\begin{eqnarray}
\mathcal{C}_{ij} =   f_{i, \uparrow} f_{j, \downarrow} - 
f_{i, \downarrow} f_{j, \uparrow}
\label{Chapter4eq57}
\end{eqnarray}

\noindent
This leads to the expression

\begin{eqnarray}
H  = \frac {|J|}{2} \underset{<i,j>}{\sum} 
\left( \mathcal{C}_{ij}^{\dagger} \mathcal{C}_{ij} 
+ \frac{ n_i n_j }{2} \right)
\label{Chapter4eq58}
\end{eqnarray}

\noindent
where $n_i = \underset{\sigma}{\sum} f^{\dagger}_{i, \sigma} f_{i, \sigma}$.
As was explained in section \ref{Chapter4Section2} for the Hamiltonian 
composed of the operator $\mathcal{D}_{ij}$ the second term in the
right hand side of equation \eqref{Chapter4eq58} is quartic in the
fermion creation and annihilation operator $f^\dagger,f$ hence one
needs to use a Hubbard-Stratonovich transformation in order to reduce 
this term to a quadratic form. Calculations show that
the second term is irrelevant. The solution of the self-consistent
equations of the auxiliary field introduced by the Hubbard-Stratonovich 
procedure lead to $\left\{ n_i = 1 \right\}$.

\subsection{Exact occupation procedure}

As in the preceding cases it is possible to implement a 
Hubbard-Stratonovich procedure on the 
$\left\{ \mathcal{C}_{ij} \right\}$ in such a way that the expresssion of 
the corresponding partition function gets quadratic in the fields 
$\left\{ f_{i, \uparrow}, f_{i, \downarrow} \right\}$.
The corresponding HS fields are $\left\{ \Gamma_{ij} \right\}$ and

\begin{eqnarray}
H  = \frac {2}{|J|} \underset{<i,j>}{\sum} \bar{\Gamma}_{ij} \Gamma_{ij}
+ \underset{<i,j>}{\sum} \left( \bar{\Gamma}_{ij} \mathcal{C}_{ij} 
+ \Gamma_{ij} \mathcal{C}_{ij}^{\dagger} \right)
\end{eqnarray}

\noindent
Introducing the homogeneous mean-fields
$\left\{ \Gamma_{ij} = \Gamma = \bar{\Gamma} \right\}$, 
one gets in Fourier space 

\begin{eqnarray}
\mathcal{H}_{MF}^{(PFP)} =
\mathcal{N} \frac{z \Gamma^2}{|J|}
- \underset{\vec{k} \in BZ,\sigma}{\sum}
\left[ \mu f^{\dagger}_{\vec{k},\sigma} f_{\vec{k},\sigma}
+ \sigma \frac{z \Gamma \gamma_{\vec{k}} }{2}
\left( f_{\vec{k},\sigma} f_{-\vec{k},-\sigma} 
- f^{\dagger}_{\vec{k},\sigma} f^{\dagger}_{-\vec{k},-\sigma}
\right) \right]
\notag \\
\label{Chapter4eq60}
\end{eqnarray}

\noindent
with

\begin{eqnarray}
\gamma_{\vec k} = \frac{1}{2} (cos k_{x} + cosk_{y})
\end{eqnarray}

\noindent
The second term in this expression is complex since $\mu = i \pi/ 2 \beta$. 
In this representation it is not possible to find a \textbf{unitary} 
Bogolioubov transformation which diagonalizes $H$ as shown in appendix
\ref{AppendixDCooperon}. Hence a rigorous implementation of 
the constraint on the particle number per site is not possible.

Reasons for this situation are the fact that the Hamiltonian contains 
terms with two particles with opposite spin created or annihilated on the same 
site which is incompatible with the fact that such configurations are not 
allowed in the present scheme. Terms of this type are typical in mean-field 
pairing Hamiltonians which lead to a non-conservation of the number of 
particles of the system as for the $BCS$-normal transition.

\subsection{Lagrange multiplier approximation}

If the sites are occupied by one electron in the average the Lagrange 
procedure works opposite to the exact procedure. 
Here a Bogoliubov transformation can be defined and used to diagonalize
the mean-field Hamiltonian.
We do not develop the 
derivation of the mean-field physical behaviour here since it has been 
done elsewhere \cite{ArovasAuerbach-88,Auerbach-94,Manousakis-91}.

\newpage

\section{Summary and conclusions}

In summary we have shown that a strict constraint on the site-occupation of 
a lattice quantum spin system described by Heisenberg-type models shows a 
sizable quantitative different localization of the critical temperature 
when compared with the outcome of an average occupation constraint.
Consequently it generates sizable effects on the behaviour of  
order parameters, see also reference \cite{Dillen-05}.
With exact site-occupation the transition temperature of
antiferromagnetic N\'eel and spin states order parameters are twice as large as
the critical temperature one gets from an average Lagrange multiplier
method.

Opposite to the average procedure the exact occupation procedure
can not be used on cooperon states mean-field Hamiltonian. 
Cooperon are $BCS$ pairs then destroy two 
quasi-particles in favor to create a new one which is a pair and then
the number of particle are not concerved opposite to the exact occupation
method. No fluctuations of the number of particles are tolerated by the
exact occupation procedure.

Due to the complexity of quantum spin systems the choice of a physically 
meaningful mean-field may depend on the coupling strengths of the model which 
describes the systems \cite{WenWilczekZee-89}. 
A $specific$ mean-field solution may even be a naive way to fix the 
"classical" contribution to the partition function which may in fact contain 
a mixture of different types of states. As already mentioned many efforts 
have been and are done in order to analyze and overcome these problems 
by means of different arguments
\cite{HermeleSenthilFisher-05,TanakaHu-05,Wen-02}. 

In a more realistic analysis one should of course take care of the 
contributions of quantum fluctuations which may be of overwhelming importance 
particularly in the vicinity of critical points. 
Chapter \ref{Chapter5} study the implications of the phase flutuations 
around the $\pi$-flux mean-field state.


\chapter{Two-dimensional Heisenberg model and
dynamical mass generation in $QED_3$ at finite temperature
\label{Chapter5}}

\minitoc
\newpage

A N\'eel ansatz is not necessarily a good candidate for the description of 
two dimensional quantum spin systems. We showed in chapter \ref{Chapter3} that
such an ansatz breaks the $SU(2)$ spin symmetry generating Goldstone bosons 
which destroy the N\'eel order. A better candidate seems to be the 
spin liquid ansatz since it keeps $SU(2)$ symmetry unaffected. 
For this reason in the following we concentrate us on the spin liquid phase.

Quantum Electrodynamics $QED_{(2+1)}$ is a common framework which can be used 
to describe stron\-gly correlated systems such as quantum spin systems in $1$ 
time and $2$ space dimensions, as well as related specific phenomena 
like high-$T_c$ superconductivity \cite{Tesanovic-02,GhaemiSenthil-05,LeeNagaosaWen-04,Morinari-05}. 
A gauge field formulation of antiferromagnetic Heisenberg 
models in $d = 2$ space dimensions maps the initial action onto 
a $QED_3$ action for spinons \cite{GhaemiSenthil-05,Morinari-05}. 
This description raises the problem of the
mean-field solution and the correlated question of the confinement of
test charges which may lead to the impossibility to determine the quantum 
fluctuation contributions through a loop expansion in this approach 
\cite{HandsKogutLucini, Herbut-02, NogueiraKleinert-05}.

We consider here the $\pi$-flux state approach introduced by Affleck and 
Marston \cite{AffleckMarston-88,MarstonAffleck-89}. 
The occupation of sites of the system by a single
particle is generally introduced by means of a Lagrange multiplier procedure
\cite{Auerbach-94,ArovasAuerbach-88}. In the present work we implement the
strict site-occupation by means of constraints imposed through a specific 
projection operator which introduces the imaginary chemical potential proposed 
by Popov and Fedotov \cite{Popov-88} for $SU(2)$ and modifies the Matsubara
frequencies as explained in chapter \ref{chapterPathIntegralPFP}.

Here we concentrate on the generation and behaviour of spinon mass which
stems from the presence of a $U(1)$ gauge field. Appelquist et al. 
\cite{Appelquist1,Appelquist2} have shown 
that at zero temperature the originally 
massless fermion can acquire a dynamically generated mass when the number $N$ 
of fermion flavors is lower than the critical value $N_c = 32/\pi^2$. Later 
Maris \cite{Maris} confirmed the existence of a critical value $N_c \simeq 3.3$
below which the dynamical mass can be generated. Since we consider only 
spin-$1/2$ systems, $N=2$ and hence $N<N_c$.
\index{Dynamical mass}
\index{$U(1)$ Gauge field}
\index{Fermion flavor}

At finite temperature Dorey and Mavromatos \cite{DoreyMavromatos} and Lee
\cite{Lee-98} showed that the dynamically generated mass vanishes at a 
temperature $T$ larger than the critical one $T_c$.

We shall show below that the imaginary chemical potential introduced by Popov 
and Fedotov \cite{Popov-88} modifies noticeably the effective potential between
two charged particles and doubles the dynamical mass transition temperature, 
in agreement with former work at the same mean-field level \cite{Dillen-05}.
This original work published in \cite{DRcondmat-06} leads to another work 
which is not directly related to the PFP and concerns the introduction of a 
Chern-Simons term in our theory \cite{DRcondmatChernSimons-06}.

The outline of the chapter is the following. In section \ref{Chapter5Section0}
we give a definition of a spinon.
In section \ref{Chapter5Section1} we derive the Lagrangian which couples 
a spinon field to a $U(1)$ gauge field. Section \ref{sectionPhotonPropagator} 
is devoted to the comparison of the effective potential constructed with and 
without strict occupation constraint. In section \ref{Chapter5Section3} 
we present the calculation of the mass term using the Schwinger-Dyson 
equation of the spinon. Section \ref{Chapter5Section3} gives an 
outlook on the spin-charge separation phenomenon and confinement problems.

\newpage

\section{Definition of a Spinon \label{Chapter5Section0}}

Electrons are particles with charge $e$ and spin $S=1/2$. Formally they can be
defined by creation and annihilation operators $C^{\dagger}_{\vec{r},\sigma}$ 
and $C_{\vec{r},\sigma}$ at position $\vec{r}$ and spin projection 
$\sigma= \pm 1/2$. In this description the $t-J$ model which was largely 
studied for low-dimensional systems after the discovery of High-$T_c$ 
superconductivity reads \cite{Anderson-87,Gros-87,Hirsch-85,Zhang-88}
\index{Spinon!Definition}

\begin{eqnarray}
H = - \underset{ij}{\sum} \underset{\sigma}{\sum} t_{ij}
C^{\dagger}_{i,\sigma} C_{j,\sigma} 
+ \underset{ij}{\sum} J_{ij} \vec{S}_i.\vec{S}_j
\end{eqnarray}

\noindent
where $t_{ij}$ is the so called hopping energy when the electrons jumps from
$j$ to site $i$ and $J_{ij}$ is the spin coupling matrix as defined in section 
\ref{Chapter2Section1}.
\index{Hoping energy}
\index{$t-J$ model}

It is believed that for a linear chain the $t-J$ model leads to the phenomenon 
of \emph{spin-charge separation} for strong electronic correlations 
\cite{Kivelson-87,MudryFradkin-94}. Observation of spin-charge separation in 
one-dimensional materials such as $Sr Cu O_2$ confirms this belief
\cite{KimAll-96}. In this framework the creation and annihilation operators
$C^{\dagger}_{\vec{r},\sigma}$ and $C_{\vec{r},\sigma}$
are represented in terms of spinon $f^\dagger_{\vec{r},\sigma}$, 
$f_{\vec{r},\sigma}$, holon $h^\dagger_{\vec{r}}$, $h_{\vec{r}}$ and
doublon $d^\dagger_{\vec{r}}$ , $d_{\vec{r}}$ creation and annihilation 
operators and read \cite{NagaosaBook2}
\index{Spin-charge separation}
\index{Spinon}
\index{Holon}
\index{Doublon}

\begin{eqnarray}
C^{\dagger}_{\vec{r},\sigma} &=& f^\dagger_{\vec{r},\sigma} h_{\vec{r}}
+ \varepsilon_{\sigma \sigma^{'}} f_{\vec{r} \sigma^{'}} d^\dagger_{\vec{r}}
\notag \\
C_{\vec{r} \sigma} &=& f_{\vec{r},\sigma} h_{\vec{r}}^\dagger
+ \varepsilon_{\sigma \sigma^{'}} f^\dagger_{\vec{r} \sigma^{'}} d_{\vec{r}}
\label{HolonSpinonDoublon}
\end{eqnarray}

\noindent
where $f$ fulfils fermion anticommutation, $h$ and $d$
fulfil commutation relations. $\varepsilon_{\sigma \sigma^{'}}$ is the
antisymmetric tensor.
The transformation \eqref{HolonSpinonDoublon} is exact when
the occupation lattice sites of these entities is controlled
by a constraint and reads

\begin{eqnarray}
h^\dagger_{\vec{r}} h_{\vec{r}} + f^\dagger_{\vec{r},\uparrow}
f_{\vec{r},\uparrow} + f^\dagger_{\vec{r},\downarrow} f_{\vec{r},\downarrow}
+ d^\dagger_{\vec{r}} d_{\vec{r}} = 1
\label{Chapter5Slave1}
\end{eqnarray}

\noindent
corresponding to the fact that each lattice site can be occupied by a hole, 
a spin $\sigma=\uparrow$, a spin $\sigma=\downarrow$ or double occupancy.
This description is the so called \emph{slave-boson} method. One sees
that holons and doublons are charge excitations while spinons are spin
excitations of the electonic systems.
\index{Slave-boson method}

If the Popov and Fedotov procedure (PFP) is adopted holons and doublons are not
allowed to live on the lattice and the constraint \eqref{Chapter5Slave1} 
reduces to

\begin{eqnarray}
f^\dagger_{\vec{r},\uparrow}f_{\vec{r},\uparrow} 
+ f^\dagger_{\vec{r},\downarrow} f_{\vec{r},\downarrow} = 1
\end{eqnarray}

\noindent
In this chapter since one considers the PFP only spinon entities are under
consideration in our Heisenberg models describing cuprates in their
insulating phase as explained in section \ref{Chapter2Section1}.
\index{High-Tc superconductivity!Cuprate}
\index{High-Tc superconductivity!Insulating phase}

\section{The $\pi$-flux Dirac action of spinons 
\label{Chapter5Section1}}

\begin{figure}[h]
\center
\begin{tabular}{cc}
\epsfig{file=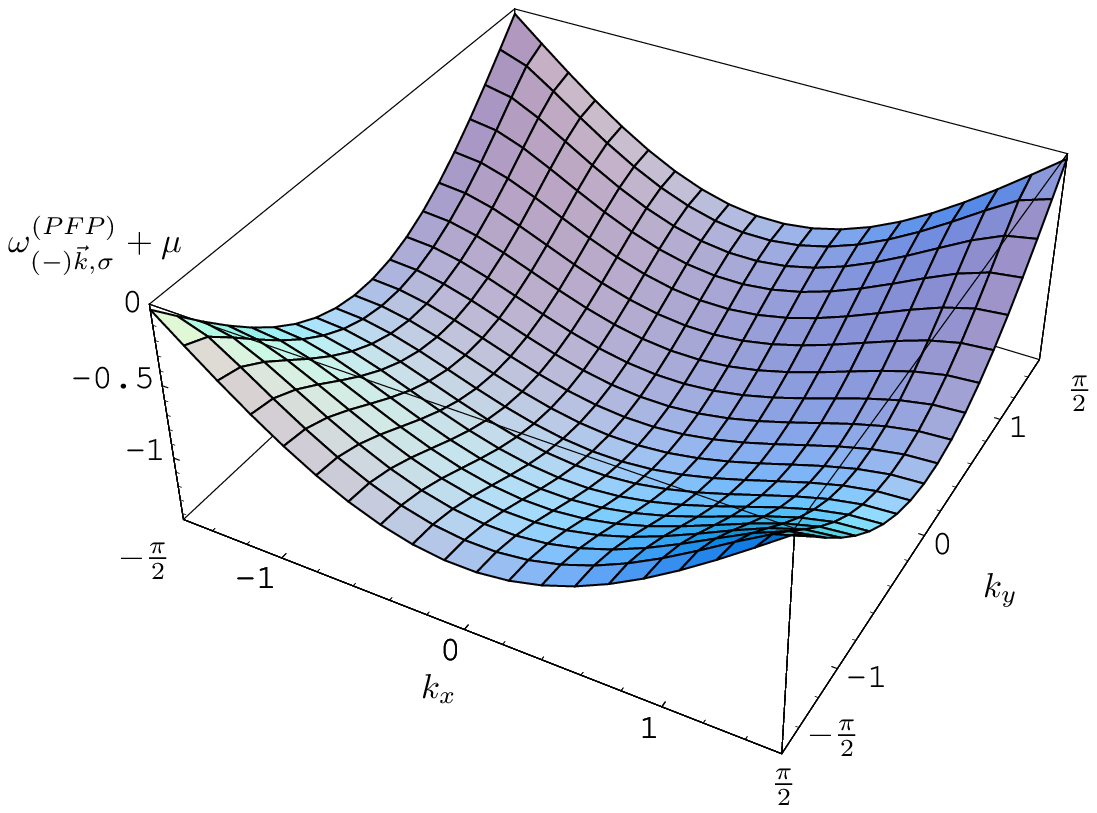,width=6cm}
 &
\epsfig{file=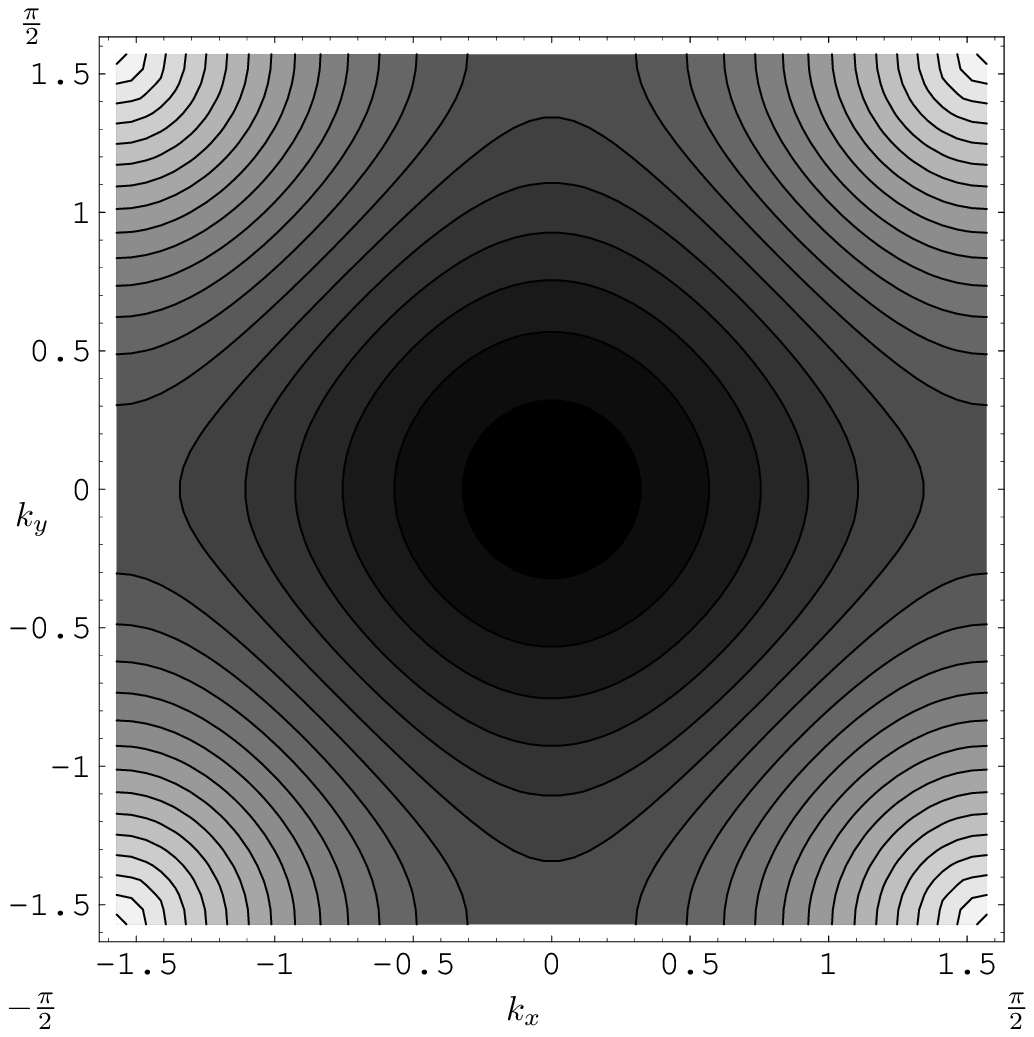,width=4.5cm}
 \\
(a) & (b)
\end{tabular}
\caption{(a) Representation of the energy spectrum 
$\omega^{(PFP)}_{(-)\vec{k},\sigma} + \mu = - 2\Delta
\sqrt{\cos^2(k_x) + \cos^2(k_y)}$ for $k_x$ and $k_y$ belonging to 
$\left[-\frac{\pi}{2},\frac{\pi}{2}\right]$ and (b) the contour 
representation of the energy spectrum showing the presence of the nodal
points $(\pm \frac{\pi}{2},\pm \frac{\pi}{2})$ where the energy is equal to
zero.}
\label{Chapter5figEnergy}
\end{figure}

We have seen in section \ref{Chapter4Section2} that the Heisenberg Hamiltonian
$H = - J \underset{<i,j>}{\sum} \vec{S}_i.\vec{S}_j$ can be expressed in
terms of the diffuson operator $\left\{\mathcal{D}_{ij} \right\}$ and leads
to the $\pi$-flux mean-field Hamiltonian
\index{Spinon!$\pi$-flux Dirac action}

\begin{eqnarray}
\mathcal{H}_{MF}^{(PFP)} &=& \mathcal{N} z \frac{\Delta^2}{|J|}
\notag \\
+ \underset{\vec{k} \in SBZ}{\sum} \underset{\sigma}{\sum}
&&\left(
f^\dagger_{\vec{k},\sigma} \,
f^\dagger_{\vec{k}+\vec{\pi},\sigma}
\right)
\left[
\begin{array}{cc}
-\mu + \Delta \cos \frac{\pi}{4} z \gamma_{k_x,k_y} &
-i \Delta \sin \frac{\pi}{4} z \gamma_{k_x,k_y+\pi} \\
+i \Delta \sin \frac{\pi}{4} z \gamma_{k_x,k_y+\pi} &
-\mu - \Delta \cos \frac{\pi}{4} z \gamma_{k_x,k_y} 
\end{array}
\right]
\left(
\begin{array}{c}
f_{\vec{k},\sigma} \\
f_{\vec{k}+\vec{\pi},\sigma}
\end{array}
\right)
\notag \\
\label{Chapter5eq1}
\end{eqnarray}

\noindent
for which the eigenvalues read
$\omega^{(PFP)}_{(\pm),\vec{k},\sigma} = - \mu \pm 2\Delta
\sqrt{\cos^2(k_x) + \cos^2(k_y)}$ and are 
shown in figure \ref{Chapter5figEnergy}.


We are interested in the low energy behaviour of the quantum spin system 
described by the Hamiltonian \eqref{Chapter5eq1} in the neighbourhood of the 
nodal points $\left( k_x = \pm \frac{\pi}{2},\pm \frac{\pi}{2} \right)$ where
the energy gap $\left( \omega^{(PFP)}_{(+),\vec{k},\sigma} -
\omega^{(PFP)}_{(-),\vec{k},\sigma} \right)$ vanishes as shown in
figure \ref{Chapter5figEnergy}.
\index{Spinon!Dispersion relation}
\index{Spinon!Nodal points}

As already shown in earlier work by Ghaemi and Senthil 
\cite{GhaemiSenthil-05} and Morinari \cite{Morinari-05} the spin liquid
Hamiltonian \eqref{Chapter5eq1} for spin systems at low energy can be 
described by four-component Dirac spinons and the corresponding Dirac 
Hamiltonian reads

\begin{eqnarray}
H = \underset{\vec{k} \in SBZ}{\sum} \underset{\sigma}{\sum}
\psi^\dagger_{\vec{k} \sigma}
\Bigg[
- \mu \Unitmatrix 
+ \widetilde{\Delta} k_{+}
\left(
\begin{array}{cc}
\tau_1 & 0 \\
0 & \tau_2
\end{array}
\right)
- \widetilde{\Delta} k_{-}
\left(
\begin{array}{cc}
\tau_2 & 0 \\
0 & \tau_1
\end{array}
\right)
\Bigg] \psi_{\vec{k} \sigma}
\label{Chapter5eq2}
\end{eqnarray}

\noindent
$k_{+} = k_x + k_y$ and $k_{-} = k_x - k_y$, 
$\widetilde{\Delta} = 2 \Delta \cos \frac{\pi}{4}$ and

\begin{eqnarray}
\psi_{\vec{k} \sigma} = \left(
\begin{array}{c}
f_{1 a, \vec{k} \sigma} \\
f_{1 b, \vec{k} \sigma} \\
f_{2 a \vec{k} \sigma} \\
f_{2 b \vec{k} \sigma}
\end{array}
\right)
\label{Chapter5eq3}
\end{eqnarray}

\noindent
Indices $1$ and $2$ in the definition of $\psi$ given by equation 
\eqref{Chapter5eq3} refers to the two independent nodal points 
$\left(\frac{\pi}{2},\frac{\pi}{2} \right)$ and 
$\left(-\frac{\pi}{2},\frac{\pi}{2} \right)$, indices $a$ and $b$ originate
from a linear transformation between $f_{\vec{k}}$ and $f_{\vec{k}+\vec{\pi}}$
as detailed in appendix \ref{AppendixEAction}.
\index{Nodal points}

The Dirac action of a spin liquid in (2+1) dimensions is derived in appendix
\ref{AppendixEAction}.
In Euclidean space the action reads
\index{Spinon!$QED_3$ Euclidean action}
\index{Spin liquid states}

\begin{eqnarray}
S_{E} = \int_0^\beta d\tau \int d^2\vec{r} \underset{\sigma}{\sum}
\bar{\psi}_{\vec{r} \sigma} \left[ \gamma^0 \left( \partial_\tau - \mu \right)
+ \widetilde{\Delta} \gamma^k \partial_k \right] \psi_{\vec{r} \sigma}
\label{SpinonAction}
\end{eqnarray}

\noindent
where $\widetilde{\Delta} = 2 \Delta \cos \frac{\pi}{4}$ is the ``light 
velocity'' space component of the three ``light velocity'' vector
$v_{\mu} = \left( 1,\widetilde{\Delta},\widetilde{\Delta} \right)$, and the
$\{\gamma^{\mu}\}$'s are the Dirac gamma matrices in (2+1) dimensions. 
\index{Light velocity}
\index{Gamma matrices}

\subsection{``Gravitational'' effects 
\label{Chapter5Section2SubsectionGravitation}}

In the action \eqref{SpinonAction} the ``light velocity'' 
$v_\mu$ which is affected by the parameter $\widetilde{\Delta}$ can be seen as 
a space-time curvature parameter. This term must modify the covariant 
derivative due to the ``gravitational'' effect induced by the ``light 
velocity'' term. We show now in which way one must modify the theory in order
to include this gravitational effect which finally is no longer important in 
our spinon action.

In a curved space-time the Dirac action of a relativistic fermion reads
\cite{Ramond}
\index{Gravitational effects!Curved space}

\begin{eqnarray}
S_{Dirac} = \int d^3 x E \bar{\psi} \gamma^p e_p^\mu \left(\partial_\mu
+ \frac{1}{8} \omega_{\mu,a b} \left[\gamma^a,\gamma^b \right] \right)\psi
\label{CurvedDirac}
\end{eqnarray}

\noindent
where $e_p^\mu$ are triads (also called \emph{dreibein}) 
\index{Gravitational effects!\emph{Dreibein}}
\index{Gravitational effects!Triads|see{\emph{Dreibein}}}
and $E = \det e_p^\mu$ relates the ``flat'' space (with ``flat'' index $m$, 
where there is no gravitational effect) and the curved space (with ``curvy'' 
index $\mu$, where there is a gravitational effect). The second term in 
brackets (equation \eqref{CurvedDirac} , where $\omega_{\mu , a b}$ is the 
connection coefficient) comes from the preservation of the \emph{local} 
invariance under Lorentz transformations in the presence of curved space-time 
\cite{Ramond} (it is similar to the gauge invariance transformation of the 
covariant derivative $\partial_\mu \rightarrow D_\mu$).
\index{Gravitational effects!Flat space}
\index{Gravitational effects!Curved space}
\index{Gravitational effects!Covariant derivation}
\index{Lorentz invariance}

Comparing equation \eqref{SpinonAction} with \eqref{CurvedDirac} one
shifts the imaginary time derivation $\partial_\tau \rightarrow \partial_\tau 
+ \mu$ in \eqref{SpinonAction} which leads to a new definition of the 
Matsubara frequencies only for the fermion fields $\psi$ \cite{Popov-88}.
These frequencies read

\begin{eqnarray}
\widetilde{\omega}_{F,n} = \omega_{F,n} - \mu/i = \frac{2 \pi}{\beta} (n + 1/4)
\label{Chapter5Matsubara}
\end{eqnarray}

\noindent
Further we define the \emph{dreibein} as
\index{Gravitational effects!\emph{Dreibein}}

\begin{eqnarray*}
e_p^\mu = \left(
\begin{array}{ccc}
\frac{1}{\sqrt{\widetilde{\Delta}}} & 0 & 0 \\
0 & \sqrt{\widetilde{\Delta}} & 0 \\
0 & 0 & \sqrt{\widetilde{\Delta}}
\end{array}
\right)
\end{eqnarray*}

\noindent
Hence spinons move in a ``gravitational'' field which can be characterized by 
the metric
\index{Gravitational effects!Metric tensor}

\begin{eqnarray}
g_{\mu \nu} = \delta^{mn} e_m^\mu e_n^\nu =
\left[
\begin{array}{ccc}
\frac{1}{\widetilde{\Delta}} & 0 & 0 \\
0 & \widetilde{\Delta} & 0 \\
0 & 0 & \widetilde{\Delta}
\end{array}
\right]
\label{Chapter5eq5}
\end{eqnarray}

\noindent
One should add the term 

\begin{eqnarray}
\int_0^\beta d\tau \int d^2 \vec{r} E
\underset{\sigma}{\sum} \bar{\psi}_{\vec{r}\sigma} \gamma^p 
\left[e_p^\mu \frac{1}{8} \omega_{\mu,a b} 
\left[\gamma^a,\gamma^b \right] \right] \psi_{\vec{r}\sigma}
\label{CurvedOmega}
\end{eqnarray}

\noindent
in equation \eqref{SpinonAction} in order to verify the \emph{local} 
invariance of the Lorentz transformation under the ``gravitational'' field.
The metric can be handled \cite{Volovik} assuming $\widetilde{\Delta} = 1$ 
without altering the physics of the problem as we will explain below. 

Since the metric \eqref{Chapter5eq5} is no longer that of a flat space-time 
the connection coefficient $\omega_{\mu , a b}$ is defined by

\begin{eqnarray}
\omega_{\mu, a b} = -\left( \partial_\mu e_{a b}
- \Gamma_{\mu a}^{\phantom{\mu a} \gamma} e_{\gamma b} \right)
\end{eqnarray}

\noindent
The $\Gamma$'s are Christoffel symbols 
($\Gamma_{\mu \nu \gamma} = \frac{1}{2}
\left(g_{\mu \nu,\gamma} + g_{\mu \gamma,\nu} - g_{\nu \gamma,\mu } \right)$ 
with $g_{\mu \nu, \gamma} = \partial_\gamma g_{\mu \nu}$) 
and $e_a^\mu$ are the \emph{dreibein}.
Since $\widetilde{\Delta}$ can be considered as constant in space-time
we see clearly that the \emph{dreibein} are also constant with respect to the
space-time coordinates. Hence $\omega_{\alpha, a b} = 0$ in a dilated flat 
space-time with the Euclidean metric \eqref{Chapter5eq5}. 
Since we showed that \eqref{CurvedOmega} is equal to zero
one can put $\widetilde{\Delta}=1$ in equation \eqref{SpinonAction} 
and finally equation \eqref{SpinonAction} reduces to
\index{Spinon!$QED_3$ Euclidean action}

\begin{eqnarray}
S_{E} = \int_0^\beta d\tau \int d^2\vec{r}
\underset{\sigma}{\sum}
\bar{\psi}_{\vec{r} \sigma}\left( \tau \right)
\gamma_\mu \partial_\mu \psi_{\vec{r} \sigma}\left( \tau \right)
\label{Chapter5eq14}
\end{eqnarray}

\noindent
Recall that we shifted the time derivative $\partial_\tau$ by the
imaginary chemical potential $\mu$ modifying the Mastubara frequencies 
\eqref{Chapter5Matsubara}.
More details concerning the modification of the Matsubara frequencies are 
given in subsection \ref{SubsectionMatsubara}.
This modification will induce substantial consequences as it will be shown 
in the following.

\subsection{Quantum Electrodynamic Spinon action in (2+1) dimensions}

Since the Heisenberg Hamiltonian \eqref{Chapter5eq2} is gauge invariant 
in the $U(1)$ transformation $\psi \rightarrow e^{i g \theta} \psi$,
where $\theta$ is a space-dependent scalar function, the Dirac 
action can be written in the form
\index{$U(1)$ Gauge field}

\begin{eqnarray}
S_{E} = \int_0^\beta d\tau \int d^2\vec{r}
&\Bigg\{&
- \frac{1}{2} a_\mu\left(\vec{r},\tau\right) 
\left[ \left(\Box \delta^{\mu \nu}
+ (1 - \lambda) \partial^\mu \partial^\nu \right) \right] 
a_\nu \left( \vec{r},\tau \right)
\notag \\
&+& \underset{\sigma}{\sum}
\bar{\psi}_{\vec{r} \sigma}\left( \tau \right) \left[
\gamma_\mu \left( \partial_\mu - i g a_\mu \right)
\right] \psi_{\vec{r} \sigma}\left( \tau \right)
\Bigg\}
\label{SpinonGaugeAction}
\end{eqnarray}

\noindent
Here $g$ comes as the coupling strength
between the gauge field $a_\mu$ and the Dirac spinons
$\psi$. 
The gauge field $a_\mu$ is related to the phase $\theta \left( \vec{r} \right)$
of the spinon at site $\vec{r}$ through the gauge transformation 
$f_{\vec{r},\sigma} \rightarrow e^{i g \theta \left( \vec{r} \right)} 
f_{\vec{r},\sigma}$ which keeps the Heisenberg Hamiltonian
invariant. From the definition of $\psi$ one gets $\psi_{\vec{r} \sigma} 
\rightarrow e^{i g \theta \left( \vec{r} \right)}\psi_{\vec{r} \sigma}$.
It is clear that $\theta \left( \vec{r} \right)$ is the phase at the lattice 
site $\vec{r}$ and that $a_\mu\left( \vec{r}\right) = \partial_\mu \theta 
\left( \vec{r} \right)$. Hence the fluctuations of the flux $\phi$ through
the plaquette (see figure \ref{Chapter4Fig2}) are directly related to the 
circulation of the gauge field $a_\mu$ around the plaquette
\index{Plaquette}

\begin{eqnarray*}
\phi &=& g \underset{<i,j> \in \Box}{\sum} 
\left(
\theta \left( \vec{r_i} \right) - \theta \left( \vec{r_j} \right) 
\right) 
\notag \\
&=& g \int_{\Box} d\vec{l}.\vec{a}
\end{eqnarray*}

\noindent
Phase fluctuations of lattice sites are measured by the gauge field $a_\mu$.
In \eqref{SpinonGaugeAction} the first term corresponds to the 
``Maxwell'' term $-\frac{1}{4}f_{\mu \nu} f^{\mu \nu}$ of the gauge field 
$a_\mu$ where $f^{\mu \nu} = \partial_\mu a_\nu - \partial_\nu a_\mu$,
$\lambda$ is the parameter of the Faddeev-Popov gauge fixing term
$-\lambda \left(\partial^\mu a_\mu \right)^2$ \cite{Itzykson},  
$\delta^{\mu \nu}$ the Kronecker $\delta$ and
$\Box = \partial_\tau^2 + \vec{\nabla}^2$ is the Laplacian in Euclidean 
space-time.
\index{Faddeev-Popov gauge fixing}
\index{Laplacian}

Since the spinons $\psi$ are minimally coupled to the gauge field $a_{\mu}$, 
materializing the effect of fluctuations around the mean-field spin liquid 
ansatz described by the ``Diffuson'', the energy-momentum conservation
leads to consider the gauge-invariant and symmetric energy-momemtum tensor
\index{Diffuson}
\index{Energy-momentum tensor}

\begin{eqnarray}
\Theta_{\mu \nu} = \frac{\partial \mathcal{L}}{\partial \left(\partial_{\mu}
a_{\delta} \right)}\partial_{\nu}a_{\delta} - g_{\mu \nu} \mathcal{L}
+ \partial_{\rho}\left(f_{\mu \rho} a_{\nu} \right)
\end{eqnarray}

\noindent
where $\mathcal{L}$ is the $QED_3$ Lagrangian of the spinon deriving from
$S_{E} = \int_0^\beta d\tau \mathcal{L}$ in the Euclidean space.
The Maxwell Lagrangian $-\frac{1}{4} f_{\mu \nu}f^{\mu \nu} = 
\frac{1}{2} a_{\mu} \left(\Box \delta^{\mu \nu} + (1 - \lambda) \partial^\mu 
\partial^\nu \right)a_{\nu}$ is added to \eqref{Chapter5eq14} in order to 
verify the energy-momentum conservation \cite{Itzykson}

\begin{eqnarray}
\partial_{\mu} \Theta_{\mu \nu} = 0
\label{EMLaw}
\end{eqnarray}

\noindent
Notice that ``photon'' described by the gauge field $a_{\mu}$ are not present
in nature and thus the dynamical Maxwell term above looks inadequate as it
stands. In order to be coherent one should do the change of variable

\begin{eqnarray}
a_{\mu} = a_{\mu}/g
\end{eqnarray}

\noindent
where $g$ is the coupling constant between the spinon and the gauge field.
The spinon action \eqref{SpinonGaugeAction} reads
\index{Spinon!$QED_3$ Euclidean action}

\begin{eqnarray}
S_{E} = \int_0^\beta d\tau \int d^2\vec{r}
&\Bigg\{&
- \frac{1}{2 g^2} a_\mu\left(\vec{r},\tau\right) 
\left[ \left(\Box \delta^{\mu \nu}
+ (1 - \lambda) \partial^\mu \partial^\nu \right) \right] 
a_\nu \left( \vec{r},\tau \right)
\notag \\
&+& \underset{\sigma}{\sum}
\bar{\psi}_{\vec{r} \sigma}\left( \tau \right) \left[
\gamma_\mu \left( \partial_\mu - i a_\mu \right)
\right] \psi_{\vec{r} \sigma}\left( \tau \right)
\Bigg\}
\end{eqnarray}

\noindent
The dynamical term can now be removed in the limit of $g \rightarrow \infty$
verifying the energy-momentum conservation law \eqref{EMLaw}.
In this limit we are treating a \emph{\bfseries strongly correlated} 
electron system. Keeping this in mind we go further in the development using 
\eqref{SpinonGaugeAction}.
\index{Strongly correlated systems}

\section{The ``Photon'' propagator at finite temperature
\label{sectionPhotonPropagator}}

Integrating over the fermion fields $\psi$, using relation 
\eqref{AppendixGaussianGrassmannIntegral} in 
appendix \ref{AppendixGrassmann}, the partition function
$\mathcal{Z}\left[ \psi,a \right] 
= \int \mathcal{D}\left(\psi,a\right) e^{-S_{E}}$ with
action $S_{E}$ given by equation \eqref{SpinonGaugeAction} leads 
to the pure gauge partition function

\begin{eqnarray}
\mathcal{Z}\left[ a \right] = 
\int \mathcal{D}\left( a \right) e^{ - S_{eff}\left[ a \right] }
\end{eqnarray}

\noindent
where the effective pure gauge field action $S_{eff}\left[ a \right]$ comes
in the form

\begin{eqnarray}
S_{eff} \left[ a \right] =
\int_0^\beta d\tau \int d^2\vec{r} 
\Bigg\{
- \frac{1}{2} a_\mu \left[ \left(\Box \delta^{\mu \nu}
+ (1 - \lambda) \partial^\mu \partial^\nu \right) \right] a_\nu
\Bigg\}
- \ln \det \left[  \gamma_\mu \left( \partial_\mu - i g a_\mu \right) \right]
\notag \\
\label{Chapter5eq10}
\end{eqnarray}

\noindent
Here one recognizes similarity between the second term in \eqref{Chapter5eq10}
and the \emph{log-det} term in \eqref{EffectiveAction}. Following the
same steps as in section \ref{Chapter3Section4} one can develop the 
second term in the effective gauge field action $S_{eff}\left[ a \right]$ 
into a series and write

\begin{eqnarray}
\ln \det \left[  \gamma_\mu \left( \partial_\mu - i g a_\mu \right) \right]
=
\ln \det G_{F}^{-1} - \overset{\infty}{\underset{n = 1}{\sum}}
\frac{1}{n} Tr \left[ i G_{F} \gamma^\mu a_\mu \right]^n
\label{Chapter5eq11}
\end{eqnarray}

\noindent
where $G_{F}^{-1}(k-k^{'}) = i \frac{\gamma^\mu k_\mu}{(2\pi)^2\beta} 
\delta(k-k^{'})$ is the fermion Green function in the Fourier space-time with
$k = \left( \widetilde{\omega}_{F,n}, \vec{k} \right)$, hence 
$G_{F} =-i \frac{\gamma^\mu k_\mu}{k^2} (2\pi)^2\beta 
\delta\left( k - k^{'} \right)$. The first term on the right hand side of 
equation \eqref{Chapter5eq11} being independent of the gauge field 
$\{ a_\mu \}$ can be removed from the series since we focus our attention
on pure gauge field terms. The first term proportional to the gauge field
$n=1$ in the sum vanish since $tr \gamma_\mu = 0$. 
Keeping only second order terms in the gauge field in order to treat
gaussian fluctuations one gets the pure gauge action 
\index{Green function}

\begin{eqnarray}
S_{eff}^{(2)}\left[ a \right] &=&
\int_0^\beta d\tau \int d^2\vec{r} 
\Bigg\{
- \frac{1}{2} a_\mu \left[ \left(\Box \delta^{\mu \nu}
+ (1 - \lambda) \partial^\mu \partial^\nu \right) \right] a_\nu
\Bigg\}
\notag \\ 
&+& g^2 \frac{1}{2 \beta} \underset{\sigma}{\sum} 
\underset{\omega_{F,1}}{\sum} \int \frac{d^2\vec{k}_1}{(2\pi)^2}.
\frac{1}{\beta} \underset{\omega_{F}^{''}}{\sum} \int \frac{d^2\vec{k^{''}}}
{(2\pi)^2}
\notag \\
&&
tr \Bigg[
\frac{\gamma^\rho k_{1,\rho}}{k_1^2}.\gamma^\mu a_\mu(k_1-k^{''}).
\frac{\gamma^{\eta} k^{''}_\eta}{{k^{''}}^2}.
\gamma^\nu a_\nu \left(-(k_1 - k^{''}) \right)
\Bigg]
\label{Chapter5eq13}
\end{eqnarray}

\noindent
The second term in equation \eqref{Chapter5eq13} is worked out in details
in appendix \ref{AppendixEPolarization} and one gets

\begin{eqnarray}
S_{eff}^{(2)} =
- \frac{1}{2 \beta}  
\underset{\omega_{B}}{\sum} \int \frac{d^2\vec{q}}{(2\pi)^2}
a_\mu(-q) \left[ {\Delta^{(0)}_{\mu \nu}}^{-1} 
+ \Pi_{\mu \nu}(q) \right] a_\nu(q)
\end{eqnarray}

\noindent
${\Delta^{(0)}_{\mu \nu}}^{-1} = \left[ \left(\Box \delta^{\mu \nu}
+ (1 - \lambda) \partial^\mu \partial^\nu \right) \right]$ is the bare
photon propagator. 
The detailed calculation of the polarization function $\Pi_{\mu \nu}$
is given in appendix \ref{AppendixEPolarization}. Equations 
\eqref{AppendixEeq24} and \eqref{AppendixEeq25} give the components of 
$\Pi_{\mu \nu}$. The finite-temperature dressed photon propagator in Euclidean 
space (imaginary time formulation) verifies the Dyson equation

\begin{eqnarray}
\Delta_{\mu \nu}^{-1} &=& {\Delta_{\mu \nu}^{(0)}}^{-1} + \Pi_{\mu \nu}
\label{Dyson}
\end{eqnarray}

\noindent
Finally the gauge effective action reads 
$S_{eff}^{(2)} = - \frac{1}{2 \beta}  
\underset{\omega_{B}}{\sum} \int \frac{d^2\vec{q}}{(2\pi)^2}
a_\mu(-q) \Delta_{\mu \nu}^{-1}\left( q \right) a_\nu(q)$.

\subsection{Comparison of the Popov and Fedotov procedure 
with the Lagrange multiplier method}

\begin{table}[h]
\center
\caption{Comparison of polarization function components obtained
from the Lagrange multiplier approximation and the Popov and Fedotov
procedure}
\label{Chapter5Table1}
\begin{tabular}{|c||c|c|}

\hline
&&\\
$\alpha = 2.g^2$
 &
Lagrange multiplier method
 &
PFP
 \\

&&\\
\hline
\hline
&&\\

$\widetilde{\Pi}_1$
 &  
$\frac{\alpha q}{\pi} \int_0^1 dx \sqrt{x(1-x)}
\frac{\sinh \beta q \sqrt{x(1-x)} }{D(X,Y)}$
 &  
$\frac{\alpha q}{\pi} \int_0^1 dx \sqrt{x(1-x)}
\frac{\sinh \beta q \sqrt{x(1-x)} }{D(X,Y)}$
 \\

&&\\
\hline
&&\\

$\widetilde{\Pi}_2$
 &
$\frac{\alpha m}{\beta}
\int_0^1 dx (1-2x) \frac{\sin 2 \pi x m}{D(X,Y)}$
 &
$\frac{\alpha m}{\beta}
\int_0^1 dx (1-2x) \frac{\cos 2 \pi x m}{D(X,Y)}$
 \\

&&\\
\hline
&&\\

$\widetilde{\Pi}_3$
 &
$\frac{\alpha}{\pi \beta} \int_0^1 dx \ln 2 D(X,Y)$
 &
$\frac{\alpha}{\pi \beta} \int_0^1 dx \ln 2 D(X,Y)$
 \\

&&\\
\hline
&&\\

$D(X,Y)$
 &
$\cosh \left( \beta q \sqrt{x(1-x)} \right) + \cos (2\pi x m)$
 &
$\cosh \left( \beta q \sqrt{x(1-x)} \right) + \sin (2\pi x m)$
 \\

&&\\
\hline
\end{tabular}
\end{table}

Table \ref{Chapter5Table1} compares the polarization function
components $\widetilde{\Pi}_1$, $\widetilde{\Pi}_2$ and $\widetilde{\Pi}_3$
obtained by means of the Lagrange multiplier approximation for which 
$\lambda = 0$ as shown in section \ref{Chapter4Section2} with the PFP
as computed in appendix \ref{AppendixEPolarization}. Differences appear in the
denominator term $D(X,Y)$ and in the numerator of the integrant in 
$\widetilde{\Pi}_2$ where the cosine terms are replaced by sine terms.

We now push the comparison further and show below how the PFP
modifies the effective potential between two test particles and also
affects the dynamically generated mass of the spinons.

\subsection{Covariant description of the polarization function
\label{Chapter5Section2Subsection1}}

One may believe that a system at finite temperature breaks Lorentz invariance
since the frame described by the heat bath already selects out a specific 
Lorentz frame. However this is not true and one can formulate the statistical 
mechanics in a Lorentz covariant form \cite{Das}.
\index{Lorentz invariance}

We consider the three dimensional Euclidean space.
Define the proper $3$-velocity $u^{\mu}$ of the heat bath.
In the rest frame of the heat bath the three velocity
has the form $u^{\mu} = \left( 1,0,0 \right)$ and the inverse temperature
$\beta$ characterizes the thermal property of the heat bath.
\index{Rest frame}
\index{Heat bath}

Given the $3$-velocity vector $u^{\mu}$ one can decompose any three vector 
into parallel and orthogonal components with respect to the proper velocity
of the heat bath, the velocity $u^{\mu}$. In particular the parallel and 
transversal components of the three momentum $q^{\mu}$ with respect to 
$u^{\mu}$ read

\begin{eqnarray}
q^{\mu}_{\parallel} = \left( q . u \right) u^{\mu}
\end{eqnarray}
\begin{eqnarray}
\widetilde{q}^{\mu} = q^{\mu} - q^{\mu}_{\parallel}
\end{eqnarray}

\noindent
Similarly one can decompose any vector and tensor 
into components which is parallel
and transverse to a given momentum vector $q^{\mu}$

\begin{eqnarray}
\bar{u}_\mu &=& u_\mu - \frac{(q.u)}{q^2} q_\mu \\
\bar{\eta}_{\mu \nu} &=& \delta_{\mu \nu} - \frac{q_\mu q_\nu}{q^2}
\end{eqnarray}

\noindent
It is now easy to define second rank symmetric tensors $T_{\mu \nu}$
constructed at finite temperature from $q^{\mu}$, $u^{\mu}$ and 
$\delta_{\mu \nu}$ which verifies $q^{\mu} T_{\mu \nu}=0$
\index{Photon propagator!Covariant tensor $A_{\mu \nu}$ and $B_{\mu \nu}$}

\begin{eqnarray}
A_{\mu \nu} &=& \delta_{\mu \nu} - u^{\mu} u^{\nu} 
- \frac{\widetilde{q}_\mu \widetilde{q}_\nu} {\widetilde{q}^2} \\
B_{\mu \nu} &=& \frac{q^2}{\widetilde{q}^2} \bar{u}_\mu \bar{u}_\nu \\
C_{\mu \nu} &=& \delta_{\mu \nu} - \frac{q_{\mu} q_{\nu}}{q^2}
\end{eqnarray}

\noindent
where $A_{\mu \nu}$ and $B_{\mu \nu}$ verifies the relation

\begin{eqnarray}
A_{\mu \nu} + B_{\mu \nu} &=& C_{\mu \nu}
\end{eqnarray}

Since one considers a spin system
at finite temperature and ``relativistic'' covariance
should be preserved the polarization function may be put in the general 
form \cite{Das}
\index{Photon propagator!Polarization function}

\begin{eqnarray}
\Pi_{\mu \nu} = \Pi_A A_{\mu \nu} + \Pi_B B_{\mu \nu}
\label{Chapter5eq23}
\end{eqnarray}

\noindent
and the Dyson equation \eqref{Dyson} can now be expressed in a covariant
form if one uses relation \eqref{Chapter5eq23}.
\index{Photon propagator!Dyson equation}

\subsection{Dressed ``photon'' propagator \label{Chapter5Section2Subsection3}}

Inverting the Dyson equation \eqref{Dyson}
the dressed photon propagator $\Delta_{\mu \nu}$
is obtained by summation of the geometric series 
\index{Photon propagator!Dyson equation}

\begin{eqnarray}
\Delta_{\mu \nu} = \Delta_{\mu \nu}^{(0)} + 
\Delta_{\mu \alpha}^{(0)}.(-\Pi^{\alpha \beta}).\Delta_{\beta \nu}^{(0)}
+ \Delta_{\mu \alpha}^{(0)}.(-\Pi^{\alpha \beta}).\Delta_{\beta \delta}^{(0)}.
(-\Pi^{\delta \rho}).\Delta_{\rho \nu}^{(0)} + \cdots
\label{Chapter5eq24}
\end{eqnarray}

\begin{figure}
\centering
\epsfig{file=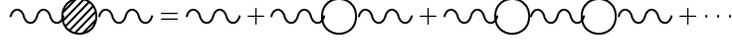}
\caption{The dressed photon propagator. Wavy lines correspond to the photon
and solid loops to the fermion insertions}
\label{FeynmannGraphs1}
\end{figure}

\noindent
Figure \ref{FeynmannGraphs1} shows the Feynman diagrammatic representation
of the Dyson series \eqref{Chapter5eq24}. The dressed photon propagator reads
\index{Feynman diagram}

\begin{eqnarray}
\Delta_{\mu \nu} = \frac{A_{\mu \nu}}{q^2 + \Pi_A}
+ \frac{B_{\mu \nu}}{q^2 + \Pi_B} 
- (1-1/\lambda) \frac{q_\mu q_\nu}{(q^2)^2}
\notag \\
\label{PhotonPropagator}
\end{eqnarray}

\noindent
where $\Pi_A$ and $\Pi_B$ are related to $\widetilde{\Pi}_k$ by

\begin{eqnarray}
\Pi_A &=& \widetilde{\Pi}_1 + \widetilde{\Pi}_2 \\ 
\Pi_B &=& \widetilde{\Pi}_3
\end{eqnarray}

\noindent
The expressions of $\widetilde{\Pi}_1$, $\widetilde{\Pi}_2$ and
$\widetilde{\Pi}_3$ are explicitly worked out in appendix 
\ref{AppendixEPolarization}. Here $q^0 = \frac{2 \pi}{\beta} m$
is the boson Matsubara frequency energy component 
of the photon three-vector $q^{\mu}$.
\index{Matsubara frequencies!Bosonic}

The dressed photon propagator is clearly expressed in a covariant form since
the basis tensor $A_{\mu \nu}$ and $B_{\mu \nu}$ are as shown in subsection
\ref{Chapter5Section2Subsection1}. Remarkably the dressed photon propagator
is composed of a longitudinal and a transverse part with respect to the photon
momentum $q_{\mu}$ unlike the bare photon propagator
${\Delta^{(0)}_{\mu \nu}}^{-1} = \frac{1}{q^2}
\left[ \delta_{\mu \nu} - \left(1 - 1/\lambda \right) 
\frac{q_\mu q_\nu}{q^2} \right]$ when one takes the Landau gauge fixing
condition $\lambda \rightarrow \infty$. Even $\Pi_{\mu \nu}$ is transverse
to the photon momentum $q_\mu$. In the Landau gauge one gets
\index{Photon propagator!Bare photon propagator}
\index{Landau gauge}

\begin{eqnarray}
\Delta_{\mu \nu} = \frac{A_{\mu \nu}}{q^2 + \Pi_A}
+ \frac{B_{\mu \nu}}{q^2 + \Pi_B} 
- \frac{q_\mu q_\nu}{(q^2)^2}
\end{eqnarray}

\noindent
showing clearly that the dressed photon propagator 
\index{Photon propagator!Dressed photon propagator}
presents a longitudinal part with respect to the
photon momentum since $q_\mu A_{\mu \nu} = q_\mu B_{\mu \nu} = 0$.

\subsection{Effective potential between test particles}

The questions to which one answers here concern the derivation of the 
effective potential interacting between two spinons and the impact of PFP 
site-occupation constraint on it.

The spinon action \eqref{SpinonGaugeAction} describes a gas of spinons
and photons coupled together by a coupling constant $g$ which can
be interpreted as the charge of the spinon.
Paying attention to the analogy with a plasma made of electrons
and the spinon gas one can identify the interaction between two spinon like 
one would identify the interaction between two electrons.
The term proportional to the product $q_a . q_b$ where $q_a$ and $q_b$ are
the charges of particles $a$ and $b$ correspond to the Coulomb interaction 
between two charge and reads
\index{Plasma}
\index{Coulomb interaction}

\begin{eqnarray}
V(R = |\vec{r}_a - \vec{r}_b|) = - 2\pi q_a q_b \ln 
\left(|\vec{r}_a - \vec{r}_b|  \right)
\end{eqnarray}

\noindent
in a two-dimensional system where particles are at position $\vec{r}_a$ with
charge $q_a$ and $\vec{r}_b$ with charge $q_b$. The corresponding action comes
as

\begin{eqnarray}
S_{plasma} = \frac{\beta}{2} \left\{ \underset{a \neq b}{\sum}
- q_a q_b 2\pi \ln \left(|\vec{r}_a - \vec{r}_b|  \right) + \text{const}.
\underset{a}{\sum} q_a^2 \right\}
\end{eqnarray}

\noindent
for a plasma two-dimensional system at temperature $1/\beta$.
Define the spinon three-current $j_{\mu} = - i g \underset{\sigma}{\sum}
\bar{\psi}_{\sigma} \gamma_\mu \psi_{\sigma}$ where $\psi$ is the Dirac spinor 
of the spinon defined in section \ref{Chapter5Section1}. In a classical
picture the time component of the current vector reads $j_0(x) = \underset{a}
{\sum} q_a \delta\left(x - x_a\right)$.
Identifying the charge $q$ with the component $j_0$ of the three-vector $j_\mu$
the Coulomb interaction comes as

\begin{eqnarray}
V\left( R \right) = q_a . q_b \frac{\partial^2}
{\partial j_0 \left( R, \tau \right) 
\partial j_0 \left( 0, \tau \right) } \ln \mathcal{Z}
\label{Chapter5eq31}
\end{eqnarray}

\noindent
where $\mathcal{Z} \propto e^{-S_{plasma}}$ 
is the partition function of the corresponding action
$S_{plasma}$. Now using relation \eqref{Chapter5eq31} with the partition
function $\mathcal{Z}\left[ \psi,a \right] 
= \int \mathcal{D}\left(\psi,a\right) e^{-S_{E}}$ with the spinon action 
$S_{E}$ given by equation \eqref{SpinonGaugeAction} and rewritten as

\begin{eqnarray}
S_{E} = \int_0^\beta d\tau \int d^2\vec{r}
&\Bigg\{&
- \frac{1}{2} a_\mu\left(\vec{r},\tau\right) 
\left[ \left(\Box \delta^{\mu \nu}
+ (1 - \lambda) \partial^\mu \partial^\nu \right) \right] 
a_\nu \left( \vec{r},\tau \right)
\notag \\
&+& \underset{\sigma}{\sum}
\bar{\psi}_{\vec{r} \sigma}\left( \tau \right) 
\gamma_\mu  \partial_\mu \psi_{\vec{r} \sigma}\left( \tau \right)
+ j_\mu\left(\vec{r},\tau\right) a_\mu \left(\vec{r},\tau\right)
\Bigg\}
\end{eqnarray}

\noindent
Considering the term proportional in $j_{\mu}$, one finally gets from deriving 
the effective potential

\begin{eqnarray}
V\left( R \right) = q_a q_b \int_0^\beta d\tau 
< a_0 \left( R, \tau \right) a_0 \left( 0, \tau \right) >
\end{eqnarray}

\noindent
Here $< a_0 \left( R, \tau \right) a_0 \left( 0, \tau \right) >$ is to
identify with the dressed photon propagator time components $\Delta_{00}$.
The effective static potential $V(R)$ between two test particles (spinons) of 
opposite chages $q_a=-q_b=g$ at distance $R$ is given by
\index{Test particles}

\begin{eqnarray}
V(R) &=& - g^2 \int_0^\beta d\tau \Delta_{00}(\tau,R)
\label{Chapter5eq33}
\end{eqnarray}

\noindent
After a Fourier transformation on $\Delta_{00}$ the time components of the
dressed propagator $\Delta_{\mu \nu}$ in equation \eqref{Chapter5eq33} one gets

\begin{eqnarray}
V(R) = -g^2 \frac{1}{2\pi} \int \frac{d^2\vec{q}}{(2\pi)^2} \Delta_{00}
(q^0=0,\vec{q}) e^{i\vec{q}.\vec{R}}
\end{eqnarray}

\noindent
with

\begin{eqnarray}
\Delta_{00}\left( q^0 = 0, \vec{q} \right) =
\frac{1}{q^2 + \widetilde{\Pi}_3 \left(q^0 = \frac{2\pi}{\beta}m=0 \right)}
\end{eqnarray}

\noindent
The effective potential then reads

\begin{eqnarray}
V(R) = -\frac{g^2}{2\pi} \int_0^\infty dq q J_0(qR).\frac{1}{q^2 +
\widetilde{\Pi}_3(m=0)}
\end{eqnarray}

\noindent
where $J_0(qR)$ is the zero order Bessel function of the first kind. 
\index{Bessel function}

The polarization contribution $\widetilde{\Pi}_3(q^0=0,\vec{q})$ is equal to 
$\frac{\alpha}{\pi \beta}
\int_0^1 dx \ln \left(2 \cosh \beta q \sqrt{x(1-x)} \right)$ when taking
the Popov and Fedotov imaginary chemical potential into account. 
This has to be compared to the expression
$\frac{2 \alpha}{\pi \beta} \int_0^1 dx \ln \left(2 \cosh \frac{\beta}{2} q 
\sqrt{x(1-x)} \right)$ when the Lagrange multiplier method for
which $\lambda=0$ is used \cite{DoreyMavromatos}, a detailed comparaison is
given in table \ref{Chapter5Table1}.
\index{Popov and Fedotov procedure!Imaginary chemical potential}
\index{Lagrange multiplier}

\begin{figure}
\centering
\epsfig{file=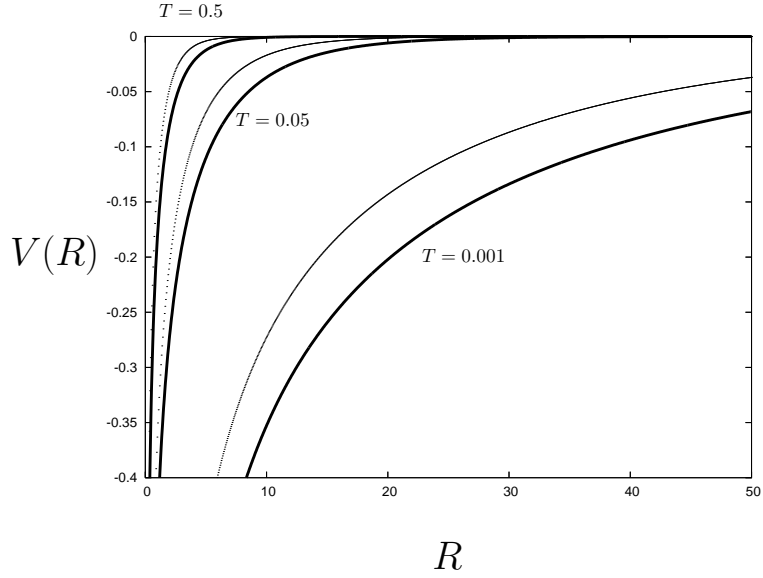,width=10cm}
\caption{Effective static potential with (fat line) and without (dotted line)
the Popov-Fedotov imaginary chemical potential for the temperature 
$T=\{0.001,0.05,0.5\}$.}
\label{StaticPotential}
\end{figure}

For small momentum $q \rightarrow 0$, $\widetilde{\Pi}_3(m=0)$ can be 
identified as a mass term $(M_0^{(PFP)}(\beta))^2$ and reads

\begin{eqnarray}
\underset{q \rightarrow 0}{\lim} \widetilde{\Pi}_3(m=0) =
(M_0^{(PFP)}(\beta))^2 = \frac{\alpha}{\pi \beta} \ln 2
\end{eqnarray}

\noindent
For $R \gg (M_0^{(PFP)})^{-1}$ the effective potential reads

\begin{eqnarray*}
V(R,\beta) &\simeq& - \frac{g^2}{2\pi} \int_0^\infty dq \frac{q J_0(qR)}{q^2 +
\left(M_0^{(PFP)}\right)^2} 
\notag \\
&=& - \frac{\alpha}{N} \sqrt{\frac{1}{8 \pi R M_0^{(PFP)}}} 
e^{-M_0^{(PFP)} R}
\end{eqnarray*}
where $N=2$ since we consider only $S=1/2$ spins.

Figure \ref{StaticPotential} shows the effective potential between two
opposite test charges at distance $R \gg (M_0^{(PFP)})^{-1}$. 
The screening effect is smaller when the imaginary PFP chemical potential $\mu$
is implemented rather than the Lagrange multiplier $\lambda$.
By inspection one sees that $(M_0^{(PFP)})^{-1} = \sqrt{2} 
(M_0^{(\lambda)})^{-1}$. The main effect of combinating of the PFP
and thermal fluctuations is to increase the range of effective
static interaction, which is infinite at zero temperature, between two test 
particles of opposite charges compared to the case where the Lagrange 
multiplier method is used.
\index{Screening effect}
\index{Thermal fluctuations}

\section{Dynamical mass generation \label{Chapter5Section3}}
\index{Dynamical mass|(}

\begin{figure}
\centering
\epsfig{file=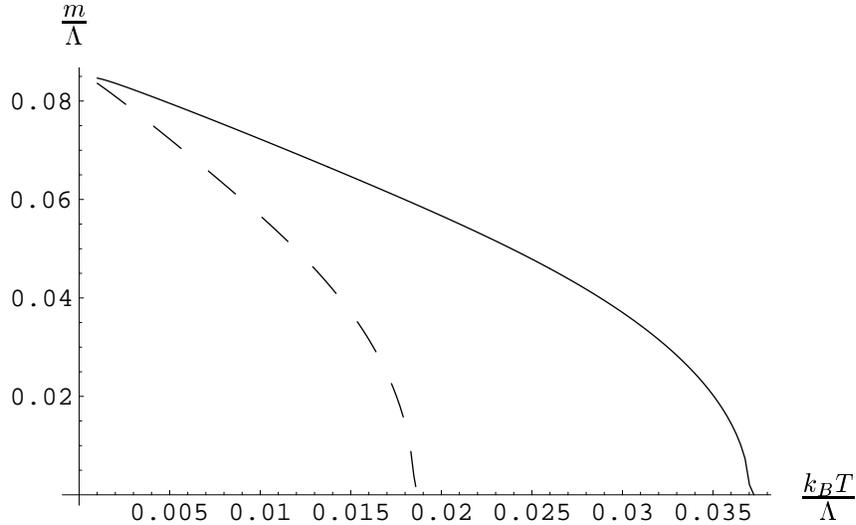}
\caption{Temperature dependence of the dynamical mass generated
with (full line) and without (dashed line) the use of the Popov-Fedotov 
procedure.}
\label{Chapter5Fig3}
\end{figure}

Appelquist et al. 
\cite{Appelquist1,Appelquist2} have shown that at zero temperature the 
originally massless fermion can acquire a dynamically generated mass when the 
number $N$ of fermion flavors is lower than the critical value 
$N_c = 32/\pi^2$. Later Maris \cite{Maris} confirmed the existence of a 
critical value $N_c \simeq 3.3$ below which a dynamical mass can be 
generated. Since we consider only spin-$1/2$ systems, $N=2$ and hence $N<N_c$.

At finite temperature Dorey and Mavromatos \cite{DoreyMavromatos} and Lee
\cite{Lee-98} have shown that the dynamically generated mass vanishes at a 
temperature $T$ larger than the critical one $T_c$.

We now show how the Popov and Fedotov procedure doubles the ``chiral'' 
restoring transition temperature $T_c$ of the dynamical mass generation. 
The Schwinger-Dyson equation for the spinon propagator at finite temperature 
reads
\index{Chiral symmetry}
\index{Schwinger-Dyson equation}

\begin{eqnarray}
G^{-1}(k) = {G^{(0)}}^{-1}(k) 
- \frac{g}{\beta} \underset{\widetilde{\omega}_{F,n}}
{\sum} \int \frac{d^2 \vec{P}}{(2 \pi)^2} \gamma_\mu G(p) \Delta_{\mu \nu}
(k-p) \Gamma_\nu
\label{SchwingerDyson}
\end{eqnarray}

\noindent
where $p=(p_0=\widetilde{\omega}_{F,n},\vec{P})$,
$G$ is the spinon propagator, $\Gamma_\nu$ the spinon-''photon''
vertex which will be approximated here by its bare value $g \gamma_\nu$ and
$\Delta_{\mu \nu}$ the dressed photon propagator \eqref{PhotonPropagator}.
The second term in \eqref{SchwingerDyson} is the fermion self-energy $\Sigma$,
($G^{-1} = {G^(0)}^{-1} - \Sigma$).
Performing the trace over the $\gamma$ matrices in equation 
\eqref{SchwingerDyson} leads to a self-consistent equation for the self-energy

\begin{eqnarray}
\Sigma(k) = \frac{g^2}{\beta} \underset{\widetilde{\omega}_{F,n}}{\sum} 
\int \frac{d^2 \vec{P}}{(2 \pi)^2}
\Delta_{\mu \mu}(k-p) \frac{\Sigma(p)}{p^2 + \Sigma(p)^2}
\label{selfconsitent}
\end{eqnarray}

\noindent
$\Sigma \left( k \right)$ corresponds to a mass term which can be estimated
at low energy and momentum limit $m(\beta) = \Sigma(k) \simeq \Sigma(0)$.
In this regime equation \eqref{selfconsitent} simplifies to

\begin{eqnarray}
1 = \frac{g^2}{\beta} \underset{\widetilde{\omega}_{F,n}}{\sum} 
\int \frac{d^2 \vec{P}}{(2 \pi)^2}
\Delta_{\mu \mu}(-p). \frac{1}{p^2 + m(\beta)^2}
\label{Mass1}
\end{eqnarray}

\noindent
Admitting that the main contribution to \eqref{Mass1} comes from the 
longitudinal part $\Delta_{00}(0,-\vec{P})$ of the photon propagator 
\eqref{Mass1} goes over to

\begin{eqnarray}
1 = \frac{g^2}{\beta} \underset{\widetilde{\omega}_{F,n}}{\sum} 
\int \frac{d^2 \vec{P}}{(2 \pi)^2}
\Bigg( \frac{1}{\vec{P}^2 + \widetilde{\Pi}_3(m=0)}. \frac{1}
{\widetilde{\omega}_{F,n}^2 + \vec{P}^2 + m(\beta)^2} \Bigg)
\end{eqnarray}

\noindent
Performing the summation over the fermion Matsubara frequencies 
$\widetilde{\omega}_{F,n}$ the self-consistent equation takes the form

\begin{eqnarray}
1 = \frac{\alpha}{4 \pi N} \int_0^\infty d P
\Bigg(
\frac{P \tanh \beta \sqrt{\vec{P}^2 + m(\beta)^2}}
{\left[\vec{P}^2 + \widetilde{\Pi}_3(m=0) \right] \sqrt{\vec{P}^2 
+ m(\beta)^2}} \Bigg)
\label{Mass2}
\end{eqnarray}

Relation \eqref{Mass2} can be solved numerically with a cutoff $\Lambda$ 
to control the ultraviolet integration limit

\begin{eqnarray}
1 = \frac{\alpha}{4 \pi N} \int_0^\Lambda d P
\Bigg(
\frac{P \tanh \beta \sqrt{\vec{P}^2 + m(\beta)^2}}
{\left[\vec{P}^2 + \widetilde{\Pi}_3(m=0) \right] \sqrt{\vec{P}^2 
+ m(\beta)^2}} \Bigg)
\label{Chapter5eq37}
\end{eqnarray}

\noindent
It has been shown elsewhere \cite{Aitchison-92} and \cite{DoreyMavromatos}
that in the limit of $\Lambda$ going to $\infty$ numerical results are stable
and finite.
Figure \ref{Chapter5Fig3} compares the temperature dependence of the
dynamical mass generated with and without the imaginary chemical potential
introduced by the Popov and Fedotov procedure where 
$\alpha / \Lambda = 10^5$.
By inspection of equation \eqref{Mass2} and the corresponding result obtained
by Dorey and Mavromatos \cite{DoreyMavromatos} and Lee \cite{Lee-98} one sees
that the imaginary chemical potential used which fixes rigorously one spin 
per lattice site of the original Hamiltonian \eqref{Chapter4eq31}
doubles the transition temperature. This result is coherent with the results
obtained elsewhere \cite{Dillen-05} and shown in section 
\ref{Chapter4Section2} where spinons are massless.

Since the mass can be related to the spinon energy gap and $m \left( T = 0 
\right)$ corresponds to the spinon-antispinon condensate 
$<\bar{\psi}_{\vec{k}=\vec{0}\sigma}  \psi_{\vec{k}=\vec{0},\sigma}>$
an amount of energy at least equal to $m(T)$ is necessary in order to break
a pair of spinon-antispinon and liberate a spinon.
Interpreting the spinon as the spin excitation part breaking a Cooper pair 
one can identify $m(T)$ with a superconducting gap and
evaluate the reduced energy gap parameter $r = \frac{2 m(0)}{k_B T_c}$ 
where $m(0)$ is the mass at zero temperature and $T_c$ the transition 
temperature for which the mass becomes zero. 
Dorey and Mavromatos \cite{DoreyMavromatos} obtained $r \simeq 10$ 
Lee \cite{Lee-98} computed the mass by taking into account the frequency
dependence of the photon propagator 
and obtained $r \simeq 6$. We have shown above that the imaginary
chemical potential doubles the transition temperature so that the
parameter $r$ is $ \simeq 4.8$ for $\alpha/\Lambda = \infty$ to be compared
with the result of Dorey and Mavromatos and $r \simeq 3$ to be compared with
Lee's result. Recall that the BCS parameter $r$ is roughly equal
to $3.5$ and the $Y Ba CuO$ parameter $r \simeq 8$ as given by the experiment
\cite{Y-Ba-Cu-O}.
\index{Dynamical mass|)}
\index{High-Tc superconductivity!Gap}
\index{High-Tc superconductivity!Cuprate}

\subsection{Antiferromagnetic N\'eel order parameter}

There is another possible physical interpretation of the dynamical generated 
mass different from the former one \cite{Aitchison-92,DoreyMavromatos,Lee-98}.
One may also consider it as the antiferromagnetic N\'eel order parameter as
will be explained belove \cite{GhaemiSenthil-05,KimLee-99,Morinari-05,NogueiraKleinert-05}.

Consider the mean-field Hamiltonian for which one takes into account both
the diffuson \eqref{Chapter4eq38} and the N\'eel ansatz
\eqref{Chapter4MeanFieldHamiltonian}. This mean-field Hamiltonian reads

\begin{eqnarray}
\mathcal{H}_{MF} &=& 
-\frac{1}{2} \underset{i,j}{\sum} 
\left( J^{-1} \right)_{ij} \left(\vec{\bar{\varphi}}_i - \vec{B}_i \right)
.\left(\vec{\bar{\varphi}}_j - \vec{B}_j \right)
+ \underset{i}{\sum} \vec{\bar{\varphi}}_i.\vec{S}_i
\notag \\
&&
+ \frac{2}{|J|}\underset{<ij>}{\sum} 
\bar{\Delta}_{ij}^{MF}.\Delta_{ij}^{MF}
+ \underset{<ij>}{\sum} \left[
\bar{\Delta}_{ij}^{MF} \mathcal{D}_{ij} + 
\Delta_{ij}^{MF} \mathcal{D}^{\dagger}_{ij} \right]  - \mu N
\label{Chapter5eq52}
\end{eqnarray}

\noindent
Following sections \ref{Chapter5Section1} and \ref{Chapter4Section1} 
and after some transformation steps on \eqref{Chapter5eq52}
one gets the Dirac action \cite{Morinari-05}

\begin{eqnarray}
S_{E} = \int_0^\beta d\tau \int d^2\vec{r} \underset{\sigma}{\sum}
\bar{\psi}_{\vec{r} \sigma} \left[ \gamma^0 \left( \partial_\tau - \mu \right)
+ \widetilde{\Delta} \gamma^k \partial_k 
- i \vec{m}_{\text{N\'eel}} \vec{\sigma}
\right] \psi_{\vec{r} \sigma}
\label{Chapter5eq58}
\end{eqnarray}

\noindent
The ``mass'' term
$\vec{m}_{\text{N\'eel}} \propto \vec{\bar{\varphi}}$ is similar to the 
dynamical generated mass $m(\beta)$ introduced above but not equal to 
$m(\beta)$. The N\'eel ansatz introduces the Pauli matrix $\vec{\sigma}$ 
affected to the mass term $\vec{m}_{\text{N\'eel}}.\vec{\sigma}$ of the
spinons. 
In the context of the insulator compound described by the Hamiltonian 
\eqref{Chapter4eq31} the ``chiral'' symmetry breaking term 
$\vec{m}_{\text{N\'eel}}.\vec{\sigma}$ corresponds to the 
development of N\'eel order \cite{KimLee-99,NogueiraKleinert-05} as can
be understood by simple comparison of \eqref{SpinonGaugeAction} with
\eqref{Chapter5eq58}. One sees that adding the N\'eel anstaz, pictured
by the term $\vec{\bar{\varphi}}_i.\vec{S}_i$ in \eqref{Chapter5eq52},
changes \eqref{SpinonGaugeAction} into \eqref{Chapter5eq58} thus 
$\vec{m}_{\text{N\'eel}}$ is clearly related to the N\'eel order parameter.

As a consequence of this derivation
the dynamical mass generation introduces naturally the concept of second
order transition from a N\'eel phase to a genuine paramagnetic spin liquid in
two dimensional quantum antiferromagnets contrary to general wisdom,
as explained in \cite{Tesanovic-02,GhaemiSenthil-05}. This general wisdom
claimed that a phase transition can only take place on one hand between
collinear magnets and \textbf{V}alence \textbf{B}ond \textbf{S}olid 
(\textbf{VBS}) paramagnets and on the other hand between non-collinear magnets 
and spin liquids.
\index{Collinear magnet}
\index{Valence Bond State (VBS)}

Finally one can interpret the dynamically generated mass $m(\beta)$ obtained
in section \ref{Chapter5Section3} as the emergence of something like a N\'eel
phase from a spinon gas describing the spin liquid state, more 
precisely from a $\pi$-flux state. One may push further and say that if this 
interpretation is correct then one confirms the results obtained in section 
\ref{Chapter4Section1} and in \cite{Azakov-01,Dillen-05}
concerning the doubling of the N\'eel order transition temperature $T_c$.
Indeed the dynamical generated mass sees its transition temperature doubled by
the PFP compared to the Lagrange multiplier method like for the N\'eel order
parameter as shown in chapter \ref{Chapter4}.

\section{The PFP and the confinement problem : outlook}

\index{Spinon!Confinement/Deconfinement|(}
\index{Compact gauge theory|(}
\index{Non-compact gauge theory|(}
\index{Instanton|(}
The $QED_3$ theory described above deals with \emph{noncompact} ``Maxwell''
theory, in other words the integration over the gauge field $a_\mu$ goes
within the limits $\left]-\infty,\infty\right[$ and the Maxwell action
can be described on a lattice by 

\begin{eqnarray}
S_{noncompact} = \frac{1}{4} \underset{x,\mu \nu}{\sum} f^2_{x,\mu \nu}
\end{eqnarray}

\noindent
In a rigorous derivation on the lattice the ``Maxwell'' action is in the
\emph{compact} form \cite{Polyakov-87}

\begin{eqnarray}
S_{compact} = \frac{1}{2} \underset{x,\mu \nu}{\sum}
\left(1 - \cos f_{x,\mu \nu} \right)
\end{eqnarray}

\noindent
with $f_{x,\mu \nu} = a_{x,\mu} + a_{x+\mu,\nu} - a_{x+\nu,\mu} - a_{x,\nu}$
and $-\pi \leq a_{x,\mu} \leq \pi$. Using this compact version of the Maxwell
theory the partition function reads

\begin{eqnarray}
\mathcal{Z} = \underset{\left\{n_{x,\mu \nu}\right\}}{\sum}
\int_{-\pi}^{\pi} \underset{x,\mu}{\prod} da_{x,\mu} 
e^{- \frac{1}{4} \underset{x,\mu \nu}{\sum} 
\left( f_{x,\mu \nu} - 2\pi n_{x,\mu \nu} \right)^2 }
\end{eqnarray}

\noindent
and takes the periodicity of the action into account. $n_{x,\mu \nu}$
is an integer coming from the periodicity of the cosine and introduces
new entities called instantons.
Polyakov showed \cite{Polyakov-87} that the compactness causes important 
changes in physical properties. Indeed a pure compact Maxwell theory confines
test charged particles in (2+1) dimensions due to the formation of an gas of
instantons. The electrostatic potential between two test particles behaves like
$V(R) \propto R$ at zero temperature and (2+1) dimensions while the 
electrostatic potential is deconfining in the non-compact form and behaves like
$V(R) \propto \ln R$ at zero temperature and in (2+1) dimensions.

However when matter fields are taken into account the system may show a 
deconfined phase even in the compact formulation of the theory 
\cite{NogueiraKleinert-05}. One may ask now whether spinons deconfine or not ?
A large amount of work has been devoted these last decades to the search of an 
answer \cite{CaseSeradjehHerbut-04,HandsKogutLucini,Herbut-02,KimLee-99,KleinertNogueiraSudbo-02,KleinertNogueiraSudbo-03,NogueiraKleinert-05}.
The possible existence of a confinement-deconfinement transition could leads to
an explanation for spin-charge separation in strongly correlated electronic
systems \cite{Nayak-99}.
In the present context one may ask how the Popov and Fedotov procedure affects 
a \emph{compact} $QED_3$ theory of spinons. As seen in chapters \ref{Chapter3}
-\ref{Chapter5} previous work has shown that the PFP probably modifies 
the physical behaviour in a quantitative way but does not produce a deep 
qualitative modification. These questions are potentially open for further 
work.

Previous works showed that deconfined spinons (minimally coupled with a
compact $U(1)$ gauge field) can appear in the region of 
``\emph{a second order quantum phase transition from a collinear 
N\'eel phase to a paramagnetic spin liquid in two dimensional square lattices
with quantum antiferromagnets and short ranged interactions}''
\cite{GhaemiSenthil-05,SenthilFisher-04}. In this case the Hamiltonian is
the one we used in our previous description with ``diffusons'' to which
we add the term $\frac{1}{g} \underset{<i,j> \in N.N.}{\sum} 
\vec{S}_i.\vec{S}_j$ where $g$ is a parameter which controls the importance of
the antiferromagnetic ordering. When $g \rightarrow \infty$ we reach the 
algebraic spin liquid phase and when $g \rightarrow 0$ one gets the N\'eel 
phase. There is a critical coupling parameter $g_c$ under which the spinons
are gapped, condense and are confined (through a Higgs mechanism).
This corresponds to the N\'eel phase. For $g$ larger than $g_c$ spinons are  
again confined but this time due to proliferation of instantons 
\cite{SenthilBalents-04}. However these results are obtained for zero
temperature and experiments have not yet proven the existence of spinons.
\index{Algebraic spin liquid}
\index{Paramagnetic spin liquid}

\newpage

\section{Summary and conclusions}

We mapped a Heisenberg $2d$ Hamiltonian describing an antiferromagnetic quantum
spin system onto a  $QED_{(2+1)}$ Lagrangian coupling a Dirac spinon field 
with a $U(1)$ gauge field. In this framework we showed that the implementation
of the constraint which fixes rigorously the site occupation in a quantum spin 
system described by a $2d$ Heisenberg model leads to a substantial 
quantitative modification of the transition temperature at which the 
dynamically generated mass vanishes in the $QED_{(2+1)}$ description. 
It modifies consequently the effective static potential which acts between 
two test particles of opposite charges \cite{DRcondmat-06}.

The imaginary chemical potential \cite{Popov-88} reduces the screening of this 
static potential between test fermions when compared to the potential obtained 
from standard $QED_{(2+1)}$ calculations by Dorey and Mavromatos 
\cite{DoreyMavromatos} who implicitly used a Lagrange multiplier procedure 
in order to fix the number of particles per lattice site 
\cite{ArovasAuerbach-88,Manousakis-91}
since $\lambda=0$ at the mean-field level.

We showed that the transition temperature to ``chiral'' symmetry 
restoration corresponding to the vanishing of the spinon mass $m(\beta)$ is 
doubled by the introduction of the Popov-Fedotov imaginary chemical 
potential. The trend is consistent with earlier results concerning the value 
of $T_c$ \cite{Dillen-05}. It reduces sizably the parameter 
$r = \frac{2 m(0)}{k_B T_c}$ determined by Dorey and Mavromatos 
\cite{DoreyMavromatos} and Lee \cite{Lee-98}. 

Marston \cite{Marston} showed that in order to remove
``forbidden'' $U(1)$ gauge configuration of the antiferromagnet Heisenberg
model a Chern-Simons term should be naturally included in the $QED_3$ action 
and fix the total flux through a plaquette. When the magnetic flux through
a plaquette is fixed the system becomes 
$2\pi$-invariant in the gauge field $a_\mu$ and instantons appear in the
system. This is the case when the present non-compact formulation of $QED_3$
is replaced by its correct compact version \cite{Polyakov-77,Polyakov-87}.

The implementation of a Chern-Simons term \cite{Dunne} in a 
non-compact formulation of the spinon system constrained by a rigorous site 
occupation has been submitted to publication \cite{DRcondmatChernSimons-06}.
\index{Chern-Simons action}
\index{Spinon!Confinement/Deconfinement|)}
\index{Compact gauge theory|)}
\index{Non-compact gauge theory|)}
\index{Instanton|)}



\chapter{Conclusions and outlook \label{Chapter6}}

There is a general consensus on the phase diagram of high-$T_c$
superconductivity. The simplest model which seems to take account of the 
strong correlation physics of the high-$T_c$ superconductors is the Hubbard 
model and its strong coupling limit the $t - J$ model. 
The schematic phase diagram of high-$T_c$ superconductors shows an
insulating antiferromagnetic phase in the underdoped regime and is well 
described by Heisenberg-like models. A superconducting phase is present
for low temperature and for a finite range of doping with holes (or electrons)
as shown in figure \ref{IntroFig1}.
Anderson presented the viewpoint of the existence of a spin 
liquid state for a possible key to understand the physics of highly 
correlated superconducting phase \cite{Anderson-87}. The concept of 
\textbf{R}esonating \textbf{V}alence \textbf{B}ond (\textbf{RVB}) states
was further developed \cite{Kivelson-87,MudryFradkin-94} and gave rise to 
the notion of spin-charge separation. The inclusion of fluctuations around
the corresponding ``diffuson'' mean-field led to a $U(1)$ gauge theory 
\cite{DRcondmat-06,GhaemiSenthil-05,LeeNagaosaWen-04,Morinari-05} which
can be treated at the one-loop level.

One of the approaches to high-$T_c$ supercontuctivity
phenomena consists in a slight doping of the superconductor materials
starting from the undoped insulating antiferromagnet phase to the underdoped
regime. The underdoped superconducting should retain part of the correlation 
features of the insulating phase. For this reason it is of high interest to 
study two dimensional antiferromagnetic strongly correlated spin systems.

\vspace*{2cm}

In chapter \ref{Chapter3} we presented and discussed applications of the
mean-field and loop expansion to the determination of physical
properties of antiferromagnetic Heisenberg-type systems (supraconductors
in the antiferromagnet insulating phase) in spatial dimension D 
\cite{DRepjb-05}.

We worked out the expression of physical 
observables (magnetization and susceptibility) starting from a specific
mean-field ansatz and including contributions up to first
order in a loop expansion in order to investigate the effects of fluctuation
corrections to mean-field contributions at the gaussian approximation.
The mean-field was chosen as a N\'eel state which is an \emph{a priori}
reasonable choice for spin systems described in terms of unfrustrated bipartite
Heisenberg model. The results were compared to those obtained in the 
framework of spin wave theory.

The number of particles per spin lattice site was fixed by means of the 
rigorous constraint imposed by the imaginary chemical potential introduced by 
Popov and Fedotov \cite{Popov-88}.

At low temperature the magnetization and the magnetic susceptibility are close 
to the spin wave value as expected, also in agreement with former work 
\cite{Azakov-01}.

At higher temperature the fluctuation contributions of quantum and thermal 
nature grow to a singularity in the neighbourhood of the critical temperature.
In addition the N\'eel order breaks $SU(2)$ symmetry of the Heisenberg 
Hamiltonian inducing low momentum fluctuations near $T_c$ which is not the 
case in the XXZ-model.

The influence of fluctuations decreases with the dimension $D$ 
of the system due to the expected fact that the mean-field contribution 
increases relatively to the loop contribution. 

In dimension $D=2$ the magnetization verifies the Mermin and Wagner theorem 
\cite{MerminWagner-66} for $T \neq 0$, the fluctuations are larger than the 
mean-field contribution for any temperature. In a more realistic description
another mean-field ansatz may be necessary in order to describe the correct 
physics. Ghaemi and Senthil \cite{GhaemiSenthil-05} introduced a specific 
model in which a second order phase transition
from a N\'eel mean-field to an ASL (algebraic spin liquid) may be at work
depending on the strength of interaction parameter which enters the Hamiltonian
of the system. These considerations enforce the belief that another mean-field 
solution like ASL may be a better starting point than a N\'eel state when the 
temperature $T$ increases.

\vspace*{0.5cm}

In chapter \ref{Chapter4} we worked out a rigorous versus average treatment 
of the occupation constraint on spin systems governed by 
Heisenberg-type Hamiltonians at the mean-field level, for different 
types of spin mean-field ansatz.

We showed that a strict constraint on the site occupation of 
a lattice quantum spin system described by Heisenberg-type models shows a 
sizable quantitative different localization of the critical temperature 
when compared with the outcome of an average occupation constraint.
With an exact site-occupation by spin the transition temperature of
antiferromagnetic N\'eel and spin liquid states order parameters are twice 
as large as the critical temperature one gets from an average Lagrange 
multiplier procedure.

The exact occupation procedure cannot be applied to a so called cooperon 
state mean-field Hamiltonian. Cooperons are $BCS$-like pairs of particles.
In this scheme the number of particles is not conserved, hence it is 
incompatible with a strict site occupation constraint.

\vspace{0.5cm}

In a further step we mapped a Heisenberg $2d$ Hamiltonian describing 
an antiferromagnetic quantum spin system onto a  $QED_{(2+1)}$ Lagrangian 
which couples a Dirac spinon field to a $U(1)$ gauge field. 
We considered the $\pi$-flux state approach introduced by Affleck and 
Marston \cite{AffleckMarston-88,MarstonAffleck-89}. 
We implemented the strict site-occupation by means of the constraint imposed 
through Popov and Fedotov's imaginary chemical potential \cite{Popov-88} 
for $SU(2)$ and used the modified Matsubara frequencies.
In this framework we showed that the implementation of the PFP constraint 
which fixes rigorously the site occupation in a quantum spin 
system described by a $2d$ Heisenberg model leads to a substantial 
quantitative modification of the transition temperature at which the 
dynamically generated mass vanishes in the $QED_{(2+1)}$ description. 
It modifies consequently the effective static potential which acts between 
two test spinons of opposite coupling charge $g$.

The imaginary chemical potential \cite{Popov-88} reduces the screening of the
static potential between test fermions when compared to the potential obtained 
from standard $QED_{(2+1)}$ calculations by Dorey and Mavromatos 
\cite{DoreyMavromatos}.

We showed that the transition temperature corresponding to the vanishing of 
the spinon mass $m(\beta)$ is doubled by the introduction of the PFP. The 
trend is consistent with earlier results concerning the value 
of $T_c$ \cite{Dillen-05}. 
It reduces sizably the ratio $r = \frac{2 m(0)}{k_B T_c}$ between the 
superconducting gap $m(0)$ and the transition temperature $T_c$
determined by Dorey and Mavromatos \cite{DoreyMavromatos} and Lee 
\cite{Lee-98}.

\vspace*{0.5cm}

We conclude from this work that the Popov and Fedotov procedure which
treats exactly the constraint of strict site-occupation does not modify 
qualitatively but quantitatively the physical results obtained by means of 
the average Lagrange multiplier method.

\vspace*{2cm}

The $QED_3$ theory described in chapter \ref{Chapter5} deals with a
non-compact version of the Abelian gauge field action. 
In order to remove ``forbidden'' $U(1)$ gauge configuration of the 
antiferromagnet Heisenberg model a Chern-Simons term should be naturally 
included in the $QED_3$ action \cite{Marston,DRcondmatChernSimons-06}.

The compact Maxwell theory confines test charged particles in (2+1) dimensions 
due to the formation of an instanton gas \cite{Polyakov-87}. When matter fields
are taken into account the system may show a confinement/deconfinement 
transition \cite{NogueiraKleinert-05}. 
The possible existence of a confinement-deconfinement transition could give
an explanation for spin-charge separation in strongly correlated electronic
systems \cite{Nayak-99}. The fundamental question concerning confinement
invalidating a loop expansion is up to now unsettled 
\cite{Herbut-02,NogueiraKleinert-05}.

It would be interesting to implement a Chern-Simons term 
\cite{Dunne,DRcondmatChernSimons-06} in a \emph{compact} $U(1)$ gauge 
formulation of the spinon system constrained by a strict site-occupation 
imposed by the Popov and Fedotov procedure.


\myappendix


\chapter{Grassmann algebra and coherent states \label{AppendixGrassmann}}

Here we review the \emph{Grassmann} algebra and some properties
of coherent states. 
\index{Path integrals!Grassmann algebra|(}
\index{Path integrals!Grassmann variables|(}

\begin{itemize}

\item The anticommutation relation of the fermion creation and annihilation
operators are

\begin{eqnarray}
\left\{f_\alpha,f_\beta^\dagger \right\} = \delta_{\alpha \beta} \\
\left\{f_\alpha,f_\beta\right\} = 0 \\
\left\{f_\alpha^\dagger,f_\beta^\dagger\right\} = 0
\end{eqnarray}

\item The anticommutation relations of the \emph{Grassmann} variables with
themselves and the creation and annihilation fermionic operators are

\begin{eqnarray}
\{ \xi_\alpha, \xi_\lambda \}= 
\xi_\alpha \xi_\lambda + \xi_\lambda \xi_\alpha &=& 0
\\
\{ \xi_\alpha^{*}, \xi_\lambda^{*} \} &=& 0
 \\
\{ \xi_\alpha, \xi_\lambda^{*} \} &=& 0
 \\
\left( \lambda \xi_\alpha \right)^{*} &=&
\lambda^{*} \xi_\alpha^{*}, \text{ $\lambda$ is a complex number}
 \\
\{ \xi_\alpha , f_\beta \} = \{ \xi_\alpha, f_\beta^\dagger \}
&=& \{ \xi_\alpha^{*} , f_\beta \} = \{ \xi_\alpha^{*} , f_\beta^\dagger \} = 0
\end{eqnarray}

\noindent
where $\xi^{*}$ is the conjugate of $\xi_\alpha$. Consequently 
the product of two identical \emph{Grassmann} variables is zero due 
to the anticommutation property

\begin{eqnarray}
\xi_\alpha^2 = \left( \xi_\alpha^{*} \right)^2 &=& 0
\end{eqnarray}

\item Conjugation rules :

\begin{eqnarray}
\left( \xi_\alpha \right)^{*} &=& \xi_\alpha^{*}
 \\
\left( \xi_\alpha^{*} \right)^{*} &=& \xi_\alpha
\end{eqnarray}

\noindent
Hermitic conjugation :

\begin{eqnarray}
\left( \xi f \right)^\dagger &=& f^\dagger \xi^{*}
\end{eqnarray}

\item Derivation :

\begin{eqnarray}
\frac{\partial}{\partial \xi} \left( \xi^{*} \xi \right)
&=&
\frac{\partial}{\partial \xi} \left( - \xi \xi^{*} \right)
\notag \\
&=&
-\xi^{*}
\end{eqnarray}

\noindent
Define the operator $\mathcal{O}$

\begin{eqnarray}
\mathcal{O} \left( \xi, \xi^{*} \right) =
\alpha + \beta \xi + \gamma \xi^{*} + \lambda \xi^{*} \xi
\label{OperatorO}
\end{eqnarray}

\noindent
where $\alpha, \beta, \gamma, \lambda$ are complex number. With this definition

\begin{eqnarray}
\frac{\partial}{\partial \xi} \mathcal{O}\left( \xi, \xi^{*} \right)
&=&
\beta - \lambda \xi^{*}
 \\
\frac{\partial}{\partial \xi^{*}} \mathcal{O}\left( \xi, \xi^{*} \right)
&=& \gamma + \lambda \xi
 \\
\frac{\partial}{\partial \xi} \frac{\partial}{\partial \xi^{*}}
\mathcal{O} \left( \xi, \xi^{*} \right) 
&=&
\lambda 
\notag \\
&=& - \frac{\partial}{\partial \xi^{*}} \frac{\partial}{\partial \xi}
\mathcal{O} \left( \xi, \xi^{*} \right)
\end{eqnarray}

\item Integration :

\begin{eqnarray}
\int d\xi \, 1 &=& \int d\xi^{*} \, 1 = 0 \\
\int d\xi \, \xi &=& \int d\xi^{*} \, \xi^{*} = 1
\end{eqnarray}

\noindent
Using the definition \eqref{OperatorO} of the operator $\mathcal{O}$

\begin{eqnarray}
\int d\xi \mathcal{O} \left( \xi, \xi^{*} \right)
&=&
\beta - \lambda \xi^{*}
 \\
\int d\xi^{*} \mathcal{O} \left( \xi, \xi^{*} \right)
&=&
\gamma + \lambda
 \\
\int d\xi^{*} d\xi \mathcal{O} \left( \xi, \xi^{*} \right)
&=&
- \lambda
\notag \\
&=& - \int d\xi d\xi^{*} \mathcal{O} \left( \xi, \xi^{*} \right)
\end{eqnarray}

\noindent
Notice that the integration operator is equivalent to an ordinary
derivation operator.

\item Gaussian integrals :

\noindent
For commuting variables

\begin{eqnarray}
\int \prod_\alpha \frac{d\phi_\alpha^{*} d\phi_\alpha}{2 \pi i}
e^{- \underset{\alpha,\lambda}{\sum} 
\phi^{*}_\alpha M_{\alpha,\lambda} \phi_\lambda 
+ \underset{\alpha}{\sum} ( z_\alpha^{*} \phi_\alpha + z_i \phi_\alpha^{*} )}
=
\left[ \det M \right]^{-\frac{1}{2}} 
e^{\underset{\alpha,\lambda}{\sum} 
z_\alpha^{*} M_{\alpha,\lambda}^{-1} z_\lambda}
\end{eqnarray}

\noindent
For Grassmann variables

\begin{eqnarray}
\int \prod_\alpha d\xi_\alpha^{*} d\xi_\alpha
e^{-\underset{\alpha,\lambda}{\sum} 
\xi^{*}_\alpha M_{\alpha,\lambda} \xi_\lambda 
+ \underset{\alpha}{\sum} 
( \eta_\alpha^{*} \xi_\alpha + \eta_\alpha \xi_\alpha^{*}) }
=
\left[ \det M \right] 
e^{\underset{\alpha,\lambda}{\sum} 
\eta_\alpha^{*} M_{\alpha,\lambda}^{-1} \eta_\lambda}
\label{AppendixGaussianGrassmannIntegral}
\end{eqnarray}

\item Coherent states : 
\index{Path integrals!Coherent states}

\noindent
The fermionic coherent states $|\xi>$ are defined as

\begin{eqnarray}
|\xi> &=& e^{- \sum_\alpha \xi f_\alpha^\dagger} |0> \\
<\xi| &=&  <0| e^{\sum_\alpha \xi^{*} f_\alpha }
\end{eqnarray}

\noindent
Application of the creation and annihilation operator on coherent states
leads to the following expressions

\begin{eqnarray}
f_\alpha |\xi> &=& \xi_\alpha |\xi>
 \\
f_\alpha^\dagger |\xi> &=& 
f_\alpha^\dagger \prod_\lambda 
\left( 1 - \xi_\lambda f_\lambda^\dagger \right) |0>
\notag \\
&=&
f_\alpha^\dagger \left( 1 - \xi_\alpha f_\alpha^\dagger \right) 
\prod_{\lambda \neq \alpha} 
\left( 1 - \xi_\lambda f_\lambda^\dagger \right) |0>
\notag \\
&=&
\left( f_\alpha^\dagger \right)
\prod_{\lambda \neq \alpha} 
\left( 1 - \xi_\lambda f_\lambda^\dagger \right) |0>
\notag \\
&=&
-\frac{\partial}{\partial \xi_\alpha} 
\left( 1 - \xi_\alpha f_\alpha^\dagger \right)
\prod_{\lambda \neq \alpha} 
\left( 1 - \xi_\lambda f_\lambda^\dagger \right) |0>
\notag \\
&=& - \frac{\partial}{\partial \xi_\alpha} |\xi>
 \\
<\xi| f_\alpha &=& \frac{\partial}{\partial \xi^{*}} <\xi| 
 \\
<\xi| f_\alpha^\dagger &=& <\xi| \xi_\alpha^{*}
\end{eqnarray}

\noindent
The \emph{overlap} of two coherent states is

\begin{eqnarray}
<\xi|\xi^{'}> 
&=&
<0| \prod_\alpha ( 1 + \xi^{*}_\alpha f_\alpha  ) 
( 1 - \xi^{'}_\alpha f_\alpha^\dagger ) |0>
 \\
&=& e^{\sum_\alpha \xi^{*} \xi^{'}}
\end{eqnarray}

\noindent
and the elements of fermionic operators between two different coherent
states $|\xi>$ and $|\xi^{'}>$ read

\begin{eqnarray}
<\xi| 
\mathcal{O} \left( \{ f_\alpha \}, \{ f_\alpha^\dagger \} \right) |\xi^{'}>
= e^{\sum_\alpha \xi^{*} \xi^{'}}
\mathcal{O} \left( \{ \xi_\alpha \}, \{ \xi_\alpha^\dagger \} \right)
\end{eqnarray}

\end{itemize}

\index{Path integrals!Grassmann algebra|)}
\index{Path integrals!Grassmann variables|)}


\index{Spin Brillouin Zone|(}
\chapter{Spin Brillouin Zone \label{AppendixSpinBrillouinZone}}

\section{Two dimensional bipartite lattices}

Here we construct the Spin Brillouin Zone (SPZ) of a two dimensional lattice.
We define the components of the wave vector $\vec{k}$ on the orthonormal
reciprocal sublattice basis $\left( \vec{\tilde{e}}_1
, \vec{\tilde{e}}_2 \right)$ which is given by
$\vec{\tilde{e}}_i.\vec{e}_j  = \delta_{ij}$ where the direct basis is
defined as
\index{Spin Brillouin Zone!Direct basis!Two dimensional}
\index{Antiferromagnetic Heisenberg model!Bipartite lattice|(}

\begin{eqnarray}
\vec{e}_1 &=& \frac{1}{\sqrt{2}} \left( \vec{e}_x + \vec{e}_y \right) \\
\vec{e}_2 &=& \frac{1}{\sqrt{2}} \left( \vec{e}_x - \vec{e}_y \right)
\end{eqnarray}

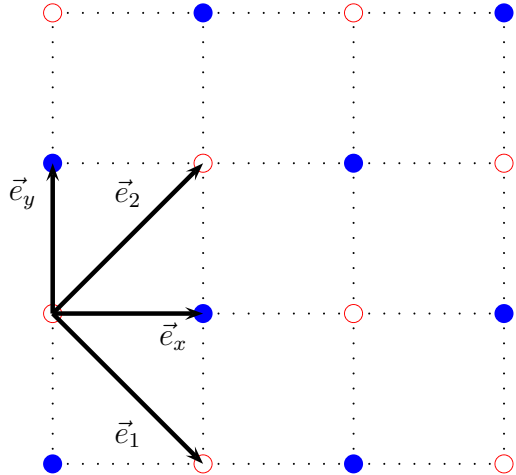
\begin{figure}[h]
\center
\begin{pspicture}(6,6)
\psset{unit=2}
\psgrid[griddots=10,gridlabels=0,subgriddiv=1](3,3)
\multips(0,0)(0,2){2}{
\multips(0,0)(2,0){2}{
\psdots[linecolor=blue,dotscale=2,dotstyle=*](0,0)
\psdots[linecolor=blue,dotscale=2,dotstyle=*](1,1)
\psdots[linecolor=red,dotscale=2,dotstyle=o](1,0)
\psdots[linecolor=red,dotscale=2,dotstyle=o](0,1)}}
\psline[linewidth=0.03]{->}(0,1)(1,1) \rput(0.8,0.85){\text{$\vec{e}_x$}}
\psline[linewidth=0.03]{->}(0,1)(0,2) \rput(-0.2,1.8){\text{$\vec{e}_y$}}
\psline[linewidth=0.03]{->}(0,1)(1,0) \rput(0.5,0.2){\text{$\vec{e}_1$}}
\psline[linewidth=0.03]{->}(0,1)(1,2) \rput(0.5,1.8){\text{$\vec{e}_2$}}
\end{pspicture}
\caption{Two dimensional bipartite lattice with the lattice basis 
$\left(\vec{e}_x, \vec{e}_y \right)$ and the spin sublattice basis 
$\left(\vec{e}_1, \vec{e}_2 \right)$ in direct space.}
\label{AppendixFig1}
\end{figure}

\noindent
Since the direct basis is already orthonormal the reciprocal basis 
is identically equal to the first one $\left(\vec{\tilde{e}}_1 = \vec{e}_1,
\vec{\tilde{e}}_2 = \vec{e}_2 \right)$.
\index{Spin Brillouin Zone!Reciprocal basis!Two dimensional}
Figure \ref{AppendixFig1} shows a two dimensional bipartite lattice where
blue points refers to one type $A$ and 
red points to an other type $B$ of sublattices. The crystal basis 
$\left( \vec{e}_x, \vec{e}_y \right)$ as well as the direct basis
$\left( \vec{e}_1,\vec{e}_2 \right)$ are shown.

\noindent
If one sets the lattice parameter $a$ to $1$ the wave vector $\vec{k}$ is 
defined by

\begin{eqnarray}
\vec{k} = k_1 \vec{\tilde{e}}_1 + k_2 \vec{\tilde{e}}_2
 \\
k_1 = 2 \pi \frac{l_1}{\mathcal{N}_{A(B),1}}
 \\
k_2 = 2 \pi \frac{l_2}{\mathcal{N}_{A(B),2}}
\end{eqnarray}

\noindent
where $\mathcal{N}_{A(B),1(2)}$
is the number of sublattice type-$A(B)$ sites in the direction $1(2)$
with $l_1 \in \left[- \frac{\mathcal{N}_{A(B),1} }{2}
,\frac{\mathcal{N}_{A(B),1} }{2} \right] \in \Z$ 
and $l_2 \in \left[- \frac{\mathcal{N}_{A(B),2} }{2}
,\frac{\mathcal{N}_{A(B),2} }{2} \right] \in \Z$
forming the Spin Brillouin Zone depicted in figure \ref{Chapter3Fig2}.
The $\gamma_{\vec{k}}$ function defined in section \ref{SubsectionSWM} becomes

\begin{eqnarray}
\gamma_{\vec{k}} &=& \frac{1}{z} \sum_{\vec{\eta}} e^{i \vec{k}.\vec{\eta}}
\notag \\
&=& \frac{1}{2} \sum_{\vec{\eta}} \cos \vec{k}.\vec{\eta}
\notag \\
&=& \cos \left( \frac{k_1 + k_2}{\sqrt{2}} \right)  
+ \cos \left( \frac{k_1 - k_2}{\sqrt{2}} \right)
\end{eqnarray}

\noindent
where $\vec{k}$ belongs to the Spin Brillouin Zone and $D=2$. The total number
of lattice sites is $\mathcal{N} = \mathcal{N}_{A,1}.\mathcal{N}_{A,2}+
\mathcal{N}_{B,1}.\mathcal{N}_{B,2}$.

\section{Three dimensional bipartite lattices}

In three dimensions the direct spin sublattice is introduced for the bipartite 
lattice. For the N\'eel state a face-centered cubic lattice is used.
The lattice basis $\left(\vec{e}_1, \vec{e}_2, \vec{e}_3 \right)$ reads
\index{Spin Brillouin Zone!Direct basis!Three dimensional}

\begin{eqnarray}
\vec{e}_1 &=& \frac{1}{\sqrt{2}} \left(\vec{e}_x + \vec{e}_y \right) \\
\vec{e}_2 &=& \frac{1}{\sqrt{2}} \left(\vec{e}_y + \vec{e}_z \right) \\
\vec{e}_3 &=& \frac{1}{\sqrt{2}} \left(\vec{e}_x + \vec{e}_z \right)
\end{eqnarray}

\noindent
The reciprocal basis $\{ \vec{\tilde{e}}_i \}$ of the lattice is defined by 
$\vec{\tilde{e}}_i.\vec{e}_j  = \delta_{ij}$ and reads
\index{Spin Brillouin Zone!Reciprocal basis!Three dimensional}

\begin{eqnarray}
\vec{\tilde{e}}_1 &=& \frac{\sqrt{2}}{2} \left( \vec{e}_x + \vec{e}_y 
- \vec{e}_z \right) \\
\vec{\tilde{e}}_2 &=& \frac{\sqrt{2}}{2} \left(- \vec{e}_x + \vec{e}_y 
+ \vec{e}_z \right) \\
\vec{\tilde{e}}_3 &=& \frac{\sqrt{2}}{2} \left( \vec{e}_x - \vec{e}_y 
+ \vec{e}_z \right)
\end{eqnarray}

\noindent
These vectors are those of a body-centered cubic lattice.
The wave vector $\vec{k}$ reads

\begin{eqnarray}
\vec{k} &=& k_1 \vec{\tilde{e}}_1 + k_2 \vec{\tilde{e}}_2 
+ k_3 \vec{\tilde{e}}_3 \\
k_i &=& 2 \pi \frac{l_i}{\mathcal{N}_{A(B),i}} \\
l_i &\in& \left[ -\frac{\mathcal{N}_{A(B),i}}{2},
\frac{\mathcal{N}_{A(B),i}}{2} \right] \in \Z
\end{eqnarray}

\noindent
where $\mathcal{N}_{A(B),i}$ is the number of site in the direction $\vec{e}_i$
of the sublattice $A(B)$.

The $\gamma_{\vec{k}}$ function for $\vec{k}$ in the three 
dimensional Spin Brillouin Zone reads

\begin{eqnarray}
\gamma_{\vec{k}} &=&
\frac{1}{z} \underset{\vec{\eta}}{\sum} e^{i \vec{k}.\vec{\eta}}
\notag \\
&=& \frac{1}{3} \left[
\cos \left( k_1 - k_2 + k_3 \right) + \cos \left(k_1 + k_2 - k_3 \right)
+ \cos \left(-k_1 + k_2 + k_3 \right) \right]
\end{eqnarray}

\noindent
The total number of lattice is $\mathcal{N}= 
\underset{i}{\prod} \mathcal{N}_{A,i}
+ \underset{i}{\prod} \mathcal{N}_{B,i}$.

\begin{figure}[h]
\center
\begin{pspicture}(0,-1.5)(8,8)
\psset{unit=3,griddots=15,gridlabels=0,subgriddiv=1,
viewpoint=1 -0.8 0.4}

\ThreeDput[normal=0 0 1](0,0,0){
\psgrid(2,2)
\psdots[linecolor=blue,dotscale=4,dotstyle=*](0,0)
\psdots[linecolor=blue,dotscale=4,dotstyle=*](1,1)
\psdots[linecolor=blue,dotscale=4,dotstyle=*](2,0)
\psdots[linecolor=blue,dotscale=4,dotstyle=*](0,2)
\psdots[linecolor=blue,dotscale=4,dotstyle=*](2,2)
\psdots[linecolor=red,dotscale=4,dotstyle=*](1,0)
\psdots[linecolor=red,dotscale=4,dotstyle=*](0,1)
\psdots[linecolor=red,dotscale=4,dotstyle=*](2,1)
\psdots[linecolor=red,dotscale=4,dotstyle=*](1,2)}
\ThreeDput[normal=0 0 1](0,0,2){
\psdots[linecolor=blue,dotscale=4,dotstyle=*](0,0)
\psdots[linecolor=blue,dotscale=4,dotstyle=*](1,1)
\psdots[linecolor=blue,dotscale=4,dotstyle=*](0,2)
\psdots[linecolor=blue,dotscale=4,dotstyle=*](2,2)
\psdots[linecolor=red,dotscale=4,dotstyle=*](0,1)
\psdots[linecolor=red,dotscale=4,dotstyle=*](2,1)
\psdots[linecolor=red,dotscale=4,dotstyle=*](1,2) }
\ThreeDput[normal=0 0 1](0,0,1){
\psdots[linecolor=red,dotscale=4,dotstyle=*](0,0)
\psdots[linecolor=red,dotscale=4,dotstyle=*](1,1)
\psdots[linecolor=red,dotscale=4,dotstyle=*](0,2)
\psdots[linecolor=red,dotscale=4,dotstyle=*](2,2)
\psdots[linecolor=blue,dotscale=4,dotstyle=*](1,0)
\psdots[linecolor=blue,dotscale=4,dotstyle=*](0,1)
\psdots[linecolor=blue,dotscale=4,dotstyle=*](2,1)
\psdots[linecolor=blue,dotscale=4,dotstyle=*](1,2)}
\ThreeDput[normal=1 0 0](0,0,0){\psgrid(2,2)}
\ThreeDput[normal=0 1 0](2,2,0){\psgrid(2,2)}

\ThreeDput[normal=0 0 1](0,0,0){
\psline[linewidth=0.04]{->}(0,0)(1,0)
\psline[linewidth=0.04]{->}(0,0)(0,1) }
\ThreeDput[normal=1 0 0](0,0,0){
\psline[linewidth=0.03]{->}(0,0)(0,1)}
\ThreeDput[normal=1 0 0](1,1,1){
\psline[linewidth=0.03]{->}(0,0)(1,1)}
\ThreeDput[normal=0 1 0](1,1,1){
\psline[linewidth=0.03]{->}(0,0)(-1,1)}
\ThreeDput[normal=0 0 1](1,1,1){
\psline[linewidth=0.05]{->}(0,0)(1,1)}
\rput(0.5,-0.3){\text{$\vec{e}_x$}}
\rput(0.5,0.3){\text{$\vec{e}_y$}}
\rput(-0.2,0.9){\text{$\vec{e}_z$}}
\rput(2.3,0.96){\text{$\vec{e}_1$}}
\rput(1.9,1.9){\text{$\vec{e}_2$}}
\rput(1.9,1.3){\text{$\vec{e}_3$}}

\ThreeDput[normal=0 1 0](2,1,1){\psframe[linestyle=dashed](0,0)(1,1)}
\ThreeDput[normal=0 1 0](2,2,1){\psframe[linestyle=dashed](0,0)(1,1)}
\ThreeDput[normal=1 0 0](1,1,1){\psframe[linestyle=dashed](0,0)(1,1)}
\ThreeDput[normal=1 0 0](2,1,1){\psframe[linestyle=dashed](0,0)(1,1)}

\ThreeDput[normal=0 1 0](1,0,0){\psframe[linestyle=dashed](0,0)(1,1)}
\ThreeDput[normal=0 1 0](1,1,0){\psframe[linestyle=dashed](0,0)(1,1)}
\ThreeDput[normal=1 0 0](0,0,0){\psframe[linestyle=dashed](0,0)(1,1)}
\ThreeDput[normal=1 0 0](1,0,0){\psframe[linestyle=dashed](0,0)(1,1)}
\end{pspicture}
\caption{Three dimensional bipartite 
lattice with direct space basis $\left(\vec{e}_1,\vec{e}_2,\vec{e}_3 \right)$ }
\label{AppendixFig2}
\end{figure}

Figure \ref{AppendixFig2} shows a three dimensional bipartite lattice where
blue points refer to one type of sublattice $A$ and 
red points to the other type $B$. The crystal basis 
$\left( \vec{e}_x, \vec{e}_y, \vec{e}_z \right)$ as well as the direct basis
$\left( \vec{e}_1,\vec{e}_2, \vec{e}_3 \right)$ are indicated.

\index{Antiferromagnetic Heisenberg model!Bipartite lattice|)}
\index{Spin Brillouin Zone|)}


\chapter{Beyond the mean field : one-loop contributions
\label{AppendixOneLoopCorrections}}

In section \ref{SectionEffectiveAction} we defined the mean-field
contribution of the matrix $M$ as

\begin{eqnarray}
{G_0}_{p,q} = 
\begin{bmatrix}
-\frac{1}{\det G_p}
 \left[ i \omega_{F,p} - \frac{1}{2} \bar{\varphi}_i^z
(\omega_{F,p}-\omega_{F,q}) \right] \delta_{p,q}
 &
\frac{1}{\det G_p} \frac{1}{2} \bar{\varphi}_i^{-}
(\omega_{F,p}-\omega_{F,q}) \delta_{p,q}
 \\
\frac{1}{\det G_p} \frac{1}{2} \bar{\varphi}_i^{+}
(\omega_{F,p}-\omega_{F,q})\delta_{p,q}
 &
-\frac{1}{\det G_p} \left[ i \omega_{F,p} + \frac{1}{2} \bar{\varphi}_i^z
(\omega_{F,p}-\omega_{F,q}) \right] \delta_{p,q}
\end{bmatrix}
\notag \\
\end{eqnarray}

\noindent
where $\det G_p = - \left[ \widetilde{\omega}_{F,p}^2 + \left( 
\frac{\vec{\bar{\varphi}}_i (\frac{2\pi}{\beta}(p-q=0))}{2}
\right)^2 \right]$ and the contribution to the fluctuations 
$\vec{\delta \varphi}$ is given by

\begin{eqnarray}
{M_1}_{p,q} = 
\begin{bmatrix}
\frac{1}{2} \delta \varphi_i^z(\omega_{F,p}-\omega_{F,q})
 &
\frac{1}{2} \delta \varphi_i^{-}(\omega_{F,p}-\omega_{F,q})
 \\
\frac{1}{2} \delta \varphi_i^{+}(\omega_{F,p}-\omega_{F,q})
 &
-\frac{1}{2} \delta \varphi_i^z(\omega_{F,p}-\omega_{F,q})
\end{bmatrix}
\end{eqnarray}

\noindent
It comes out that the term $\ln \det \beta M$ can be decomposed
into a series expansion

\begin{eqnarray}
\ln \det \beta M
=-\sum_i \ln 2 \cosh \frac{\beta}{2} \| 
\vec{\bar{\varphi}}_i(\omega_B=0) \| 
+ Tr\{ \sum_{n=1}^{\infty} \frac{1}{n} (G_0 M_1)^n \}
\label{AppendixC3}
\end{eqnarray}

\noindent
where the fluctuation contributions appear in the second term of 
\eqref{AppendixC3}.

\section{First order contributions to the fluctuations and mean-field 
equation}

Here we aim to work out the $n=1$ contribution to
the sum of $\ln \det \beta M$,
$Tr \left[ G_0 M_1 \right]$, and extract the mean-field equation with
respect to the Hubbard-Stratonovich auxiliary mean-field $\bar{\varphi}$.
The first order term in the fluctuations $\delta \varphi$ reads

\begin{eqnarray}
\left[ G_0 M_1 \right]_{p,q} =
\left[
\begin{array}{cc}
A_i(p,q) & B_i(p,q) \\
C_i(p,q) & D_i(p,q)
\end{array}
\right]
\end{eqnarray}

\noindent
where $i$ stands for the position on the spin lattice and $p$ and $q$ refer
to the fermion Matsubara frequencies $\widetilde{\omega}_{F,p} \equiv 
\frac{2\pi}{\beta}(p+1/4)$ and similarly for $\widetilde{\omega}_{F,q}$. 
The matrix elements $A,B,C$ and $D$ are given by

\begin{eqnarray}
A_i(p,q) &=&
\Bigg[
-\frac{1}{2 \det G_p}\left[\widetilde{\omega}_{F,p} - 
\frac{\bar{\varphi}^z_i}{2} (\frac{2\pi}{\beta}(p-q=0) )
\right] \delta \varphi^z_i (\frac{2\pi}{\beta}(p-q)) 
\notag \\
&+&
\frac{1}{2 \det G_p}
\frac{\bar{\varphi}^{-}_i}{2} (\frac{2\pi}{\beta}(p-q=0))
\delta \varphi^{+}_i (\frac{2\pi}{\beta}(p-q))
\Bigg]
 \\
B_i(p,q) &=&
\Bigg[
-\frac{1}{2 \det G_p}\left[\widetilde{\omega}_{F,p} - 
\frac{\bar{\varphi}^z_i}{2} (\frac{2\pi}{\beta}(p-q=0) )
\right] \delta \varphi^{-}_i (\frac{2\pi}{\beta}(p-q)) 
\notag \\
&-&
\frac{1}{2 \det G_p}
\frac{\bar{\varphi}^{-}_i}{2} (\frac{2\pi}{\beta}(p-q=0))
\delta \varphi^z_i (\frac{2\pi}{\beta}(p-q))
\Bigg]
 \\
C_i(p,q) &=&
\Bigg[
\frac{1}{2 \det G_p} 
\frac{\bar{\varphi}^{+}_i}{2} (\frac{2\pi}{\beta}(p-q=0) )
\delta \varphi^z_i (\frac{2\pi}{\beta}(p-q)) 
\notag \\
&-&
\frac{1}{2 \det G_p}
\left[\widetilde{\omega}_{F,p} + 
\frac{\bar{\varphi}^z_i}{2} (\frac{2\pi}{\beta}(p-q=0) )
\right] \delta \varphi^{+}_i (\frac{2\pi}{\beta}(p-q))
\Bigg]
 \\
D_i(p,q) &=&
\Bigg[
\frac{1}{2 \det G_p} 
\frac{\bar{\varphi}^{+}_i}{2} (\frac{2\pi}{\beta}(p-q=0) )
\delta \varphi^{-}_i (\frac{2\pi}{\beta}(p-q)) 
\notag \\
&-&
\frac{1}{2 \det G_p}
\left[\widetilde{\omega}_{F,p} +
\frac{\bar{\varphi}^z_i}{2} (\frac{2\pi}{\beta}(p-q=0) )
\right] \delta \varphi^z_i (\frac{2\pi}{\beta}(p-q))
\Bigg]
\end{eqnarray}

\noindent
The first term in the sum of $\ln \det \beta M$ reads

\begin{eqnarray}
Tr \left[ G_0 M_1 \right] &=&
\underset{i}{\sum} \underset{p}{\sum}
\left[ A_i (p,p) + D_i(p,p) \right]
\notag \\
&=& \underset{i}{\sum} \underset{p}{\sum} \frac{1}{2 \det G_p}
\left[\bar{\varphi}^z_i(0) \delta \varphi^z_i(0) + \frac{1}{2} \left( 
\bar{\varphi}^{+}_i(0) \delta \varphi^{-}_i(0) +
\bar{\varphi}^{-}_i(0) \delta \varphi^{+}_i(0) \right) \right]
\notag \\
&=& \underset{i}{\sum} \left( \underset{p}{\sum} \frac{1}{2 \det G_p} \right) 
\vec{\bar{\varphi}}_i(0) \vec{\delta \varphi}_i(0)
\end{eqnarray}

\noindent
The value of $\left( \underset{p}{\sum} \frac{1}{2 \det G_p} \right)$ can
be found with the help of the sum relation in \cite{Gradshteyn} and leads to

\begin{eqnarray}
Tr \left[ G_0 M_1 \right] =
-\frac{\beta}{2} \underset{i}{\sum} \tanh \left(\frac{\beta}{2}
\|\vec{\bar{\varphi}}_i(0) \| \right) 
\frac{\vec{\bar{\varphi}}_i}{\bar{\varphi}_i(0)}.\vec{\delta \varphi}_i(0)
\label{FirstTerm}
\end{eqnarray}

\noindent
The first term in order of the fluctuations $\vec{\delta \varphi}$ of 
the auxiliary field action $S_0 \left[ \varphi \right]$ in equation 
\eqref{EffectiveAction2} reads

\begin{eqnarray}
\frac{\delta S_0 \left[ \varphi \right]}{\vec{\delta \varphi} }
\Bigg{\vert}_{\vec{\delta \varphi}=0} 
 &=&
\frac{\delta}{\vec{\delta \varphi}} \Bigg[
\frac{\beta}{2} \underset{\omega_B}{\sum} \underset{i,j}{\sum}
J_{i j}^{-1} \left[\left( \vec{\bar{\varphi}}_i - \vec{B}_i \right) 
\delta(\omega_B=0) + \vec{\delta \varphi}_i(-\omega_B) \right]
.
\notag \\
&&
\left[ 
\left( \vec{\bar{\varphi}}_j - \vec{B}_j \right) 
\delta(\omega_B=0) + \vec{\delta \varphi}_j(\omega_B) \right]
\Bigg] \Bigg{\vert}_{\vec{\delta \varphi}=0}
\notag \\
 &=& 
\frac{\beta}{2} \underset{i,j}{\sum} J_{i j}^{-1}
\left[\left(\vec{\bar{\varphi}}_i - \vec{B}_i \right) \vec{\delta \varphi}_i(0)
+ \left(\vec{\bar{\varphi}}_j - \vec{B}_j \right) \vec{\delta \varphi}_j(0) 
\right]
\end{eqnarray}

\noindent
Adding the first order term in the fluctuations $\delta
\varphi$ of $S_0 \left[ \varphi \right]$ in \eqref{FirstTerm} leads to

\begin{eqnarray}
\frac{\delta S_{eff}} {\delta \vec{\varphi}}
{\Big\vert}_{\left[\vec{\bar{\varphi}}\right]} \vec{\delta \varphi}
&=&
\frac{\delta S_0 \left[ \varphi \right]}{\vec{\delta \varphi} }
\vec{\delta \varphi}
\notag \\
&=&
\frac{\beta}{2} \underset{i}{\sum}
\left[ 2 \underset{j}{\sum} J_{i j}^{-1} \left( \vec{\bar{\varphi}}_j -
\vec{B}_j \right) - \frac{\vec{\bar{\varphi}}_i}{\| \vec{\bar{\varphi}_i} \|}
\tanh \left(\frac{\beta}{2} \| \vec{\bar{\varphi}}_i \| \right)
\right]. \vec{\delta \varphi}_i (\omega_B=0) = 0
\notag \\
\label{AppendixMeanFieldEquation}
\end{eqnarray}

\noindent
From \eqref{AppendixMeanFieldEquation} one obtains directly the mean-field 
equation of the Hubbard-Stratonovich equation \eqref{MeanFieldEquation}.

\section{Second order fluctuation contributions}

The term $n=2$ in equation \eqref{EffectiveAction2} is given by

\begin{eqnarray}
Tr \left[ G_0 M_1 G_0 M_1 \right] 
&=&
\underset{i}{\sum} \underset{p,q}{\sum} \Big[
A_i (p,q) A_i(q,p) + B_i(p,q) C_i(q,p) 
 \notag \\
&+& C_i(p,q) B_i(q,p) + D_i(p,q) D_i(q,p) \Big]
 \notag \\
&=&
\underset{i}{\sum} \underset{\omega_B}{\sum}
\frac{1}{4} \left[ \left( \bar{\varphi}^{+}/2 \right)^2 
\mathcal{K}_{i,\omega_B} \right]
{\delta \varphi}^{-}_i (-\omega_B)  {\delta \varphi}^{-}_i (\omega_B)
\notag \\
&+&
\frac{1}{4} \left[
\left( \bar{\varphi}^{-}/2 \right)^2 \mathcal{K}_{i,\omega_B} \right]
{\delta \varphi}^{+}_i (-\omega_B)  {\delta \varphi}^{+}_i (\omega_B)
\notag \\
&+&
\frac{1}{2} \left[\frac{1}{4}\left( (\bar{\varphi}^z_i)^2
- \bar{\varphi}^{+}_i \bar{\varphi}^{-}_i \right) \mathcal{K}_{i,\omega_B}
- \mathcal{I}_{i,\omega_B} \right] 
{\delta \varphi}^z_i (-\omega_B) .{\delta \varphi}^z_i (\omega_B)
\notag \\
&-&
\frac{1}{2} \left[ \frac{\bar{\varphi}^z_i}{2}
\left( \frac{\bar{\varphi}^z_i}{2} + i \omega_B \right)
\mathcal{K}_{i,\omega_B} + \mathcal{I}_{i,\omega_B} \right] 
{\delta \varphi}^{-}_i (-\omega_B) {\delta \varphi}^{+}_i (\omega_B)
\notag \\
&+&
\frac{1}{4} \left[
\bar{\varphi}^{-}_i \left(\bar{\varphi}^z_i + i \omega_B \right)
\mathcal{K}_{i,\omega_B} \right]
{\delta \varphi}^z_i(-\omega_B) {\delta \varphi}^{+}_i(\omega_B)
\notag \\
&+&
\frac{1}{4} \left[
\bar{\varphi}^{+}_i \left(\bar{\varphi}^z_i - i \omega_B \right)
\mathcal{K}_{i,\omega_B} \right]
{\delta \varphi}^z_i(-\omega_B) {\delta \varphi}^{-}_i(\omega_B)
\end{eqnarray}

\noindent
where $\omega_B = \widetilde{\omega}_{F,p} - \widetilde{\omega}_{F,q} = 
\frac{2 \pi (p-q)}{\beta}$
and we define $\mathcal{K}_{i,\omega_B}$ and $\mathcal{I}_{i,\omega_B}$ by

\begin{eqnarray}
\mathcal{K}_{i,\omega_B} &=&
\underset{\omega_{F,p}}{\sum} \frac{1}{\det G_{\omega_{F,p}} 
. \det G_{\omega_{F,p} + \omega_B}}
\notag \\
&=&
\begin{cases}
- \frac{1}{\bar{\varphi}} \frac{d}{d \bar{\varphi}_i} 
\left[\frac{\beta}{\bar{\varphi}_i} \tanh \frac{\beta}{2} \bar{\varphi}_i 
\right] 
\quad \text{, when $\omega_B = 0$,}
 \\
\frac{2 \beta}{\bar{\varphi}_i({\bar{\varphi}_i)}^2 + \omega_B^2} 
\tanh \frac{\beta}{2} \bar{\varphi}_i
\quad \text{, when $\omega_B \not= 0$.}
\end{cases}
\end{eqnarray}

\noindent
and

\begin{eqnarray}
\mathcal{I}_{i,\omega_B} &=&
\underset{\omega_{F,p}}{\sum} \frac{\omega_{F,p}(\omega_{F,p}+ \omega_B)}
{\det G_{\omega_{F,p}} . \det G_{\omega_{F,p} + \omega_B}}
\notag \\
&=&
\begin{cases}
\frac{\beta}{2 \bar{\varphi_i}} \tanh \frac{\beta}{2} \bar{\varphi}_i
+ \frac{\beta^2}{4} \tanh^{'} \frac{\beta}{2} \bar{\varphi}_i
\quad \text{, when $\omega_B = 0$,}
 \\
\frac{\beta}{2 \bar{\varphi}_i} \tanh \frac{\beta}{2} \bar{\varphi}_i
- \left( \frac{\omega_B}{2} \right)^2 \mathcal{K}_{i,\omega_B}
\quad \text{, when $\omega_B \neq 0$.}
\end{cases}
\end{eqnarray}

\noindent
Gathering all terms of second order with respect to the fluctuations 
$\delta \varphi$ from \eqref{EffectiveAction2} one gets \eqref{delta2Seff} 
if the mean-field $\bar{\varphi}$ is of a N\'eel or ferromagnetic type 
($\bar{\varphi}^{+} = \bar{\varphi}^{-} = 0$ and $\bar{\varphi}^z \not= 0$).

\section{Derivation of the free energy with fluctuation contributions}

From relations \eqref{delta2Seff} one can define

\begin{eqnarray}
\text{\textcircled{1}} 
= \left( \frac{\beta}{4} \tanh^{'} \left( \frac{\beta}{2}
\bar{\varphi}_i^z \right)\right)
\end{eqnarray}

\noindent
and

\begin{eqnarray}
\text{\textcircled{2}} = \left( \frac{1}{2} \frac{\tanh \left(\frac{\beta}{2}
\bar{\varphi}_i^z \right)}{\bar{\varphi}_i^z - i \omega_B}  \right)
\end{eqnarray}

\noindent
Using relations \eqref{Relation1} and \eqref{Relation2} with
$\bar{m}_i = (-1)^{i} \bar{m} + \Delta m B_i$,
\textcircled{1} and \textcircled{2} read

\begin{eqnarray}
\text{\textcircled{1}} &=& \left(\frac{\beta}{4} 
\tanh^{'} \left( \frac{\beta}{2} \bar{\varphi}_\alpha^z \right)\right)
\notag  \\
&=& 
\frac{\beta}{4}\left( 1 - 4 \bar{m}_i^2 \right)
\notag \\
&=&
\left[ 1a \right] + \left[ 1b \right] (-1)^{\vec{r}_i.\vec{\pi}}
\notag \\
&=&
\Bigg\{
\begin{array}{cc}
\frac{\beta}{4} \left(1 - 4 \left(\bar{m} + {\Delta \tilde{m}_0.B} 
\right)^2 \right) 
 &
\quad \text{ , when $B_i = (-1)^{\vec{r}_i.\vec{\pi}}$}
 \\
\frac{\beta}{4} \left( 1 - 4 \left(\bar{m}^2 + (\Delta \tilde{m}_0 . B)^2
\right) \right) - 2 \beta {\Delta m_{\chi 0}}. B \bar{m} 
(-1)^{\vec{r}_i.\vec{\pi}}
& \quad \text{ , when $B_i = B$.}
\end{array}
\notag \\
\end{eqnarray}

\noindent
and

\begin{eqnarray}
\text{\textcircled{2}} &=& \left( \frac{1}{2} 
\frac{\tanh \left( \frac{\beta}{2} \bar{\varphi}_\alpha^z \right) }
{\bar{\varphi}_\alpha^z - i \omega_B } \right)
\notag \\
&=& \left[ 2a \right]_{\omega_B} + (-1)^{\vec{r}_i.\vec{\pi}} 
\left[ 2b \right]_{\omega_B}
\notag
\end{eqnarray}

\noindent
where

\begin{eqnarray}
\left[ 2a \right]_{\omega_B} &=&
\frac{(\bar{m}+\Delta \tilde{m}_0.B).(B+2 D |J|(\bar{m}+\Delta \tilde{m}_0
.B))}
{\left[(B+2 D |J|(\bar{m}+\Delta \tilde{m}_0.B))^2 + {\omega_B}^2 \right]}
 \\
\left[ 2b \right]_{\omega_B} &=&
\frac{i \omega (\bar{m}+\Delta \tilde{m}_0.B)}
{\left[(B+2 D |J|(\bar{m}+\Delta \tilde{m}_0.B))^2 + {\omega_B}^2 \right]}
\end{eqnarray}

\noindent
if $B_i = (-1)^{\vec{r}_i.\vec{\pi}}$ and

\begin{eqnarray}
\left[ 2a \right]_{\omega_B} &=&
\frac{{\Delta m_{\chi 0}.B} \left(B - 2 |J|B.{\Delta m_{\chi 0}}
-i \omega_B \right)-2|J| \bar{m}^2}{\left(B - 2|J|{\Delta m_{\chi 0}}.B
\right)^2 - \left(2|J|\bar{m} \right)^2}
 \\
\left[ 2b \right]_{\omega_B} &=&
\frac{\bar{m} \left( B - 2 |J|B.{\Delta m_{\chi 0}}
-i \omega_B \right) -2|J|\bar{m}.{\Delta m_{\chi 0}}.B}
{{\left(B - 2|J|{\Delta m_{\chi 0}}.B
\right)^2 - \left(2|J|\bar{m} \right)^2}}
\end{eqnarray}

\noindent
if $B_i = B$.
After these developments in equations \eqref{delta2Seff}, the use of
a Fourier transformation and integration over the Hubbard-Stratonovich
auxiliary field $\vec{\delta\varphi}$ equations \eqref{PartitionFunctionZZ} 
and \eqref{PartitionFunctionPM} come out as

\begin{eqnarray}
\mathcal{Z}_{zz} &=& \underset{\vec{k} \in SBZ}{\prod}
\det \left[
\begin{array}{cc}
\left(1 - \left[ 1a \right] J(\vec{k}) \right) 
&
-\left[ 1b \right] J(\vec{k}+\vec{\pi})
\\
-\left[ 1b \right] J(\vec{k})
&
\left(1 - \left[ 1a \right] J(\vec{k}+\vec{\pi}) \right)
\end{array}
\right]^{-1/2}
\label{AppendixZzz}
 \\
\mathcal{Z}_{+-} &=&
\underset{\omega_B}{\prod} \underset{\vec{k} \in SBZ}{\prod}
\det \left[
\begin{array}{cc}
\left(1 - \left[ 2a \right]_{\omega_B} J(\vec{k}) \right)
&
- \left[ 2b \right]_{\omega_B} J(\vec{k}+\vec{\pi})
\\
- \left[ 2b \right]_{\omega_B} J(\vec{k})
&
\left(1 - \left[ 2a \right]_{\omega_B} J(\vec{k}+\vec{\pi}) \right)
\end{array}
\right]^{-1}
\label{AppendixZpm}
\end{eqnarray}

\noindent
Here $J(\vec{k}) \equiv - Z |J| \gamma_{\vec{k}}$ and $\vec{\pi}$ is the
Brioullin vector relative to the spin lattice as defined in section
\ref{SectionSpinWave}. Then free energy components $\mathcal{F}_{MF}$, 
$\mathcal{F}_{zz}$ and $\mathcal{F}_{+-}$ are given by

\begin{eqnarray}
\mathcal{F}_{MF} &=&
\mathcal{N} D |J| \left( \bar{m}+{\Delta \tilde{m}_0}.B \right)^2
- \frac{\mathcal{N}}{\beta} \ln \cosh \left( \frac{\beta}{2} 
\left[ B+2 D |J|(\bar{m}+{\Delta \tilde{m}_0.B})\right] \right)
\notag \\
\label{AppendixFmfMagnetization}
\end{eqnarray}

\begin{eqnarray}
\mathcal{F}_{zz} &=&
\frac{1}{2\beta} \underset{\vec{k} \in SBZ}{\sum}
\ln \left[ 1 - \left( \frac{\beta D |J| \gamma_{\vec{k}} }{2} \right)^2
\left[ 1- 4(\bar{m}+{\Delta \tilde{m}_0.B})^2 \right]^2 \right]
\label{AppendixFzzMagnetization}
\end{eqnarray}

\begin{align}
\mathcal{F}_{+-} =&
\frac{2}{\beta} \underset{\vec{k} \in SBZ}{\sum}
\ln \Bigg(
\notag \\
&
\frac{\sinh \left(\frac{\beta}{2}
\left(\left[B + 2 D |J|(\bar{m}+{\Delta \tilde{m}_0.B})\right]^2
- \left[2 D |J| \gamma_{\vec{k}} (\bar{m}+{\Delta \tilde{m}_0.B})\right]^2 
\right)^{1/2} \right)}
{\sinh \left(\frac{\beta}{2} 
\left[ B+2 D |J|(\bar{m}+{\Delta \tilde{m}_0.B})\right] \right)}
\Bigg)
\notag \\
\label{AppendixFpmMagnetization}
\end{align}

\noindent
if the magnetic field $B_i$ is equal to $(-1)^{\vec{r}_i.\vec{\pi}} B$ and

\begin{align}
\mathcal{F}_{MF} =&
\mathcal{N} D |J| \left( \bar{m}^2-({\Delta m_{\chi 0}}.B)^2 \right)
\notag \\
&- \frac{\mathcal{N}}{2 \beta} \ln \Bigg( 
\cosh \left[ \frac{\beta}{2} \left((1-2D|J| {\Delta m_{\chi 0}}).B 
+ 2 D |J| \bar{m} \right) \right]
\notag \\
&. \cosh \left[ \frac{\beta}{2} \left((1-2D|J| {\Delta m_{\chi 0}}).B 
- 2 D |J| \bar{m} \right)
\right] \Bigg)
\notag \\
\label{AppendixFmfSusceptibility}
\end{align}

\begin{eqnarray}
\mathcal{F}_{zz} &=&
\frac{1}{2 \beta} \underset{\vec{k} \in SBZ}{\sum}
\ln \left[ 1 -  
\left( \frac{\beta D |J| \gamma_{\vec{k}}}{2} \right)^2
\left[ 1 - 4 (\bar{m} - {\Delta m_{\chi 0}}.B)^2\right]
\left[ 1 - 4 (\bar{m} + {\Delta m_{\chi 0}}.B)^2\right]
\right]
\notag \\
\label{AppendixFzzSusceptibility}
\end{eqnarray}

\begin{align}
\mathcal{F}_{+-} =&
\frac{1}{\beta} \underset{\vec{k} \in SBZ}{\sum}
\ln \Bigg[
\notag \\
&
\frac{
\sinh^2 \left(\beta D |J| \bar{m} \sqrt{ 
1- \gamma_{\vec{k}}^2 
\left[1 - (\frac{{\Delta m_{\chi 0}}.B}{\bar{m}})^2 \right] 
} \right)
- \sinh^2 \frac{\beta}{2}((1-2D|J| {\Delta m_{\chi 0}}).B)
}
{
\sinh^2 \beta D |J| \bar{m}
- \sinh^2 \frac{\beta}{2}((1-2D|J| {\Delta m_{\chi 0}}).B)
}
\Bigg]
\notag \\
\label{AppendixFpmSusceptibility}
\end{align}

\noindent
if the magnetic field $B_i$ is unform and equal to $B$. Here $\mathcal{N}$
is the number of lattice sites (sublattices $A$ and $B$).
The magnetization of the spin system is obtained by means of
\eqref{AppendixFmfMagnetization}, \eqref{AppendixFzzMagnetization} and
\eqref{AppendixFpmMagnetization}. The magnetic spin susceptibility can be 
worked out with \eqref{AppendixFmfSusceptibility}, 
\eqref{AppendixFzzSusceptibility} and \eqref{AppendixFpmSusceptibility}.

\section{The free energy of the XXZ-model with a staggered magnetic field}

The free energy of the XXZ-model in a staggered magnetic field 
$B_i = (-1)^{i} B$
is given by the three contributions $\mathcal{F}_{MF}$, $\mathcal{F}_{zz}$ and
$\mathcal{F}_{+-}$ where

\begin{eqnarray}
\mathcal{F}_{MF} &=&
\mathcal{N} D |J(1+\delta)| \left( \bar{m}+\Delta \tilde{m}_{XXZ}.B \right)^2
\notag \\
&-& \frac{\mathcal{N}}{\beta} \ln \cosh \left(\frac{\beta}{2} 
\left[ B+2 D |J(1+\delta)|(\bar{m}+\Delta \tilde{m}_{XXZ}.B)\right] \right)
\notag \\
\end{eqnarray}

\begin{eqnarray}
\delta\mathcal{F}_{zz} =
\frac{1}{2\beta} \underset{\vec{k} \in SBZ}{\sum}
\ln \left[ 1 - \left( \frac{\beta D |J(1+\delta)| \gamma_{\vec{k}} }{2} 
\right)^2 \left[ 1- 4(\bar{m}+\Delta \tilde{m}_{XXZ}.B)^2 \right]^2 \right]
\notag \\
\end{eqnarray}

\begin{align}
\delta \mathcal{F}_{+-} =&
\frac{2}{\beta} \underset{\vec{k} \in SBZ}{\sum}
\ln \Bigg(
\notag \\
&
\frac{\sinh \left(\frac{\beta}{2}
\left(\left[B + 2 D |J(1+\delta)|(\bar{m}+\Delta \tilde{m}_{XXZ}.B)\right]^2
- \left[2 D |J| \gamma_{\vec{k}} (\bar{m}+\Delta \tilde{m}_{XXZ}.B)\right]^2 
\right)^{1/2} \right)}
{\sinh \left(\frac{\beta}{2} 
\left[ B+2 D |J(1+\delta)|(\bar{m}+\Delta \tilde{m}_{XXZ}.B)\right] \right)}
\Bigg)
\notag \\
\end{align}


\index{Bogoliubov transformation|(}

\chapter{Diagonalization of Mean-Field Hamiltonians
\label{AppendixDiagonalization}}

\section{Bogoliubov transformation on the N\'eel mean-field Hamiltonian
\label{AppendixDNeel}}

The N\'eel mean-field Hamiltonian \eqref{Chapter4MeanFieldHamiltonian2} 

\begin{eqnarray}
\mathcal{H}_{MF} &=& -\frac{1}{2} \underset{i,j}{\sum} 
J_{ij}^{-1} \left( \vec{\bar{\varphi}}_i - B_i \vec{e}_z \right)
.\left(\vec{\bar{\varphi}}_j - B_j \vec{e}_z \right)
\notag \\
&+& \underset{i}{\sum} \left( 
\begin{array}{cc}
f^{\dagger}_{i,\uparrow} & f^{\dagger}_{i,\downarrow}
\end{array}
 \right) 
\left[
\begin{array}{cc}
\left(- \mu + \frac{\bar{\varphi}_{i}^z}{2}\right) 
& \frac{\bar{\varphi}^{-}_i}{2} \\
\frac{\bar{\varphi}^{+}_i}{2} 
& -\left(\mu + \frac{\bar{\varphi}^z_i}{2} \right)
\end{array}
\right]
\left(
\begin{array}{c}
f_{i,\uparrow} \\
f_{i,\downarrow}
\end{array}
\right)
\notag \\
\end{eqnarray}

\noindent 
can be diagonalized by means of a Bogoliubov tranformation 
\cite{Bogoliubov-58}.
Introduce the linar combination of new creation and annihilation 
fermion operators $\beta^{\dagger}_{i,(\pm)}$ and $\beta_{i,(\pm)}$

\begin{eqnarray}
\left(
\begin{array}{c}
f_{i,\uparrow} \\
f_{i,\downarrow}
\end{array}
\right)
=
U
\left(
\begin{array}{c}
\beta_{i,(+)} \\
\beta_{i,(-)}
\end{array}
\right)
\label{AppendixDeq1}
\end{eqnarray}

\noindent
where the unitary matrix $U$ is define by

\begin{eqnarray}
U = 
\left[
\begin{array}{cc}
u_i & v_i^{*} \\
-v_i & u_i^{*}
\end{array}
\right]
\label{AppendixDeq3}
\end{eqnarray}

\noindent
Coefficients $\{ u_i \}$ and $\{ v_i \}$ have to verify $\det U = 1$ and
$U^{-1} = U^{\dagger}$ in such a way that in the transformation 
\eqref{AppendixDeq1} the creation and annihilation operator 
$\{ \beta^\dagger_{i,(\pm)} \}$ and $\{ \beta_{i,(\pm)} \}$ anticommute.
To diagonalize the Hamiltonian \eqref{Chapter4MeanFieldHamiltonian} one
should find the coefficients $\{ u_i \}$ and $\{ v_i \}$ for which

\begin{eqnarray}
U^{-1} \mathcal{H}_{MF} U &=& U^{\dagger} \mathcal{H}_{MF} U 
\notag \\
&=& 
-\frac{1}{2} \underset{i,j}{\sum} 
J_{ij}^{-1} \left( \vec{\bar{\varphi}}_i - B_i \vec{e}_z \right)
.\left(\vec{\bar{\varphi}}_j - B_j \vec{e}_z \right)
\notag \\
&+& \underset{i}{\sum}
\left(
\begin{array}{cc}
\beta^{\dagger}_{i,(+)} & \beta^{\dagger}_{i,(-)}
\end{array}
\right)
\left[
\begin{array}{cc}
\omega^{(PFP)}_{i,(+)} & 0 \\
0 & \omega^{(PFP)}_{i,(-)}
\end{array}
\right]
\left(
\begin{array}{c}
\beta_{i,(+)} \\
\beta_{i,(-)}
\end{array}
\right)
\notag \\
\end{eqnarray}

\noindent
By means of the transformation

\begin{eqnarray}
\begin{cases}
u_i = e^{i \alpha_i} \cos \theta_i \\
v_i = \sin \theta_i
\end{cases}
\end{eqnarray}

\noindent
where $\alpha_i$ and $\theta_i$ are angles, $U^{\dagger}\mathcal{H}_{MF}U$
is diagonal for

\begin{eqnarray}
\tan \alpha_i &=& -\frac{\bar{\varphi}_i^y}{\bar{\varphi}_i^x} 
\\
\tan 2 \theta_i &=& -\frac{\sqrt{\left(\bar{\varphi}_i^{x}\right)^2
+\left(\bar{\varphi}_i^y\right)^2 }}{\bar{\varphi}_i^z}
\end{eqnarray}

\noindent
And then

\begin{eqnarray}
U^{\dagger}\mathcal{H}_{MF}U = 
\left[
\begin{array}{cc}
\omega^{(PFP)}_{i,(+)} & 0 \\
0 & \omega^{(PFP)}_{i,(-)}
\end{array}
\right]
\end{eqnarray}

\noindent
with $\omega^{(PFP)}_{i,(\pm)} = \mu \pm \frac{\| \vec{\bar{\varphi}} \|}{2}$,
 ($PFP$ refers to the Popov and Fedotov procedure).

\section{Bogoliubov transformation on Diffuson mean-field ansatz
\label{AppendixDDiffuson}}

The mean-field Hamiltonian \eqref{Chapter4eq39} with the imaginary
chemical potential fixing exactly the number of spin per lattice site
given by

\begin{eqnarray}
\mathcal{H}_{MF}^{(PFP)} &=& \mathcal{N} z \frac{\Delta^2}{|J|}
+ \underset{\vec{k} \in SBZ}{\sum} \underset{\sigma}{\sum}
\left(
f^\dagger_{\vec{k},\sigma} \,
f^\dagger_{\vec{k}+\vec{\pi},\sigma}
\right)
\notag \\
&&\left[
\begin{array}{cc}
-\mu + \Delta \cos \frac{\pi}{4} z \gamma_{k_x,k_y} &
-i \Delta \sin \frac{\pi}{4} z \gamma_{k_x,k_y+\pi} \\
+i \Delta \sin \frac{\pi}{4} z \gamma_{k_x,k_y+\pi} &
-\mu - \Delta \cos \frac{\pi}{4} z \gamma_{k_x,k_y} 
\end{array}
\right]
\left(
\begin{array}{c}
f_{\vec{k},\sigma} \\
f_{\vec{k}+\vec{\pi},\sigma}
\end{array}
\right)
\end{eqnarray}

\noindent
can be diagonalized by the transformation \eqref{AppendixDeq3}

\begin{eqnarray}
\left(
\begin{array}{c}
f_{\vec{k},\sigma} \\
f_{\vec{k}+\vec{\pi},\sigma}
\end{array}
\right)
=
\left[
\begin{array}{cc}
u_{\vec{k}} & v^{*}_{\vec{k}} \\
- v_{\vec{k}} & u^{*}_{\vec{k}}
\end{array}
\right]
\left(
\begin{array}{c}
\beta_{(+),\vec{k},\sigma} \\
\beta_{(-),\vec{k},\sigma}
\end{array}
\right)
\end{eqnarray}

\noindent
The transformation matrix

\begin{eqnarray}
U =
\left[
\begin{array}{cc}
u_{\vec{k}} & v^{*}_{\vec{k}} \\
- v_{\vec{k}} & u^{*}_{\vec{k}}
\end{array}
\right]
\end{eqnarray}

\noindent
is unitary and

\begin{eqnarray}
|u_{\vec{k}}|^2 + |v_{\vec{k}}|^2 = 1
\end{eqnarray}

\noindent
so that $\beta^{\dagger}$ and $\beta$ are fermion creation and annihilation 
operator. Following the section \ref{AppendixDNeel} we define

\begin{eqnarray}
u_{\vec{k}} &=& e^{i \varphi_{\vec{k}}} \cos \theta_{\vec{k}} \\
v_{\vec{k}} &=& \sin \theta_{\vec{k}}
\end{eqnarray}

\noindent
The mean-field Hamiltonian is diagonal if

\begin{eqnarray}
\varphi_{\vec{k}} &=& \frac{\pi}{2} \\
\tan 2 \theta_{\vec{k}} &=& \frac{\gamma_{k_x,k_x+\pi}}{\gamma_{k_x,k_y}}
\end{eqnarray}

\noindent
where $\gamma_{\vec{k}} = \frac{1}{2} \left(\cos k_x + \cos k_y \right)$ and
reads

\begin{eqnarray}
\mathcal{H}_{MF}^{(PFP)} = \mathcal{N} \frac{z \Delta^2}{|J|}
+ \underset{\vec{k} \in SBZ}{\sum} \underset{\sigma}{\sum}
\left[ \omega^{(PFP)}_{(+),\vec{k},\sigma} 
\beta^{\dagger}_{(+),\vec{k},\sigma} \beta_{(+),\vec{k},\sigma}
+ \omega^{(PFP)}_{(-),\vec{k},\sigma} 
\beta^{\dagger}_{(-),\vec{k},\sigma} \beta_{(-),\vec{k},\sigma}
\right]
\notag \\
\end{eqnarray}

\noindent
with

\begin{eqnarray}
\omega^{(PFP)}_{(+),\vec{k},\sigma} &=&
- \mu + 2 \Delta \sqrt{\cos^2 k_x + \cos^2 k_y}
\\
\omega^{(PFP)}_{(-),\vec{k},\sigma} &=&
- \mu - 2 \Delta \sqrt{\cos^2 k_x + \cos^2 k_y}
\end{eqnarray}

\section{Bogoliubov transformation on Cooperon mean-field ansatz
\label{AppendixDCooperon}}

The mean-field Hamiltonian \eqref{Chapter4eq60} describing the cooperon 
mean-field ansatz with exact occupation parameter $\mu$ is given by

\begin{eqnarray}
\mathcal{H}_{MF}^{(PFP)} =
\mathcal{N} \frac{z \Gamma^2}{|J|}
- \underset{\vec{k} \in BZ,\sigma}{\sum}
\left[ \mu f^{\dagger}_{\vec{k},\sigma} f_{\vec{k},\sigma}
+ \sigma \frac{z \Gamma \gamma_{\vec{k}} }{2}
\left( f_{\vec{k},\sigma} f_{-\vec{k},-\sigma} 
- f^{\dagger}_{\vec{k},\sigma} f^{\dagger}_{-\vec{k},-\sigma}
\right) \right]
\end{eqnarray}

\noindent
and rewritten as

\begin{eqnarray}
\mathcal{H}_{MF}^{(PFP)} =
\mathcal{N} \frac{z \Gamma^2}{|J|} - \mathcal{N} \mu
+ \underset{\vec{k} \in BZ}{\sum}
\left(
\begin{array}{cc}
f^{\dagger}_{\vec{k},\sigma} &
f_{-\vec{k},-\sigma}
\end{array}
\right)
\left[
\begin{array}{cc}
H_{11,\vec{k},\sigma} & H_{12,\vec{k},\sigma} \\
H_{21,\vec{k},\sigma} & H_{22,\vec{k},\sigma}
\end{array}
\right]
\left(
\begin{array}{c}
f_{\vec{k},\sigma} \\
f^{\dagger}_{-\vec{k},-\sigma}
\end{array}
\right)
\notag \\
\end{eqnarray}

\noindent
where 

\begin{eqnarray}
H_{11,\vec{k},\sigma} &=& \frac{\mu}{2} \\
H_{12,\vec{k},\sigma} &=& -\sigma \frac{z \Gamma \gamma_{\vec{k}}}{2}  \\
H_{21,\vec{k},\sigma} &=& -\sigma \frac{z \Gamma \gamma_{\vec{k}}}{2}  \\
H_{22,\vec{k},\sigma} &=& -\frac{\mu}{2}
\end{eqnarray}

\noindent
Following the same step as in section \ref{AppendixDDiffuson} one define
the Bogoliubov transformation

\begin{eqnarray}
\left(
\begin{array}{c}
f_{\vec{k},\sigma} \\
f^{\dagger}_{-\vec{k},-\sigma}
\end{array}
\right)
=
U
\left(
\begin{array}{c}
\beta_{\vec{k},\sigma} \\
\beta^{\dagger}_{-\vec{k},-\sigma}
\end{array}
\right)
\end{eqnarray}

\noindent
where $U$ is the unitary matrix

\begin{eqnarray}
U =
\left[
\begin{array}{cc}
u_{\vec{k}} & v^{*}_{\vec{k}} \\
- v_{\vec{k}} & u^{*}_{\vec{k}}
\end{array}
\right]
\end{eqnarray}

\noindent
The coefficients verify $|u_{\vec{k}}|^2 + |v_{\vec{k}}|^2 = 1$ and

\begin{eqnarray}
u_{\vec{k}} &=& e^{i \varphi_{\vec{k}}} \cos \theta_{\vec{k}}
\notag \\
v_{\vec{k}} &=& \sin \theta_{\vec{k}}
\label{AppendixDeq28}
\end{eqnarray}

\noindent
In order to diagonalize the mean-field Hamiltonian one looks for angles
$\theta_{\vec{k}}$ and $\varphi_{\vec{k}}$ such that 

\begin{eqnarray}
U^{-1} \mathcal{H}_{MF}^{(PFP)} U = 
\left(
\begin{array}{cc}
\beta^{\dagger}_{\vec{k},\sigma} & 
\beta_{-\vec{k},-\sigma}
\end{array}
\right)
\left[
\begin{array}{cc}
\omega^{(PFP)}_{(+),\vec{k},\sigma} & 0 \\
0 & \omega^{(PFP)}_{(-),\vec{k},\sigma}
\end{array}
\right]
\left(
\begin{array}{c}
\beta_{\vec{k},\sigma} \\
\beta^{\dagger}_{-\vec{k},-\sigma}
\end{array}
\right)
\end{eqnarray}

\noindent
Developing $U^{-1} \mathcal{H}_{MF}^{(PFP)} U$ in detail one gets the
system of equations

\begin{eqnarray}
\begin{cases}
\text{\textcircled{1}} : \quad
H_{11} u^{*}_{\vec{k}} u_{\vec{k}} - H_{12} u^{*}_{\vec{k}} v_{\vec{k}}
- H_{21} u_{\vec{k}} v^{*}_{\vec{k}} + H_{22} v^{*}_{\vec{k}} v_{\vec{k}}
= \omega^{(PFP)}_{(+),\vec{k},\sigma}
 \\
\text{\textcircled{2}} : \quad
H_{11} v^{*}_{\vec{k}} v_{\vec{k}} + H_{12} u^{*}_{\vec{k}} v_{\vec{k}}
+ H_{21} u_{\vec{k}} v^{*}_{\vec{k}} + H_{22} v^{*}_{\vec{k}} v_{\vec{k}}
= \omega^{(PFP)}_{(-),\vec{k},\sigma}
 \\
\text{\textcircled{3}} : \quad
H_{11} u^{*}_{\vec{k}} v^{*}_{\vec{k}} + H_{12} (u^{*}_{\vec{k}})^2
- H_{21} (v^{*}_{\vec{k}})^2 - H_{22} v^{*}_{\vec{k}} u^{*}_{\vec{k}}
= 0
 \\
\text{\textcircled{4}} : \quad
H_{11} u_{\vec{k}} v_{\vec{k}} - H_{12} (v_{\vec{k}})^2
+ H_{21} (u_{\vec{k}})^2 - H_{22} v_{\vec{k}} u_{\vec{k}}
= 0
\end{cases}
\end{eqnarray}

\noindent
Introducing \eqref{AppendixDeq28} 
in \textcircled{3} and \textcircled{4} leads to

\begin{eqnarray}
\text{\textcircled{3}} &:& \quad
H_{11} e^{-i \varphi_{\vec{k}}}.2 \cos \theta_{\vec{k}} \sin \theta_{\vec{k}}
+
H_{12} \left( e^{-2i \varphi_{\vec{k}}}\cos^2 \theta_{\vec{k}} 
- \sin^2 \theta_{\vec{k}} \right) = 0
\\
\text{\textcircled{4}} &:& \quad
H_{11} e^{i \varphi_{\vec{k}}}. 2 \cos \theta_{\vec{k}} \sin \theta_{\vec{k}}
+ H_{12} \left( e^{2i \varphi_{\vec{k}}}\cos^2 \theta_{\vec{k}} 
- \sin^2 \theta_{\vec{k}} \right) = 0
\end{eqnarray}

\noindent
Separating real and imaginary part one gets

\begin{eqnarray}
\begin{array}{ccc}
 & \text{$\mathcal{R}$e part} & \text{$\mathcal{I}$m part} \\
\text{\textcircled{3}} :
& H_{12} \cos 2 \theta_{\vec{k}} \cos \varphi_{\vec{k}} = 0
& \text{and } \frac{\mathcal{I}m(\mu)}{2} \sin 2 \theta_{\vec{k}} 
- H_{12} \sin \varphi_{\vec{k}} = 0
 \\
\text{\textcircled{4}} :
& H_{12} \cos 2 \theta_{\vec{k}} \cos \varphi_{\vec{k}} = 0
& \text{and } \frac{\mathcal{I}m(\mu)}{2} \sin 2 \theta_{\vec{k}} 
+ H_{12} \sin \varphi_{\vec{k}} = 0
\end{array}
\end{eqnarray}

\noindent
Equation \textcircled{3} and \textcircled{4} cannot be verified simultaneously.
As a consequence there is no unitary matrix $U$ and hence no Bogoliubov 
transformation which can diagonalize the mean-field Hamiltonian
\eqref{Chapter4eq60}. The mean-field Hamiltonian of the cooperon does not
conserve the number of particles and hence gets in conflict with the Popov and
Fedotov procedure which fixes the number of particles strictly.

\index{Bogoliubov transformation|)}


\chapter{Derivation of the $QED_3$ action and the polarization function at
finite temperature \label{AppendixE}}

\section{Derivation of the Euclidean QED action in (2+1) dimensions 
\label{AppendixEAction}}

At low energy near the two independent points 
$\vec{k}= \left( \pm \frac{\pi}{2},\frac{\pi}{2} \right) + \vec{k}$ 
of the Spin Brillouin Zone (see figure \ref{Chapter3Fig2}) the Hamiltonian 
\eqref{Chapter5eq1} can be rewritten in
the form

\begin{eqnarray}
H &=& \underset{\vec{k} \in SBZ}{\sum} \underset{\sigma}{\sum}
\left( f^\dagger_{1, \vec{k}, \sigma} \, f^\dagger_{1, \vec{k}
+\vec{\pi} , \sigma} \, f^\dagger_{2,\vec{k},\sigma} \, 
f^\dagger_{2,\vec{k}+\vec{\pi},\sigma} \right) 
\notag \\
&\Bigg\{&
-\mu \Unitmatrix + \sqrt{2} \Delta \left[ - k_x 
\left(
\begin{array}{cc}
\tau_3 & 0 \\
0 & \tau_3
\end{array}
\right) 
- k_y \Unitmatrix
\right]
\notag \\
&+&
\sqrt{2} \Delta \left[
-k_x
\left(
\begin{array}{cc}
\tau_2 & 0 \\
0 & -\tau_2
\end{array}
\right)
+ k_y . i \Unitmatrix
\right]
\Bigg\}
\left(
\begin{array}{c}
f_{1, \vec{k}, \sigma} \\
f_{1, \vec{k}+\vec{\pi}, \sigma} \\
f_{2,\vec{k},\sigma} \\
f_{2,\vec{k}+\vec{\pi},\sigma}
\end{array}
\right)
\end{eqnarray}

\noindent
with $\vec{\pi} = (\pi,\pi)$ the Brillouin vector. $\tau_1,\tau_2$ and 
$\tau_3$ are Pauli matrices

\begin{eqnarray}
\tau_1 =
\left( 
\begin{array}{cc}
0 & 1 \\
1 & 0
\end{array}
\right)
, \,
\tau_2 =
\left(
\begin{array}{cc}
0 & -i \\
i & 0
\end{array}
\right)
, \,
\tau_3 =
\left(
\begin{array}{cc}
1 & 0 \\
0 & -1
\end{array}
\right)
\end{eqnarray}

\noindent
$f^\dagger_{1,\vec{k},\sigma}$ and $f_{1,\vec{k},\sigma}$ 
($f^\dagger_{2,\vec{k},\sigma}$ and $f_{2,\vec{k},\sigma}$) 
are fermion creation and annihilation operators near the point 
$(\frac{\pi}{2},\frac{\pi}{2})$ ($(-\frac{\pi}{2},\frac{\pi}{2})$).

\noindent
Rotating the operators 

\begin{eqnarray}
\begin{cases}
f_{\vec{k}} = \frac{1}{\sqrt{2}} 
\left(f_{a,\vec{k}} + f_{b,\vec{k}} \right) \\
f_{\vec{k}+\vec{\pi}} = \frac{1}{\sqrt{2}} 
\left(f_{a,\vec{k}} - f_{b,\vec{k}} \right)
\end{cases}
\end{eqnarray}

\noindent
leads to

\begin{eqnarray}
H = \underset{\vec{k} \in SBZ}{\sum} \underset{\sigma}{\sum}
\psi^\dagger_{\vec{k} \sigma}
\Bigg[
- \mu \Unitmatrix 
+ \widetilde{\Delta} k_{+}
\left(
\begin{array}{cc}
\tau_1 & 0 \\
0 & \tau_2
\end{array}
\right)
- \widetilde{\Delta} k_{-}
\left(
\begin{array}{cc}
\tau_2 & 0 \\
0 & \tau_1
\end{array}
\right)
\Bigg] \psi_{\vec{k} \sigma}
\end{eqnarray}

\noindent
where $k_{+} = k_x + k_y$ and $k_{-} = k_x - k_y$, 
$\widetilde{\Delta} = 2 \Delta \cos \frac{\pi}{4}$ and

\begin{eqnarray}
\psi_{\vec{k} \sigma} = \left(
\begin{array}{c}
f_{1 a, \vec{k} \sigma} \\
f_{1 b, \vec{k} \sigma} \\
f_{2 a \vec{k} \sigma} \\
f_{2 b \vec{k} \sigma}
\end{array}
\right)
\end{eqnarray}

\noindent
In the Euclidean metric the action reads

\begin{eqnarray}
S_{E} &=&
\int_0^\tau d\tau \underset{\vec{k} \in SBZ}{\sum} 
\underset{\sigma}{\sum}
\psi^\dagger_{\vec{k} \sigma}
\left(
\begin{array}{cc}
\tau_3 & 0 \\
0 & \tau_3
\end{array}
\right)
\Bigg[
\left(\partial_\tau - \mu \right)
\left(
\begin{array}{cc}
\tau_3 & 0 \\
0 & \tau_3
\end{array}
\right)
\notag \\
&+& i \widetilde{\Delta} k_{+}
\left(
\begin{array}{cc}
\tau_2 & 0 \\
0 & -\tau_1
\end{array}
\right)
+ i \widetilde{\Delta} k_{-}
\left(
\begin{array}{cc}
\tau_1 & 0 \\
0 & -\tau_2
\end{array}
\right)
\Bigg] \psi_{\vec{k} \sigma}
\end{eqnarray}

\noindent
Through the unitary transformation

\begin{eqnarray}
\psi_{\vec{k} \sigma} \rightarrow 
\left(
\begin{array}{cc}
1 & 0 \\
0 & e^{i \frac{\pi}{4} \tau_3}
\end{array}
\right)
. \left(
\begin{array}{cc}
1 & 0 \\
0 & -\tau_1
\end{array}
\right) \psi_{\vec{k} \sigma}
\end{eqnarray}

\noindent
and writing $k_{+}=k_2$ and $k_{-} = k_1$

\begin{eqnarray}
S_{E} = \int_0^\beta d\tau \underset{\vec{k} \in SBZ}{\sum} 
\underset{\sigma}{\sum}
\bar{\psi}_{\vec{k} \sigma} \Big[ 
\gamma^0 \left( \partial_\tau - \mu \right)
+ \widetilde{\Delta} i k_1 \gamma^1 + \widetilde{\Delta} i k_2 \gamma^2 \Big]
\psi_{\vec{k} \sigma}
\end{eqnarray}

\noindent
where $\bar{\psi} = \psi^\dagger \gamma^0$ and the $\gamma$ matrices are 
defined as

\begin{eqnarray}
\gamma^0 =
\left(
\begin{array}{cc}
\tau_3 & 0 \\
0 & -\tau_3
\end{array}
\right)
 ,\,
\gamma^1 =
\left(
\begin{array}{cc}
\tau_1 & 0 \\
0 & -\tau_1
\end{array}
\right)
 ,\,
\gamma^2 =
\left(
\begin{array}{cc}
\tau_2 & 0 \\
0 & -\tau_2
\end{array}
\right)
\end{eqnarray}

\noindent
Using the inverse Fourier transform 
$\psi_{\vec{k} \sigma} = \int d^2\vec{r} \psi_{\vec{r} \sigma} 
e^{i \vec{k}.\vec{r}}$ the Euclidean action reads finally

\begin{eqnarray}
S_{E} = \int_0^\beta d\tau \int d^2\vec{r} \underset{\sigma}{\sum}
\bar{\psi}_{\vec{r} \sigma} \left[ \gamma^0 \left( \partial_\tau - \mu \right)
+ \widetilde{\Delta} \gamma^k \partial_k \right] \psi_{\vec{r} \sigma}
\end{eqnarray}

\noindent
With a ``light velocity'' $v_\mu = (1,\widetilde{\Delta},\widetilde{\Delta})$
leading to a curved metric as explained in subsection 
\ref{Chapter5Section2SubsectionGravitation}.

\index{Photon propagator!Polarization function|(}

\section{Derivation of the photon polarization function at finite temperature
\label{AppendixEPolarization}}

The Fourier transformation of the second term of the spinon action given by 
equation \eqref{SpinonGaugeAction} reads

\begin{eqnarray}
S_{E}^{(2)}\left[ \psi,a \right] &=& \underset{\sigma}{\sum}
\underset{\widetilde{\omega}_{F,1},\widetilde{\omega}_{F,2}}{\sum}
\int \frac{d^2\vec{k}_1}{(2 \pi)^2} \int \frac{d^2\vec{k}_2}{(2 \pi)^2}
\bar{\psi}_\sigma \left( k_1 \right) 
\notag \\
&\Bigg[& \frac{i \gamma^\mu k_\mu}
{(2\pi)^2 \beta} \delta \left(k_1 - k_2 \right) - \frac{i g \gamma^\mu 
a_\mu(k_1-k_2)}{(2 \pi)^2 \beta)^2} \Bigg] \psi_\sigma \left( k_2 \right)
\end{eqnarray}

\noindent
with $k = (\widetilde{\omega}_{F} \equiv \frac{2 \pi}{\beta}(n+1/4),\vec{k})$. 
Integrating over the fermion 
field $\psi$ and keeping the second order in the gauge field leads to the 
effective gauge action 

\begin{eqnarray}
S_{eff}^{(2)} \left[a\right] 
= \frac{1}{2} Tr \left[G_{F}. i g \gamma^\mu a_\mu \right]^2
\end{eqnarray}

\noindent
with $Tr = \underset{\omega^{'}_{F}}{\sum} \int \frac{d^2\vec{k^{'}}}{(2\pi)^2}
. \underset{\omega^{''}_{F}}{\sum} \int \frac{d^2\vec{k^{''}}}{(2\pi)^2} 
tr$. The trace $tr$ extends over the $\gamma$ matrix space, and
$G_{F}^{-1}(k_1-k_2) = i \frac{\gamma^\mu k_\mu}{(2\pi)^2\beta} 
\delta(k_1-k_2)$. The pure gauge action comes as

\begin{eqnarray}
S_{eff}^{(2)}\left[ a \right] &=& - g^2 \frac{1}{2 \beta} \underset{\sigma}{\sum} 
\underset{\omega_{F,1}}{\sum} \int \frac{d^2\vec{k}_1}{(2\pi)^2}.
\frac{1}{\beta} \underset{\omega_{F}^{''}}{\sum} \int \frac{d^2\vec{k^{''}}}
{(2\pi)^2}
\notag \\
&&
tr \Bigg[
\frac{\gamma^\rho k_{1,\rho}}{k_1^2}.\gamma^\mu a_\mu(k_1-k^{''}).
\frac{\gamma^{\eta} k^{''}_\eta}{{k^{''}}^2}.
\gamma^\nu a_\nu \left(-(k_1 - k^{''}) \right)
\Bigg]
\end{eqnarray}

\noindent
With the change of variables $k_1-k^{''}=q$ and $k_1 = k$

\begin{eqnarray}
S_{eff}^{(2)} =
- \frac{1}{2 \beta}  
\underset{\omega_{B}}{\sum} \int \frac{d^2\vec{q}}{(2\pi)^2}
a_\mu(-q) \Pi^{\mu \nu}(q) a_\nu(q). 
\end{eqnarray}

\noindent
where $q = (\omega_{B}=\frac{2\pi}{\beta} m, \vec{q})$ and
the polarization function is given by

\begin{eqnarray}
\Pi^{\mu \nu}(q) = \frac{g^2}{\beta} \underset{\sigma}{\sum}  
\underset{\omega_{F}}{\sum} \int \frac{d^2\vec{k}}{(2\pi)^2}
tr \left[
\frac{\gamma^\rho k_\rho}{k^2} . \gamma^\mu. \gamma^\eta
\frac{\left(k_\eta + q_\eta \right)}{\left(k + q \right)}. \gamma^\nu
\right]
\end{eqnarray}

\noindent
Then using the Feynmann identity $\frac{1}{a b} = \int_0^1 dx 
\frac{1}{\left(a x + (1-x)b \right)^2}$ $\Pi^{\mu \nu}$ can be rewritten as

\begin{eqnarray}
\Pi^{\mu \nu}(q) = \frac{g^2 }{\beta} \underset{\sigma}{\sum}  
\underset{\omega_F}{\sum} \int \frac{d^2\vec{k}}{(2\pi)^2}
tr \left[\gamma^\rho \gamma^\mu \gamma^\eta \gamma^\nu \right].
\int_0^1 dx \frac{k_\rho (k_\eta + q_\eta )}
{\left[(k+q)^2 x + (1-x) k^2 \right]^2}
\end{eqnarray}

By means of a change of variables
$k \rightarrow k^{'} - x q$ and using the identity 
$tr \left[\gamma^\rho \gamma^\mu \gamma^\eta \gamma^\nu \right]=
4.\left[\delta_{\rho \mu}.\delta_{\eta \nu}-
\delta_{\rho \eta}.\delta_{\mu \nu} + \delta_{\rho \nu}.\delta_{\mu \eta} 
\right]$ one obtains

\begin{eqnarray}
\Pi^{\mu \nu}(q) &=& 4 \alpha \int_0^1 dx \frac{1}{\beta}
\underset{\omega_F^{'}}{\sum} \int \frac{d^2\vec{k^{'}}}{(2\pi)^2}
\Bigg\{
\Big[
2 k^{'}_\mu k^{'}_\nu 
\notag \\
&+& (1-2x)(k^{'}_\mu q_\nu + q_\mu k^{'}_\nu)
-x(1-x) 2 q_\mu q_\nu 
\notag \\
&-& \delta_{\mu \nu} \underset{\eta}{\sum} \left( 
{k^{'}_\eta}^2 + (1-2x) k^{'}_\eta q_\eta - x(1-x) {q_\eta}^2
\right)
\notag \\
&& \Big]/\left[{k^{'}}^2 + x(1-x)q^2 \right]^2
\Bigg\}
\end{eqnarray}

\noindent
where $\alpha = g^2 \overset{N=2}{\underset{\sigma = 1}{\sum} 1 }$.
Following Dorey and Mavromatos \cite{DoreyMavromatos}, Lee \cite{Lee-98},
Aitchison \emph{et al.} \cite{Aitchison-92} and Gradshteyn \cite{Gradshteyn}
we define

\begin{eqnarray}
S_1 &=& \overset{\infty}{\underset{n = -\infty}{\sum}}
\frac{1}{\left[{k^{'}}^2 + x(1-x)q^2 \right]}
\notag \\
&=&
\frac{\beta^2}{4 \pi Y} \left[ \frac{\sinh (2\pi Y)}{\cosh(2\pi Y)
-\cos(2\pi X)} \right]
 \\
S_2 &=& \overset{\infty}{\underset{n = -\infty}{\sum}}
\frac{1}{\left[{k^{'}}^2 + x(1-x)q^2 \right]^2}
= -\frac{\beta^2}{8 \pi^2}.\frac{1}{Y} \frac{\partial S_1}{\partial Y}
 \\
S^{*} &=& \overset{\infty}{\underset{n = -\infty}{\sum}}
\frac{\omega^{'}_F}{\left[{k^{'}}^2 + x(1-x)q^2 \right]^2}
= - \frac{\beta}{4\pi} \frac{\partial S_1}{\partial X}
\end{eqnarray}

\noindent
with $X= x.m + 1/4$ and $Y = \frac{\beta}{2\pi} 
\sqrt{\vec{k^{'}}^2 + x(1-x)q^2}$.
The polarization can be expressed in terms of these sums and reads

\begin{eqnarray}
\Pi^{00} &=& \frac{\alpha}{\beta} \int_0^1 dx 
\int \frac{d^2\vec{k^{'}}}{(2\pi)^2}
\Big[ S_1 - 2\left[\vec{k^{'}}^2 + x(1-x)q_0^2 \right]S_2 
\notag \\
&+& (1-2x)q_0 S^{*} \Big]
\end{eqnarray}

\noindent
for the temporal component and

\begin{eqnarray}
\Pi^{ij} &=& \frac{\alpha}{\beta} \int_0^1 dx 
\int \frac{d^2\vec{k^{'}}}{(2\pi)^2}
\Big[ 2x(1-x)(q^2 \delta_{ij} - q_i q_j)S_2 
\notag \\
&-& (1-2x) q_0 \delta_{ij}. S^{*} \Big]
\end{eqnarray}

\noindent
for the spatial components.

Integrating over the fermion momentum $\vec{k^{'}}$ one gets

\begin{eqnarray}
\Pi^{00} &=& \widetilde{\Pi}_3 - \frac{q_0^2}{q^2} \widetilde{\Pi}_1 - 
\widetilde{\Pi}_2
\label{AppendixEeq24}
 \\
\Pi^{ij} &=& \widetilde{\Pi}_1 \left( \delta_{ij} - \frac{q_i q_j}{q^2} \right)
+ \widetilde{\Pi}_2 \delta_{ij}
\label{AppendixEeq25}
\end{eqnarray}

\noindent
where

\begin{eqnarray}
\widetilde{\Pi}_1 &=& \frac{\alpha q}{\pi} \int_0^1 dx \sqrt{x(1-x)}
\frac{\sinh \beta q \sqrt{x(1-x)} }{D(X,Y)}
 \\
\widetilde{\Pi}_2 &=& \frac{\alpha m}{\beta}
\int_0^1 dx (1-2x) \frac{\cos 2 \pi x m}{D(X,Y)}
 \\
\widetilde{\Pi}_3 &=& \frac{\alpha}{\pi \beta} \int_0^1 dx \ln 2 D(X,Y)
\end{eqnarray}

\noindent
and $D(X,Y)= \cosh \left( \beta q \sqrt{x(1-x)} \right) + \sin (2\pi x m) $.

\index{Photon propagator!Polarization function|)}



\listoffigures
\addcontentsline{toc}{chapter}{List of Figures}
\fancyhead{}
\fancyhead[LE,RO]{\textbf{\thepage}}
\fancyhead[RE,LO]{\textbf{List of Figures}}



\backmatter


\addcontentsline{toc}{chapter}{Bibliography}
\fancyhead{}
\fancyhead[LE,RO]{\textbf{\thepage}}
\fancyhead[RE,LO]{\textbf{Bibliography}}
\bibliographystyle{plain}
\bibliography{bibthez}


\newpage
\addcontentsline{toc}{chapter}{Index}
\fancyhead{}
\fancyhead[LE,RO]{\textbf{\thepage}}
\fancyhead[RE,LO]{\textbf{Index}}
\printindex

\end{document}